\documentclass[onecolumn]{aastex631}
\usepackage{url}
\usepackage{multirow}
\usepackage{booktabs}
\usepackage{amsmath}
\usepackage{natbib}
\usepackage{enumitem}

\newcommand{\STAB}[1]{\begin{tabular}{@{}c@{}}#1\end{tabular}}

\begin{document}

\title{COSINE (Cometary Object Study Investigating their Nature and Evolution) I. Project Overview and General Characteristics of Detected Comets}
\shorttitle{COSINE I. Comets in WISE/NEOWISE Mission Data}

\received{\today}
\revised{---}
\accepted{---}

\author[0000-0002-8122-3606]{Yuna G. Kwon}
\affiliation{California Institute of Technology/IPAC, Pasadena, CA, USA}

\author[0000-0003-1876-9988]{Dar W. Dahlen}
\affiliation{Technische Universit{\"a}t Braunschweig, Braunschweig, Germany}
\affiliation{California Institute of Technology/IPAC, Pasadena, CA, USA}

\author[0000-0003-2638-720X]{Joseph R. Masiero}
\affiliation{California Institute of Technology/IPAC, Pasadena, CA, USA}

\author[0000-0001-9542-0953]{James M. Bauer}
\affiliation{University of Maryland, College Park, MD, USA}

\author[0000-0003-1156-9721]{Yanga R. Fern{\'a}ndez}
\affiliation{University of Central Florida, Orlando, FL, USA}

\author[0000-0002-9347-8753]{Adeline Gicquel}
\affiliation{University of Maryland, College Park, MD, USA}

\author[0000-0002-4676-2196]{Yoonyoung Kim}
\affiliation{University of California, Los Angeles, CA, USA}

\author[0000-0002-5736-1857]{Jana Pittichov{\'a}}
\affiliation{Jet Propulsion Laboratory, California Institute of Technology, Pasadena, CA, USA}

\author[0000-0002-8532-9395]{Frank Masci}
\affiliation{California Institute of Technology/IPAC, Pasadena, CA, USA}

\author[0000-0002-0077-2305]{Roc M. Cutri}
\affiliation{California Institute of Technology/IPAC, Pasadena, CA, USA}

\author[0000-0002-7578-3885]{Amy K. Mainzer}
\affiliation{University of California, Los Angeles, CA, USA}

\correspondingauthor{Yuna G. Kwon}
\email{ynkwontop@gmail.com}

\begin{abstract}

We present the first results from the COSINE (Cometary Object Study Investigating their Nature and Evolution) project, based on a uniformly processed dataset of 484 comets observed over the full 15-year duration of the WISE/NEOWISE mission. This compilation includes 1,633 coadded images spanning 966 epochs with signal-to-noise ratios (S/N) greater than 4, representing the largest consistently analyzed infrared comet dataset obtained from a single instrument. Dynamical classification identifies 234 long-period (LPCs) and 250 short-period comets (SPCs), spanning heliocentric distances of 0.996--10.804 au. LPCs are statistically brighter than SPCs in the W1 (3.4 $\mu$m) and W2 (4.6 $\mu$m) bands at comparable heliocentric distances. Cometary activity peaks near perihelion, with SPCs exhibiting a pronounced post-perihelion asymmetry. Multi-epoch photometry reveals that SPCs show steeper brightening and fading slopes than LPCs. The observing geometry of WISE/NEOWISE -- constrained to a fixed $\sim$90$^\circ$ solar elongation from low-Earth orbit -- introduces systematic biases in the sampling of orientation angles for extended features. Collectively, the results reveal a continuous evolutionary gradient across comet populations, likely driven by accumulated solar heating and surface processing. This study establishes a foundation for subsequent COSINE analyses, which will separate nucleus and coma contributions and model dust dynamics to further probe cometary activity and evolution.
\end{abstract}

\keywords{comets: general --- Methods: observational, numerical --- techniques: photometric}

% 1. Introduction %%%%%%%%%%%%%%%%%%%%%%%%%%%%%%%%%%%%%%%%%%%%%%%%%%%%%%%%%%%%%%%%%%%%%%%%%%%%%%%%%%%%%%%%%%%%%%%%
\section{Introduction} \label{sec:intro}

Comets, composed of dust and volatile ice, are among the most primitive remnants of planetesimal formation in the early Solar System. As they migrate into the inner Solar System, solar heating drives the sublimation of surface and near-surface ices, releasing dust and producing observable activity. This activity is shaped by the nucleus's intrinsic properties, which encode the environmental conditions the comet has encountered since formation. Consequently, comets serve as accessible records of early Solar System processes, providing a direct link between present-day observations and the conditions that prevailed during planetary accretion.

Comets are broadly classified into two dynamical populations -- Long-Period Comets (LPCs) and Short-Period Comets (SPCs) -- based on their orbital characteristics and source reservoirs \citep{Levison1996}. SPCs predominantly originate from the scattered disk beyond Neptune \citep{Duncan1997,Levison1997,Nesvorny2017} and typically follow low-inclination orbits near the ecliptic plane. They are dynamically coupled to the giant planets and possess orbital periods of a few decades or less. In contrast, LPCs are sourced from the distant Oort cloud and spend the majority of their lifetimes in the outermost Solar System, often approaching the Sun only once or on timescales of thousands to millions of years. These rare inner Solar System passages render LPCs especially valuable for preserving primitive material. Although both populations are thought to have formed in largely overlapping regions of the protoplanetary disk, likely near the orbits of Uranus and Neptune (\citealt{Dones2015} and references therein), their subsequent dynamical evolution has diverged markedly. This divergence has shaped the present-day distribution of LPCs and SPCs and contributed to the large-scale architecture of the Solar System.

As comets orbit the Sun, solar irradiation progressively depletes surface and near-surface volatiles, leading to the formation of stratified volatile layers at increasing depths \citep{Prialnik2004}. Due to their more frequent perihelion passages, SPCs typically develop more thermally processed surface layers than LPCs. As a result, LPCs often exhibit higher activity levels (e.g., \citealt{Garcia2020}) and distinct coma structures such as jets and spirals, indicative of fresher, less thermally altered material near the surface \citep{KrishnaSwamy2010}. Dust expelled from subsurface layers during outbursts or fragmentation events, such as those seen in SPCs \citep{Yang2009,Ishiguro2010} and shortly after the \emph{Deep Impact} experiment on 9P/Tempel 1 \citep{Lisse2006,Hadamcik2007}, differs in character from dust released under nominal activity conditions in SPCs \citep{Hadamcik2009}. Polarimetric studies have further shown that the degree of linear polarization in cometary dust varies with perihelion distance, indicating a change in the characteristic size of the dust grains, reinforcing the influence of solar processing on dust properties \citep{Kolokolova2007,Kwon2021}.

Recent observations, however, have revealed substantial diversity within each dynamical class. Contrary to the traditional expectation that LPCs exhibit long dust and ion tails extending millions of kilometers, some LPCs display only low to moderate activity (e.g., \citealt{Licandro2019}), or appear dust-free, so-called \emph{Manx} comets \citep{Meech2016,Kwon2022c}, suggesting a broader range of formation environments and/or evolutionary paths among Oort cloud objects. Similarly, the deuterium-to-hydrogen (D/H) ratio in SPCs, a topic of ongoing debate \citep{Altwegg2015,Mandt2024,Biver2024}, may point to a wide spread in source regions. Additional evidence comes from observed diversity in organic parent species across both infrared and radio wavelengths, further supporting a wide spatial origin for cometary material (e.g., \citealt{Bergin2024,Biver2024}).

An era of data-driven Solar System science is unfolding with the advent of recently launched and upcoming large-scale surveys, including the \emph{Near-Earth Object Surveyor} (NEOS; \citealt{Mainzer2023}), the \emph{Legacy Survey of Space and Time} (LSST; \citealt{LSST2021}), and the \emph{Spectro-Photometer for the History of the Universe, Epoch of Reionization and Ices Explorer} (SPHEREx; \citealt{Crill2020}). At this pivotal juncture, a robust understanding of comet populations -- including both their common features and intrinsic diversity -- is essential for interpreting their role in Solar System evolution. Equally important is the development of scalable, reliable frameworks for analyzing the impending influx of observational data within a coherent scientific context.

To address these needs, we initiated the \emph{COSINE} project (Cometary Object Study Investigating their Nature and Evolution), focusing on primitive, comet-like bodies that actively exhibit, or preserve signatures of, volatile-driven activity. As \citet{Gehrels1974} noted, ``It is not a lack of observations but a lack of conclusive interpretations." In this spirit, we leverage publicly available archive data to connect observable features with underlying physical and compositional properties, aiming to place comets within the broader narrative of Solar System formation and dynamical evolution. Another goal of COSINE is to deliver a consistently processed, scientifically vetted dataset to support and enable future investigations by the comet research community. 

\begin{figure*}[htb]
\centering
\includegraphics[width=0.63\textwidth]{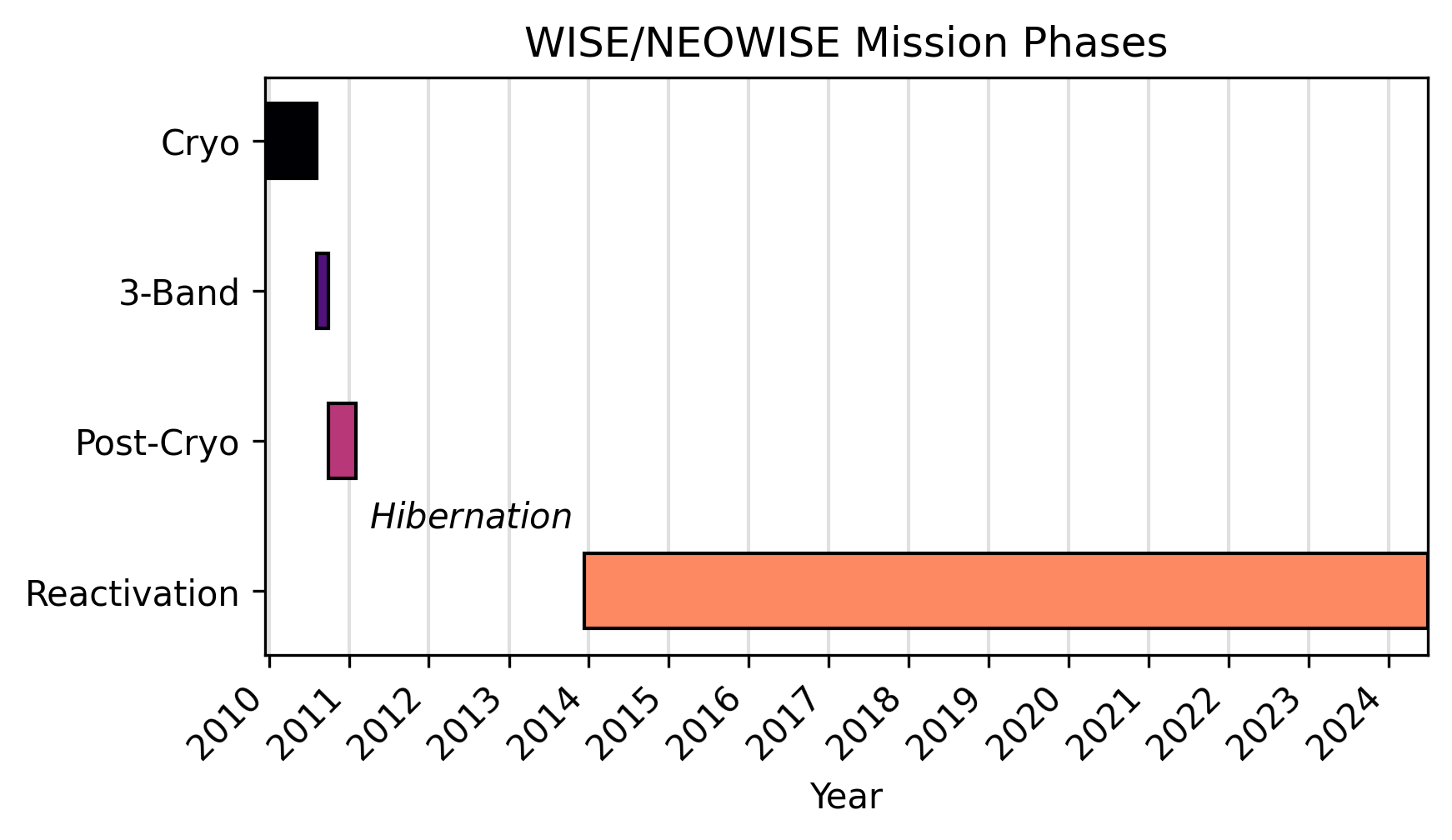}
\caption{Timeline of the WISE/NEOWISE mission phases. ``Cryo": UT 2010 January 14 to UT 2010 August 6 (bands: W1, W2, W3, W4); ``3-Band": UT 2010 August 6 to UT 2010 September 29 (bands: W1, W2, W3); ``Post-Cryo": UT 2010 September 29 to 2011 February 1 (bands: W1, W2); ``Reactivation": UT 2013 December 13 to UT 2024 August 1 (bands: W1, W2).
\label{Fig01}}
\end{figure*}

In this first paper of the series, we present a comprehensive analysis of the full 15-year dataset from the \emph{Wide-field Infrared Survey Explorer} (WISE; \citealt{Wright2010,Cutri2012}) and its planetary science successor, NEOWISE \citep{Mainzer2011,Cutri2013,Mainzer2014}. Following first light in January 2010, WISE/NEOWISE has conducted an all-sky infrared survey, significantly advancing our knowledge of small Solar System bodies. The mission concluded on July 31, 2024, due to accelerated orbital decay caused by elevated solar activity. Over its operational lifetime (Fig. \ref{Fig01}), WISE/NEOWISE observed a diverse array of comets across multiple epochs and activity levels, offering a unique dataset for investigating comet behavior under a wide range of thermal and radiative environments.  

%%%%%%%%%%%%%%%%%%%%%
%%%%%%%%%%%%%%%%%%%%%

This paper introduces the COSINE project and outlines our target selection strategy, which is based on well-constrained orbits and reliable photometric detectability. We summarize the general properties of the detected comet population and provide an overview of the dataset. Future studies in this series will separate nucleus and coma contributions (dust and/or gas), enabling detailed modeling of dust dynamics and compositional properties.
\\

% 2. Observations %%%%%%%%%%%%%%%%%%%%%%%%%%%%%%%%%%%%%%%%%%%%%%%%%%%%%%%%%%%%%%%%%%%%%%%%%%%%%%%%%%%%%%%%%%%%%%%%
\section{Data Description and Analysis} \label{sec:obsdata}

This study utilizes the complete WISE/NEOWISE dataset, spanning from the original all-sky WISE survey initiated in January 2010 through the final NEOWISE Reactivation phase, which concluded in 2024 (Fig. \ref{Fig01}). 

The mission employed a 40-cm diameter telescope that continuously scanned the sky, operating through multiple phases dictated by the status of its dual-stage solid hydrogen cryostat \citep{Mainzer2011,Mainzer2014}. During the initial ``Cryo" phase, WISE collected data in four infrared bands centered at 3.4, 4.6, 12, and 22 $\mu$m -- designated W1, W2, W3, and W4, respectively \citep{Wright2010}. Table~\ref{tab:band} summarizes key properties of the four WISE bands, based on \citet{Wright2010} and the SVO Filter Profile Service \citep{Rodrigo2012}\footnote{\url{https://svo2.cab.inta-csic.es/theory/fps3/index.php?mode=browse&gname=WISE&asttype=}}.
As the cryogen supply was exhausted, W4 and subsequently W3 were decommissioned, transitioning the mission through the ``3-Band" and ``Post-Cryo" phases. After a two-year hibernation beginning in February 2011, the spacecraft resumed observations in December 2013, initiating the decade-long ``Reactivation" phase using only the W1 and W2 bands.

Throughout all mission phases, WISE/NEOWISE maintained a 47\arcmin\ $\times$ 47\arcmin\ field of view (FoV) at a solar elongation of approximately 90\arcdeg. An angular motion of $\sim$1\arcdeg\ per day with respect to the Sun combined with a 10 \% in-scan overlap ensured robust sky coverage. Motion blur was limited to $\sim$0.05\arcsec, negligible compared to the instrument’s pixel scales: 2.75\arcsec\ per pixel for W1--W3, and 5.5\arcsec\ per pixel for the 2$\times$2-binned W4 channel. The corresponding angular resolutions are 6.1\arcsec, 6.4\arcsec, 6.5\arcsec, and 12.0\arcsec\ for W1 through W4, respectively \citep{Wright2010}. Exposure times were 7.7 secs for W1 and W2, and 8.8 secs for W3 and W4.

\begin{figure}[ht]
\centering
\includegraphics[width=0.33\textwidth]{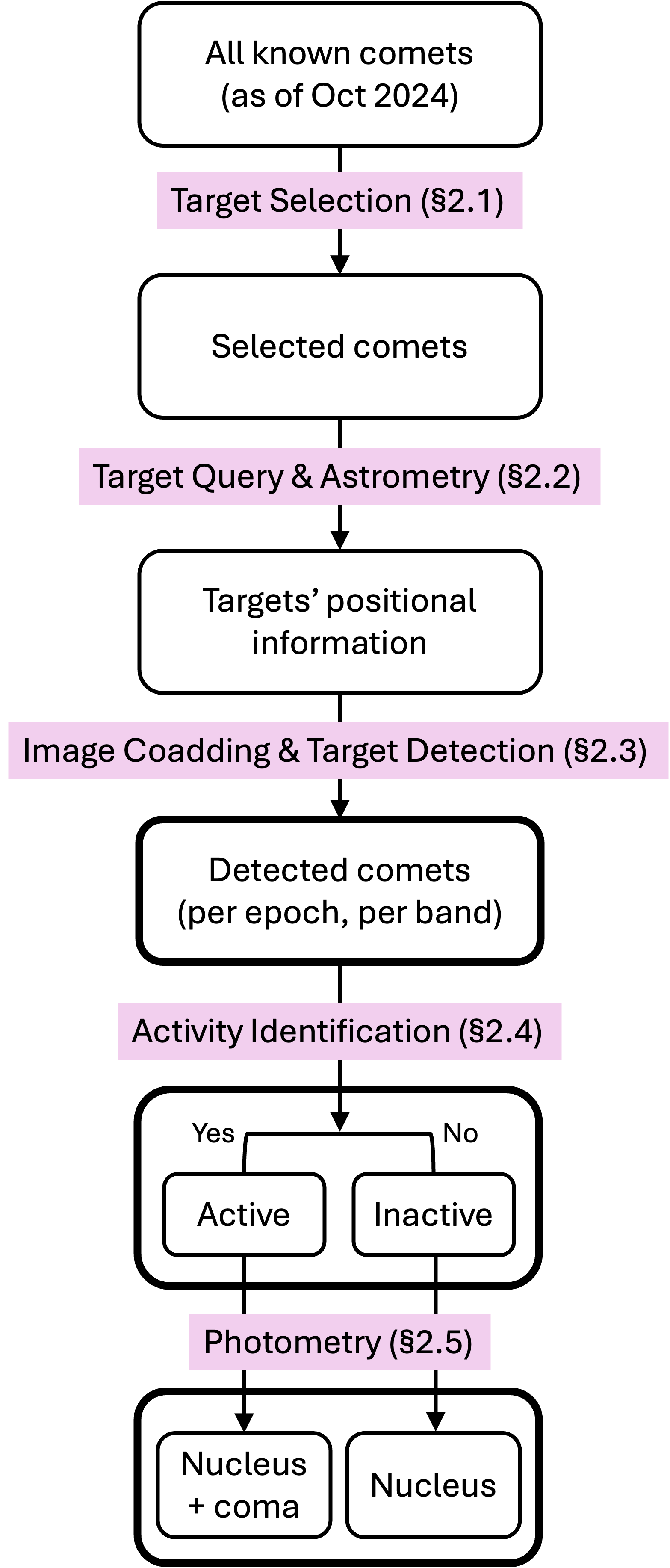}
\caption{Flowchart summarizing the WISE/NEOWISE data analysis procedure used in this study. Each highlighted module corresponds to detailed methods described in the following sections. Photometric results are based on the combined signal from the nucleus and coma.
\label{Fig02}}
\end{figure}

Comprehensive descriptions of the optical design and operational strategy can be found in \citet{Wright2010}, \citet{Mainzer2011}, and \citet{Mainzer2014}. Figure \ref{Fig02} outlines the data analysis workflow employed in this study; each major step is elaborated in the corresponding subsections. The photometric results presented here are based on the total observed signal, encompassing both nucleus and coma components.
%%%%%%%%%%%%%%%%%%%%%
\begin{deluxetable*}{ccccc}[hbt]
% \tabletypesize{\footnotesize} %\scriptsize
%\movetableright=-1in
\tablewidth{0pt} 
\tablecaption{Key Properties of WISE Bandpass Filters. \label{tab:band}}
\tablehead{
\colhead{\multirow{2}{*}{Filter ID}} &
\colhead{$\lambda_{\rm eff}$} &
\colhead{$W_{\rm eff}$} &
\colhead{$\lambda_{\rm min}$} &
\colhead{$\lambda_{\rm max}$} \\ 
\cline{2-5}
& (1) & (2) & (3) & (4)
}
\startdata
W1 & \quad3.353\quad & \quad0.663\quad & \quad2.754\quad & \quad3.872 \\
W2 & \quad4.603\quad & \quad1.042\quad & \quad3.963\quad & \quad5.341 \\
W3 & \quad11.561\quad & \quad5.506\quad & \quad7.443\quad & \quad17.261 \\
W4 & \quad22.088\quad & \quad4.102\quad & \quad19.520\quad & \quad27.911 \\
\enddata
\tablecomments{
All wavelengths are in microns. 
(1) Effective wavelength, often used interchangeably with reference wavelength. 
(2) Effective bandwidth, defined as the width of a rectangle with height equal to the peak transmission and the same area as the filter transmission curve. 
(3) Minimum wavelength, defined as the shortest wavelength where transmission exceeds 1~\% of the peak. 
(4) Maximum wavelength, the longest wavelength where transmission exceeds 1~\% of the peak.}
\end{deluxetable*}
%%%%%%%%%%%%%%%%%%%%%
%%%%%%%%%%%%%%%%%%%%%
%%%%%%%%%%%%%%%%%%%%%

% 2.1 %%%%%%%%%%%%%%%%%%%%%%%%%%%%%%%%%%%%%%%%%%%%%%%%%%%%%%%%%%%%%%%%%%%%%%%%%%%%%%%%%%%%%%%%%%%%%%%%%%%%%%%%%%%%
\subsection{Target Selection} \label{sec:sec21}

The complete set of known cometary orbits was retrieved from JPL Horizons. 
% Since Horizons typically fits a separate orbital solution for each apparition, some comets, such as 1P/Halley, have multiple entries, occasionally extending back several millennia. Many of these historical solutions are based on limited or imprecise data, resulting in low-quality orbital fits. Accordingly, a refined set of selection criteria was applied to define the target base for this study.
As of October 2024, a total of 3,768 comet designations were retrieved. This master list was then filtered using the following selection criteria to define the target base for this study:
\begin{enumerate}[itemsep=-.01in]
    \item[(1)] Orbit solutions must have an epoch post-1980.
    \item[(2)] Comets labeled ``D/'' (defunct or disintegrated) were excluded.
    \item[(3)] Only intact, non-fragmented comets were retained. For fragmented comets, we retained their parent bodies only (if they exist).
    \item[(4)] Orbital solutions must have a positional uncertainty (1$\sigma$) below 0.01 au.
    \item[(5)] The comet must have been within 11.5 au of the Sun at any point during the WISE/NEOWISE mission period. 
    \item[(6)] Five additional objects were removed:
    \subitem{$\bullet$} 483P/PANSTARRS, JPL Horizons lists it as two fragments and the Minor Planet Center (MPC) lists it as a single comet.
    \subitem{$\bullet$} A/2018 V3 and A/2020 A1, listed as trans-Neptunian objects in Horizons but classified as comets by the MPC.
    \subitem{$\bullet$} C/2010 E3 (WISE) and C/2019 L2 (NEOWISE), both observed only once by WISE/NEOWISE and associated with poor orbit solutions. 
\end{enumerate} 

\begin{figure}[tbh]
\centering
\includegraphics[width=\textwidth]{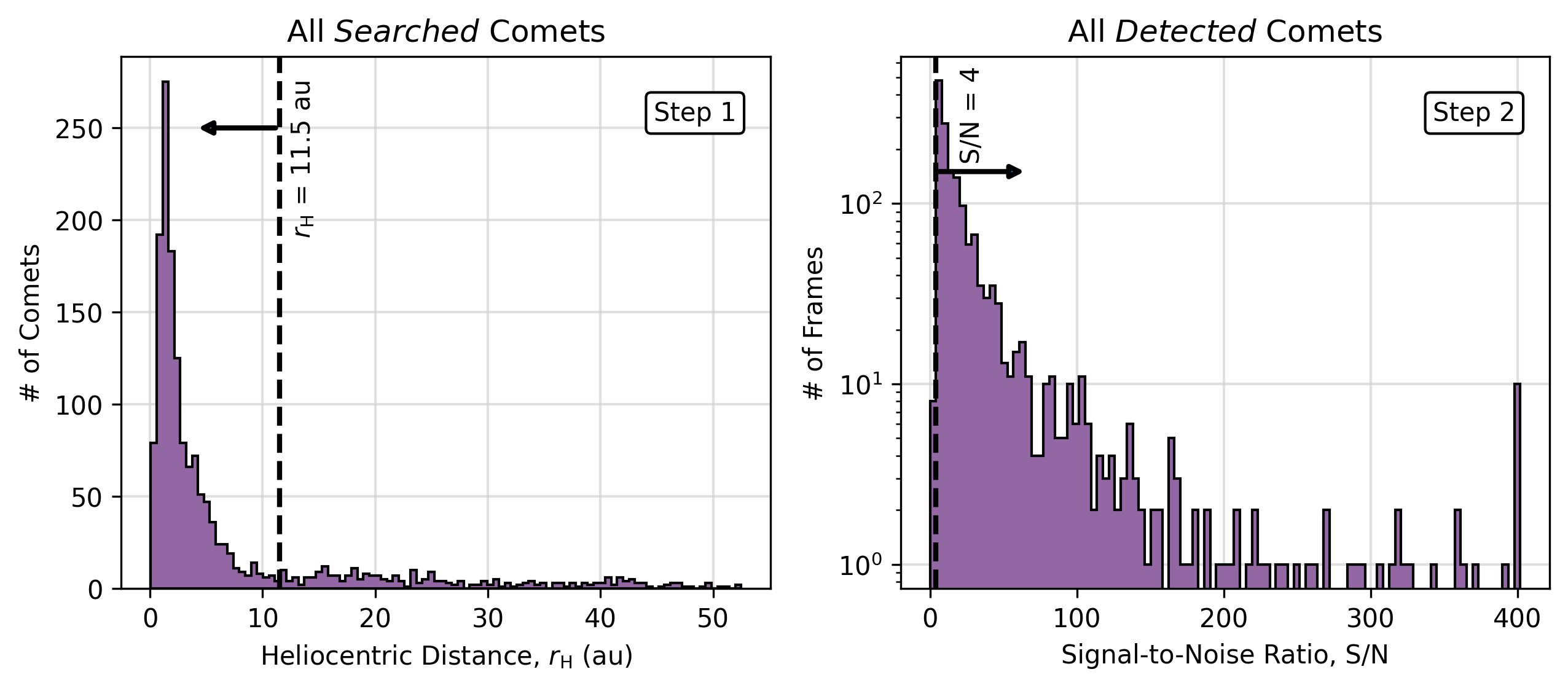}
\caption{Two-stage filtering applied to the 3,768 known comets (as of October 2024). Left: A heliocentric distance cutoff of 11.5 au (Sect. \ref{sec:sec21}) yields 1,335 queryable comets. Right: Imposing a signal-to-noise ratio threshold of S/N $>$ 4 on the stacked frames identifies 484 detected comets across 1,633 frames from 966 epochs. These detections constitute the primary dataset analyzed in this study (Sect. \ref{sec:sec23}). In the right panel, comets with \emph{stacked} S/N $>$ 400 are grouped in the highest bin.
\label{Fig03}}
\end{figure}

The criterion (2) ensured the exclusion of comets lacking orbital covariance data. The heliocentric distance cutoff of 11.5 au (Criterion 5) was determined through our visibility analysis of the full comet list during the Cryo phase, which offered the best sensitivity to cold, distant comets via the W4 band. Although over 1,000 comets were discovered by the Solar and Heliospheric Observatory (SOHO) during the late 1990s and early 2000s, many are now beyond 15 au on outbound trajectories and were not visible in WISE/NEOWISE data \citep{Bauer2024}. 
Criterion (4) is a relatively large cutoff, as there are some comets with short arcs with high positional uncertainty, which are still predictable near the times of their discovery. The two WISE/NEOWISE comets from criterion (6) are extreme examples of this.
The 11.5 au cutoff effectively excludes such objects while retaining all comets detected during the Cryo phase (Left panel in Fig. \ref{Fig03}). Applying these criteria resulted in a final sample of 1,335 candidate comets for downstream analysis.

%%%%%%%%%%%%%%%%%%%%%
%%%%%%%%%%%%%%%%%%%%%

% 2.2 %%%%%%%%%%%%%%%%%%%%%%%%%%%%%%%%%%%%%%%%%%%%%%%%%%%%%%%%%%%%%%%%%%%%%%%%%%%%%%%%%%%%%%%%%%%%%%%%%%%%%%%%%%%%
\subsection{Target Query and Astrometry} \label{sec:sec22}

Following the selection of queryable comets outlined in Section \ref{sec:sec21}, we retrieved SPICE\footnote{SPICE (Spacecraft, Planet, Instrument, C-matrix, Events) kernels are data files developed by NASA's Navigation and Ancillary Information Facility (NAIF) that provide essential information, such as spacecraft trajectory, orientation, and planetary positions, needed to interpret space mission observations accurately.} kernels for all targets. An exception was comet C/2013 A1 (Siding Spring), which was not available via the Horizons web interface and thus required manual acquisition through direct request. This object is notable for its exceptionally close encounter with Mars in 2014, approaching within 0.00094 au\footnote{\url{https://ssd.jpl.nasa.gov/tools/sbdb_lookup.html\#/?sstr=c\%2F2013\%20a1&view=OPC}} \citep{Stevenson2015}. For each comet, we conducted a comprehensive SPICE-based visibility analysis across all WISE/NEOWISE FoVs during every mission phase. This process was fully automated using \texttt{Kete}\footnote{\url{https://github.com/dahlend/kete/}} (\citealt{Dahlen2025}; Dahlen et al. 2025, under review), an open-source tool designed to simulate observations of Solar System objects by both space- and ground-based surveys. This involved checking the position of all comets in 29.6 million frames captured during the WISE/NEOWISE mission.

% His comment is a bit confusing to me, but I think he is confusing us download a spice kernel with us querying horizons. Despite that he made me relook at which parts of my code actually ended up in this paper. We used kete propagation extensively during development and the dynamical analysis. It was used for diagnostic plots and analysis that you and I did, however for the image stacking *specifically* we relied on the spice kernels themselves. 
% I am very grumpy, paper 2 will be like 99% of this, and almost no spice.
% I think we should remove this:

%Cometary positions were propagated using orbital elements from Horizons, with full n-body integration. The calculations accounted for general relativistic corrections, the solar J2 gravitational term (due to the Sun's oblateness), and the perturbations from all major planets, optionally including the five most massive asteroids. For comets, non-gravitational acceleration terms (A1, A2, and A3 terms) provided by Horizons were also incorporated. 

Observations were computed from the location of the WISE spacecraft (observatory code \texttt{C51}). For each epoch, we determined whether the comet fell within the instrument's FoV and calculated its on-sky coordinates in x and y pixel space, including corrections for light-travel time (``States" of the object). A detailed description of \texttt{Kete} and its applications to Solar System science is in preparation (Dahlen et al. 2025, under review).

This process identified all comets that entered the WISE/NEOWISE FoV at least once during the mission. To validate astrometric accuracy, we cross-referenced our predicted positions with all WISE/NEOWISE comet detections reported to the MPC. The predicted positions agreed with the MPC-reported astrometry to within 1.5\arcsec\ in nearly all cases -- well within the instrument's pixel scale. Small discrepancies, primarily from early 2010 data, are attributed to timing and calibration uncertainties during the initial mission optimization phase.

% 2.3 %%%%%%%%%%%%%%%%%%%%%%%%%%%%%%%%%%%%%%%%%%%%%%%%%%%%%%%%%%%%%%%%%%%%%%%%%%%%%%%%%%%%%%%%%%%%%%%%%%%%%%%%%%%%
\subsection{Image Coadding and Target Detection} \label{sec:sec23}

For comets observed on multiple occasions during the mission, we grouped their ``States" into distinct observational epochs, sorted chronologically. An epoch was defined as a sequence of observations separated from adjacent groups by at least seven days, ensuring a smooth, counterclockwise progression in Right Ascension (RA) at the Center Reference Pixel (as indicated by header keyword \texttt{WCROTA2}) within a continuous range smaller than 2\arcdeg. The number of epochs per comet ranges from one to several dozen.

In addition to the 11.5 au heliocentric cutoff applied, we performed statistical quality assessments on each image to winnow out frames unsuitable for coaddition. This filtering process removed frames with saturated or irregular backgrounds caused by moonlight, planetary glare, stellar contamination, detector annealing events, overlapping sources, or radiation exposure from the South Atlantic Anomaly (SAA)\footnote{\url{https://wise2.ipac.caltech.edu/docs/release/neowise/expsup/sec2\_1a.html\#saa_sep}}. Such frames typically exhibit significantly elevated pixel count standard deviations compared to nominal observations. Coadding was skipped for epochs with fewer than two high-quality frames or for comets whose photocenter landed near image edges. Only images passing all criteria were retained for stacking. Appendix \ref{sec:app0} provides a detailed account of the filtering methodology.

Image coadding was performed using the Python package \texttt{reproject}\footnote{\url{https://reproject.readthedocs.io/en/stable/\#}} \citep{Robitaille2020}, widely used for stacking diffuse astronomical sources. We made minor source code modifications to accommodate moving objects. A new FITS file was generated for each stack, adopting a uniform pixel scale of 2.75\arcsec\ for all bands. Although the W4 input data were originally binned to half this resolution, they were projected onto the same grid as the other bands for consistency, following the approach validated by the \texttt{unWISE} coadding framework \citealt{Lang2014}. The final World Coordinated System (WCS) was centered on the midpoint of the remaining ``States." 

Since the original \texttt{reproject} assumes static sky positions, we applied shifts to align comet positions across input frames relative to the first image in the epoch. \verb|NaN|-valued pixels were masked, with corresponding footprints set to zero. Pixel values were averaged to form the stacked image, and the standard deviation across frames was recorded to generate an uncertainty map. Reprojection was then performed using flux-conserving spherical polygon intersection instead of the simple interpolation option, and background levels were normalized across input images. Final stacks were trimmed to 800 $\times$ 800 pixels, centered on the target.

We assessed the signal-to-noise ratio (S/N) of all coadded images using the \texttt{photutils.aperture} module from the \texttt{photutils} Python package \citep{Bradley2024}. Photometry was performed with circular apertures centered on the expected source location, with radii of 6/9/11.5/23\arcsec\ for the W1/W2/W3/W4 bands, respectively. These aperture sizes follow those used by \citet{Bauer2011} for W1 and W2, while the W3 and W4 apertures were enlarged by 0.5\arcsec\ and 1\arcsec, respectively, to better accommodate background variations in edge cases. Background statistics were calculated using 20 evenly (angularly) spaced $10 \times 10$-pixel patches positioned 105 pixels from the nucleus center, spanning 360\arcdeg. The background level was defined as the median of the patch means (excluding \verb|NaN|s), and the associated noise was estimated from the median of the Median Absolute Deviation (MAD)-based standard deviations. 

The resulting S/N distribution is shown in the right panel of Figure~\ref{Fig03}. We excluded comets with S/N $<$ 4, yielding 1,633 valid coadded frames corresponding to 484 comets across 966 epochs. These detections comprise the baseline dataset for the COSINE project. Slight positional offsets for a few comets are attributed to orbital uncertainties in their SPICE kernels; in such cases, a broader search region was used to identify the photocenter. A complete log of all detections is presented in Table~\ref{apt01}.

%%%%%%%%%%%%%%%%%%%%%
\begin{figure}[htp]
\centering
\includegraphics[width=0.57\textwidth]{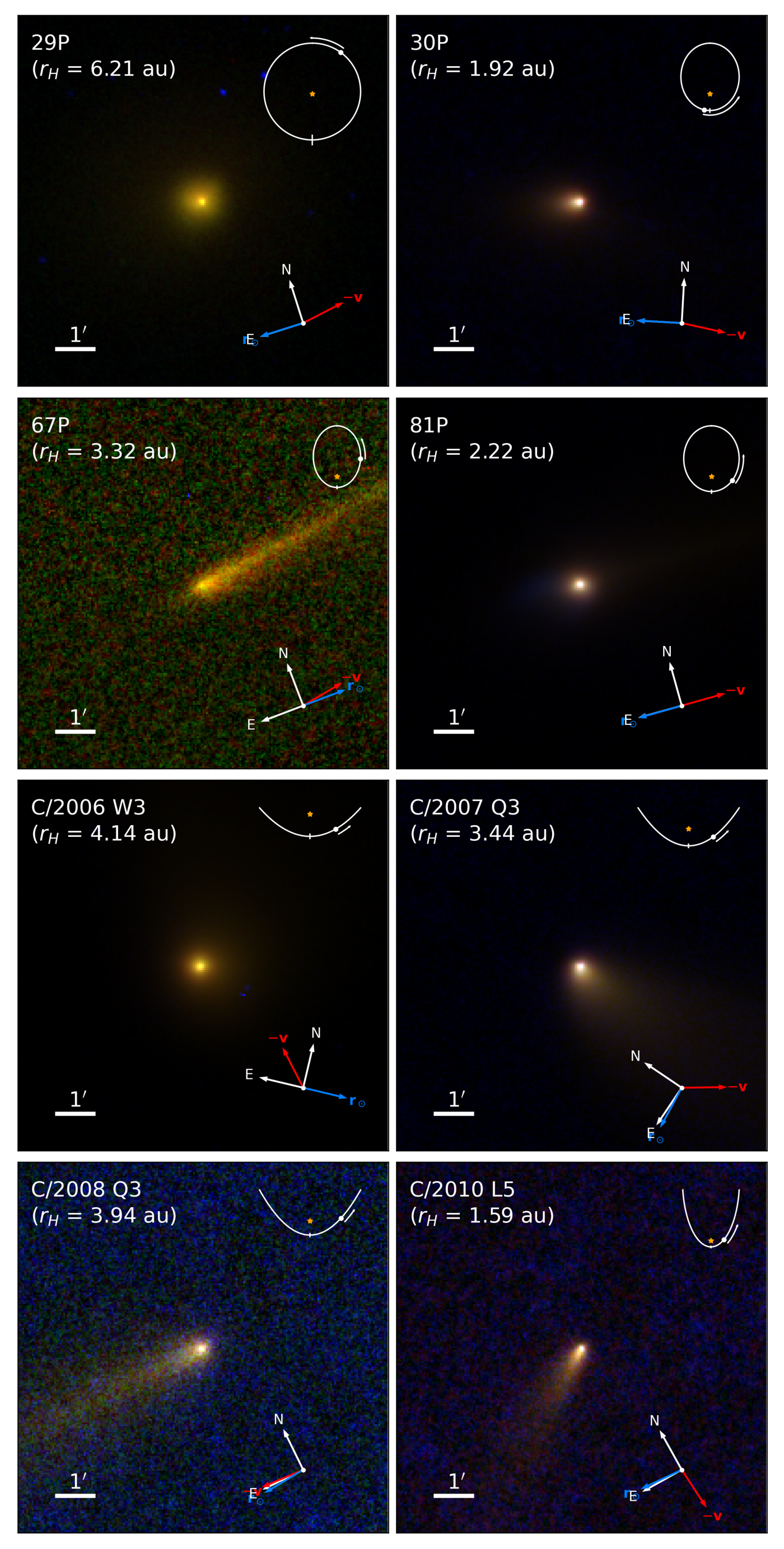}
\caption{False-color image example of detected comets from the Cryo phase. W2, W3, and W4 bands are mapped to blue, green, and red, respectively. Pixel values within the [10, 99.999] percentile were normalized to [0, 1] to suppress dominant background features, then rescaled to the [0.1, 99.9] range for visualization. Each panel includes the comet's heliocentric distance ($r_{\rm H}$, in au) at the time of observation, along with orientation vectors: ecliptic North (N), ecliptic East (E), the negative velocity vector ($-{\bf v}$), and the anti-solar vector (${\bf r}_\odot$). A 1\arcmin\ scale bar and orbital true anomaly diagram are also included.
\label{Fig04}}
\end{figure}
%%%%%%%%%%%%%%%%%%%%%

Figure~\ref{Fig04} presents an example of false-color composite images from the Cryo phase, with W2, W3, and W4 bands mapped to blue, green, and red, respectively. Additional examples are shown in Figure~\ref{Figap01}. The photometric methodology used to estimate S/N is detailed in Section~\ref{sec:app2}.
\\

% 2.4 %%%%%%%%%%%%%%%%%%%%%%%%%%%%%%%%%%%%%%%%%%%%%%%%%%%%%%%%%%%%%%%%%%%%%%%%%%%%%%%%%%%%%%%%%%%%%%%%%%%%%%%%%%%%
\subsection{Activity Identification} \label{sec:sec24}

Cometary activity is identified when extended emission -- produced by material released from the nucleus -- deviates from a point-source profile. In ideal cases, point sources such as stars follow a well-characterized point-spread function (PSF), which reflects the imaging system’s response to a point input. While intrinsically unresolved, PSFs appear broadened due to detector effects and spacecraft tracking uncertainties (e.g., jitter). The PSF varies with wavelength and position on the detector and, for WISE/NEOWISE, is documented in 9 $\times$ 9 grids for each band in the \emph{Explanatory Supplement to the WISE All-Sky}\footnote{\url{https://wise2.ipac.caltech.edu/docs/release/allsky/expsup/sec4\_4c.html\#psf}} and the \emph{NEOWISE Data Release Products}\footnote{\url{https://wise2.ipac.caltech.edu/docs/release/neowise/expsup/sec4\_2bi.html}} \citep{Cutri2012,Cutri2013}. In this study, however, coadded images were used to enhance comet S/N, effectively averaging the PSF across multiple detector regions. Hence, instead of comparing to a theoretical PSF measured in a fixed detector location, we empirically assessed activity by analyzing radial S/N profiles -- specifically, the radial extent at which the comet signal declined to background levels.

Radial S/N profiles were constructed by sampling every three pixels from the photocenter out to 102 pixels -- a radial distance sufficient to encompass the full range of cometary S/N, from the central peak to the radius where the signal becomes indistinguishable from background noise. At each sampled radius, local flux and associated S/N were computed. In high-S/N cases, inactive comets exhibited steep declines in their profiles, with S/N values dropping to $\lesssim$0.5 well within three radial sampling points ($<$9 pixels or 25\arcsec) in W1--W3, and typically within four points ($<$12 pixels or 33\arcsec) in W4. For visual comparison, each radial profile was plotted alongside three Gaussian curves (standard deviations of 2.4, 3.0, and 4.0 pixels, or $\sim$6.6, 8.25, and 11.0\arcsec), all normalized to the comet's peak S/N. We started from the 2.4-pixel Gaussian, as it closely matches the azimuthally averaged PSF for W1--W3 (and unbinned W4), consistent with the effective FWHM values reported by \citet{Wright2010}. 

However, for the majority of comets with moderate to low brightness (4 $<$ S/N $\lesssim$ 10) cases, such strict radial thresholds led to many borderline cases due to enhanced background contribution. To address this, we iteratively refined an activity classification scheme for the COSINE dataset and set the following criteria:
\begin{enumerate}[itemsep=-0.01in]
    \item[$\bullet$] Active: The radial S/N profile reaches $\sim$1 beyond the widest Gaussian profile (4-pixel standard deviation) used for reference.
    \item[$\bullet$] Inactive: The above condition is not met, or the background fluctuations exceed the S/N of any extended signal in the profile, regardless of total S/N.
\end{enumerate}
\noindent 
The above criteria naturally exclude borderline cases, such as visibly active comets yet too faint for reliable photometric evaluation, which are conservatively labeled inactive. To standardize the interpretation, we introduced an ``activity factor" label. Comets were flagged as active (``Y") or inactive (``N"), and each was further assigned a quality grade based on S/N (from Sect. \ref{sec:sec23}): Grade A for S/N $\ge$ 20, Grade B for 10 $\le$ S/N $<$ 20, and Grade C for 4 $\le$ N $<$ 10. This resulted in six discrete categories: Y-A, Y-B, Y-C, N-A, N-B, and N-C. This system minimizes subjective bias and facilitates consistent comparisons across the dataset. Table~\ref{apt01} lists activity flags for all 1,633 coadded frames associated with the 484 detected comets.

%%%%%%%%%%%%%%%%%%%%%
\begin{figure}[!t]
\centering
\includegraphics[width=0.85\textwidth]{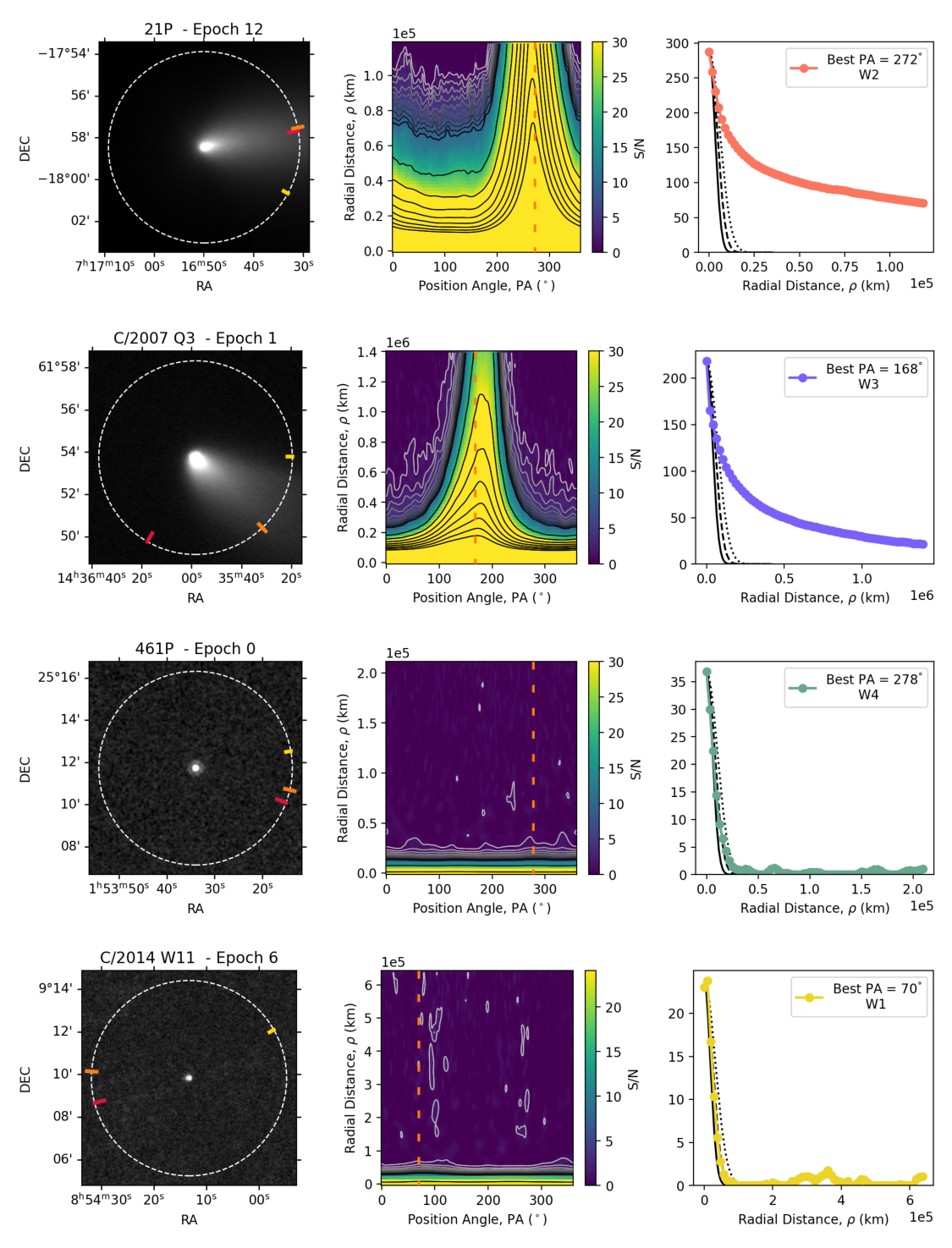}
\caption{Example of comets classified as ``Y-A" (top two rows) and ``N-A" (bottom two rows). Columns, from left to right, show: (1) the coadded image with vector overlays; (2) the azimuthal signal distribution, sampled in 1\arcdeg\ intervals, along the radial distance from the photocenter; and (3) the radial S/N profile along the primary position angle (PA), marked in the middle panel by a vertical dashed line. In the first column, vectors denote the anti-solar vector (red), best PA (salmon), and negative velocity vector (yellow). In the second column, gray and black contours indicate S/N levels in intervals of 1 (for S/N $<$ 10) and 10 (for S/N $>$ 10), respectively. In the third column, Gaussian profiles with standard deviations of 2.4, 3.0, and 4.0 pixels (6.6, 8.25, and 11.0\arcsec) are overplotted as solid, dashed, and dotted lines, respectively, with different colors representing individual bands.
\label{Fig05}}
\end{figure}
%%%%%%%%%%%%%%%%%%%%%
%%%%%%%%%%%%%%%%%%%%%
\begin{figure}[!t]
\centering
\includegraphics[width=0.85\textwidth]{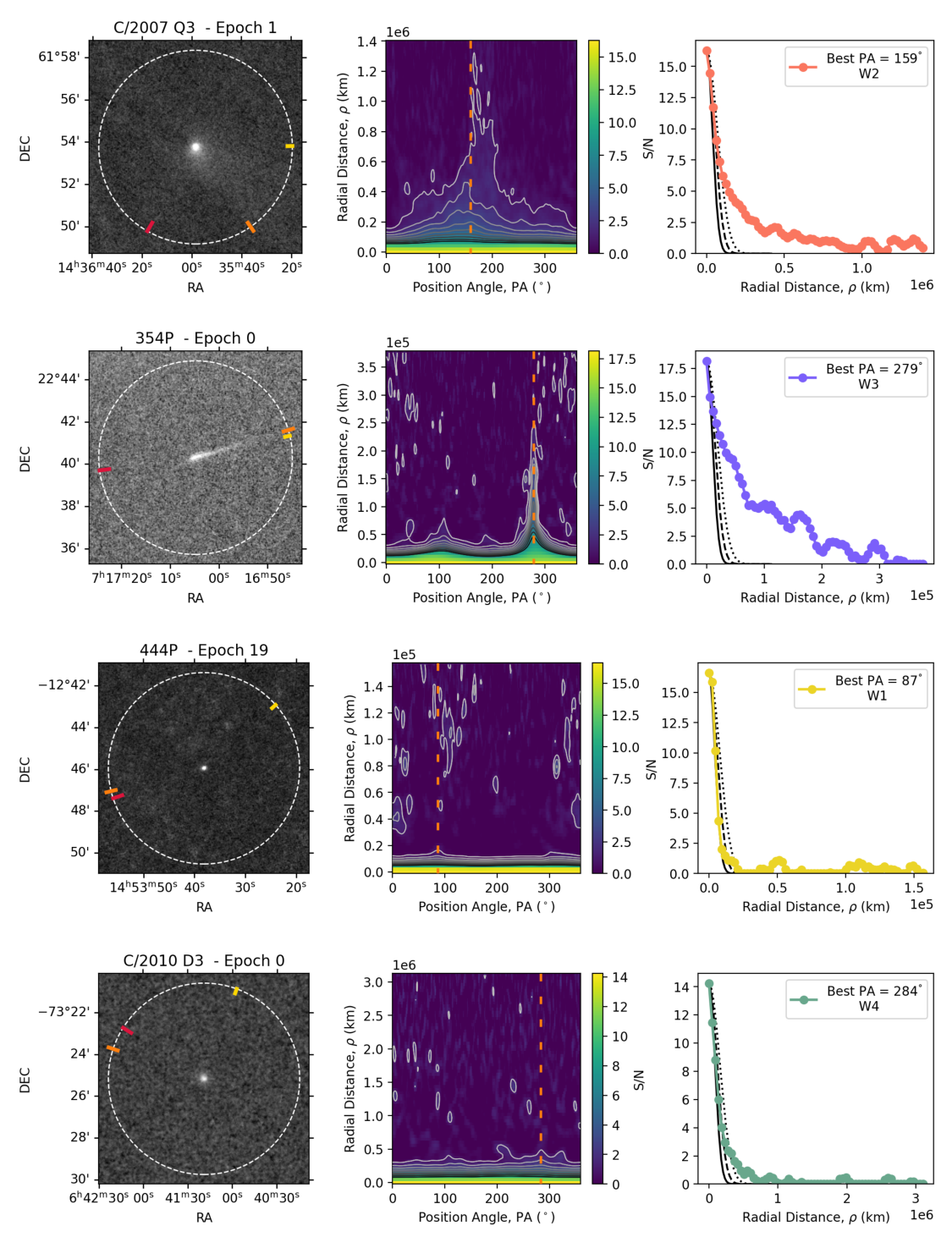}
\caption{Same as Figure \ref{Fig05}, but for comets classified as ``Y-B" (top two rows) and ``N-B" (bottom two rows).
\label{Fig06}}
\end{figure}
%%%%%%%%%%%%%%%%%%%%%
%%%%%%%%%%%%%%%%%%%%%
\begin{figure}[!t]
\centering
\includegraphics[width=0.85\textwidth]{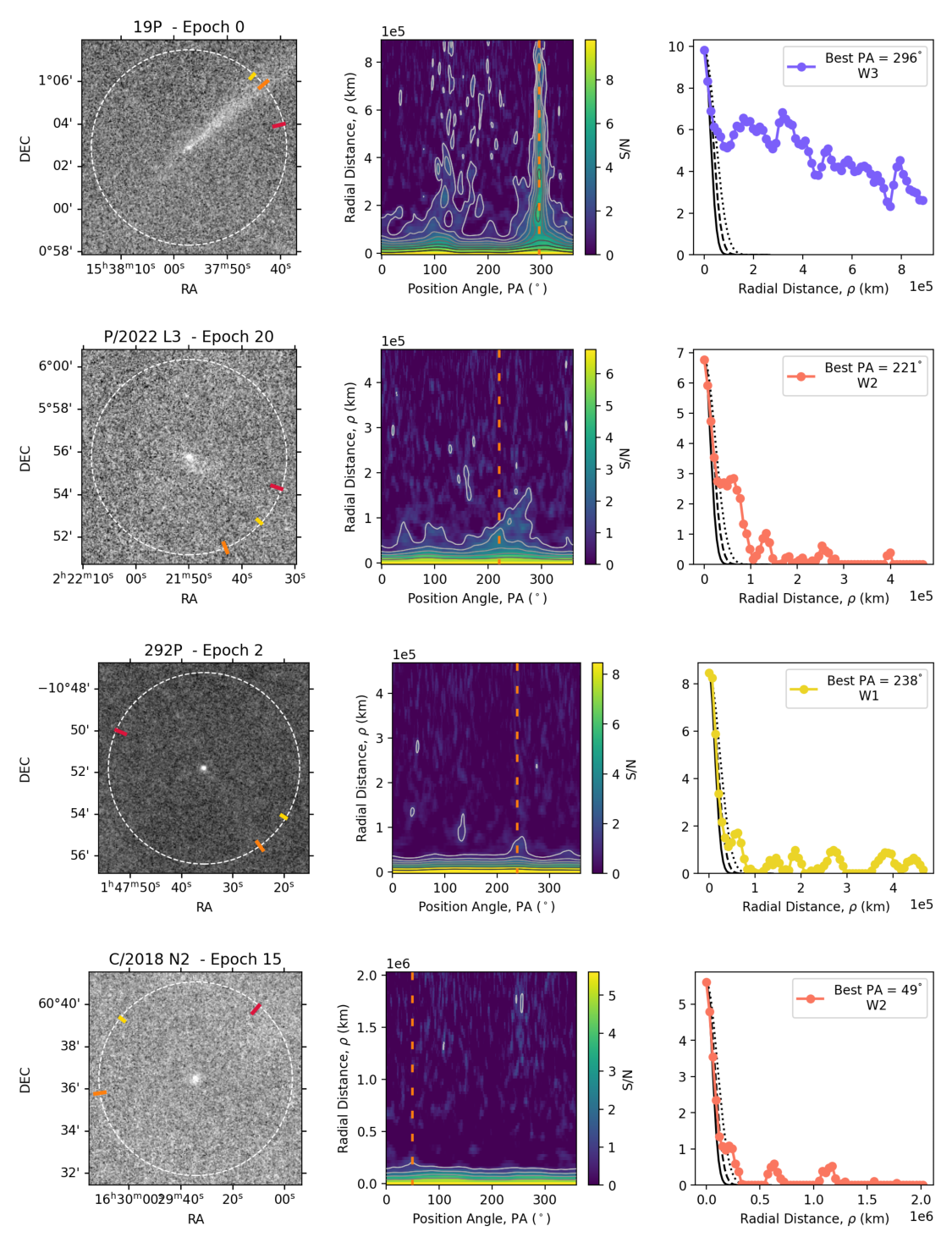}
\caption{Same as Figure \ref{Fig05}, but for comets classified as ``Y-C" (top two rows) and ``N-C" (bottom two rows).
\label{Fig07}}
\end{figure}
%%%%%%%%%%%%%%%%%%%%%

Figures \ref{Fig05}--\ref{Fig07} present representative examples of cometary activity classification, with each column illustrating a different aspect of the analysis. The first column shows the coadded image overlaid with key directional vectors and enclosed by a dashed circle marking a 102-pixel radial extent. Red, salmon, and yellow ticks indicate the anti-solar vector, best position angle (Best PA), and negative velocity vector at the epoch of observation, respectively. The second column plots the integrated signal distribution along the radial direction, sampled every 1\arcdeg\ in azimuth. Contours reflect S/N levels: gray for values below 10 (in intervals of 1) and black for values above 10 (in intervals of 10). The third column shows the radial S/N profile sampled every three pixels along the best PA. For visual reference, Gaussian PSFs with widths of 2.4, 3.0, and 4.0 pixels (corresponding to 6.6, 8.25, and 11.0\arcsec) are overlaid using solid, dashed, and dotted lines, respectively, color-coded. 

The optimal PA was defined as the azimuthal direction yielding the highest mean S/N within the inner 30-pixel region centered on the photocenter, sampled at 1\arcdeg\ intervals counterclockwise from celestial north. This restricted angular analysis mitigates contamination from background sources (e.g., stars, galaxies) and enhances sensitivity to the primary tail feature likely formed during the current apparition. It also minimizes the influence of dust trails or neck-line features, which often extend across the entire FoV and trace material from earlier orbital passages \citep{Fulle2004}. While this approach effectively identifies linear tail orientations, it provides only an averaged direction in cases of curved or fan-shaped structures and may not fully capture complex mass-loss histories. Detailed dynamical modeling of tail morphology will be the topic of future work. This three-panel analysis was applied uniformly to all coadded frames across the COSINE dataset.
%% activity identification =  not using target ephemeris since we didn't want to be controlled by ephemeris uncertainties.
%% A full 3-plot set of 1,899 coadded images for all 523 detected comets, covering each band and epoch, is available in the Caltech Box\footnote{\url{https://caltech.app.box.com/collection/27692956013}}.

% The WISE/NEOWISE instruments have a moderate pixel scale: $\sim$2.75\arcsec\ for W1--W3 and unbinned W4 \citep{Wright2010}. 
The WISE/NEOWISE pixel scale corresponds to cometocentric distances ranging from approximately 230 to 51,100~km, with a median value of $\sim$10,700~km, across the observed geocentric distance range of 0.049--10.673~au (median: 2.232~au).
A recognized limitation of this moderate spatial resolution is the potential contamination from unresolved circumnuclear dust or gas, which can bias photometric measurements, particularly at large heliocentric distances \citep{Fernandez2013}. For instance, several comets observed at $r_{\rm H}$ $\gtrsim$ 10~au during the Reactivation phase were classified as ``N-C" (inactive with low S/N) in W1 or W2, as W3 or W4 were not operational during this period. 
At such large heliocentric distances, where equilibrium temperatures fall below $\sim$35~K \citep{Meech2009,Womack2017}, W1 and W2 fluxes are primarily from scattered sunlight, though W1 may additionally capture organic emissions (Appendix~\ref{sec:app4-0}) and W2 encompasses potential contributions from the CO (1--0) fundamental vibrational band at 4.7~$\mu$m and the CO$_2$ $\nu_3$ vibrational band at 4.3~$\mu$m (e.g., \citealt{Reach2009,Bauer2015}).
Unless the nucleus is exceptionally large (e.g., $\sim$100~km assuming a typical cometary albedo of $\sim$0.1), the detected flux in these bands likely includes unresolved contributions from residual near-nuclear materials. As such, the number of comets classified as inactive in this study would be regarded as an upper limit.

% 2.5 %%%%%%%%%%%%%%%%%%%%%%%%%%%%%%%%%%%%%%%%%%%%%%%%%%%%%%%%%%%%%%%%%%%%%%%%%%%%%%%%%%%%%%%%%%%%%%%%%%%%%%%%%%%%
\subsection{Photometry} \label{sec:sec25}

Photometric measurements were performed on each coadded image of detected comets and converted into infrared magnitudes. Calibration parameters were adopted from the \emph{WISE All-Sky Explanatory Supplement}\footnote{\url{https://wise2.ipac.caltech.edu/docs/release/allsky/expsup/sec4\_4h.html}}.

Photometry was performed using circular apertures matching those adopted for S/N estimation (6/9/11.5/23\arcsec\ for the W1/W2/W3/W4 bands, respectively), implemented via the \texttt{photutils.aperture} module in Python \citep{Bradley2024}. Background levels were estimated following the same procedure used in S/N calculations as well (Sect. \ref{sec:sec23} and Appendix \ref{sec:app2}), by taking the median value from 20 evenly spaced 10-by-10 pixel sky patches located away from the comet nucleus. This approach, instead of using a conventional annulus, mitigates contamination from potential extended features. After background subtraction, counts in Digital Number (DN) were converted to flux densities in Jansky (Jy) using band-specific DN-to-Jy conversion factors\footnote{\url{https://wise2.ipac.caltech.edu/docs/release/prelim/expsup/sec2_3f.html}}. These fluxes were then converted to magnitudes using zero-point flux densities determined from WISE calibration sources.

\begin{figure}[th]
\centering
\includegraphics[width=0.51\textwidth]{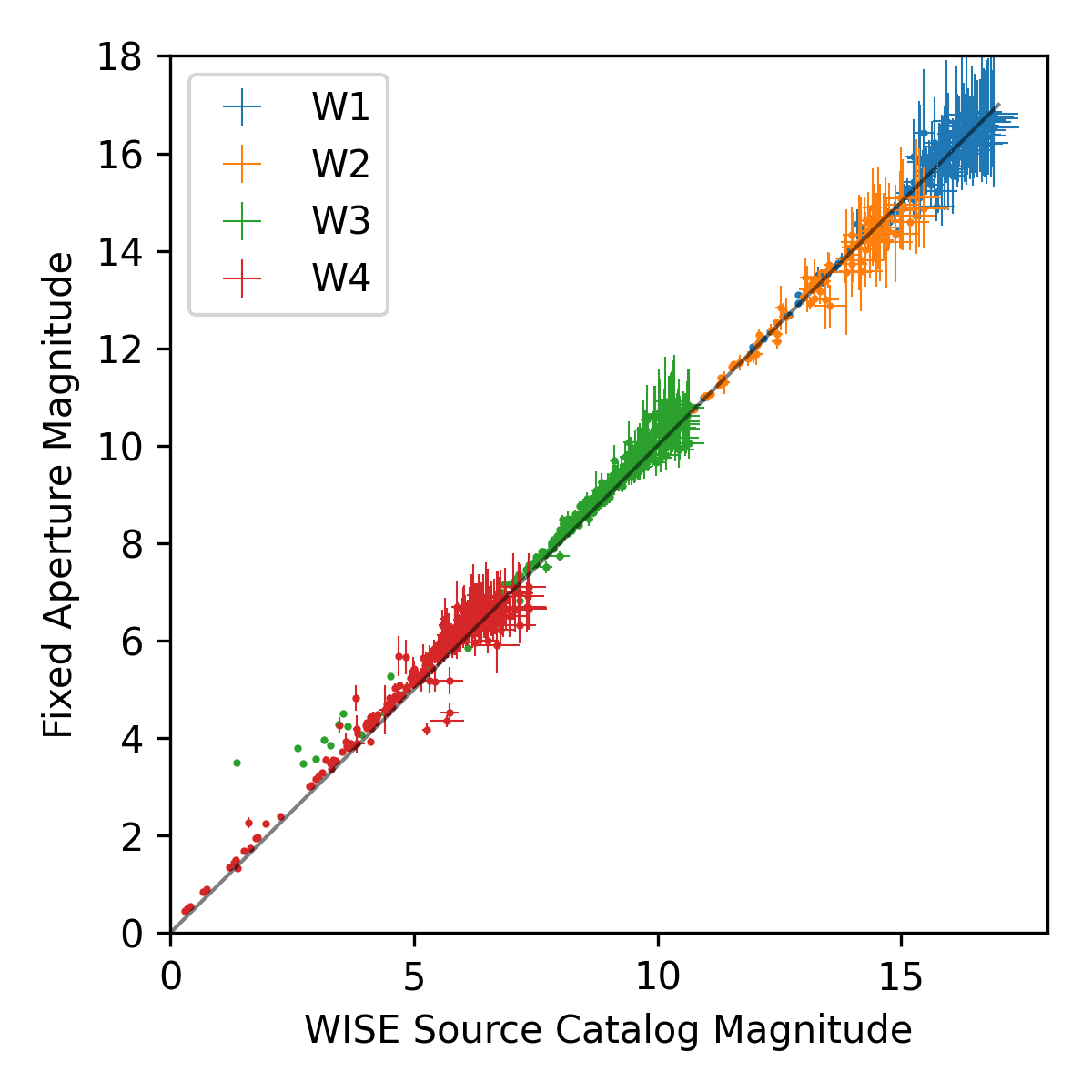}
\caption{Comparison of magnitudes derived using \texttt{photutils.aperture} (as applied in this study) with those from the WISE Source Catalogs for 1,000 randomly selected asteroids spanning a wide brightness range during the WISE Cryo phase.
\label{Fig08}}
\end{figure}

To verify the reliability of our photometric pipeline, we compared measurements taken from single point sources in individual frames with those listed in the \emph{WISE Single-Exposure Source Catalogs}\footnote{\url{https://irsa.ipac.caltech.edu/cgi-bin/Gator/nph-scan?mission=irsa\&submit=Select\&projshort=WISE}}. To isolate the performance of the photometric tool from potential contamination by faint extended emission, we selected 1,000 numbered asteroids at random with a wide brightness range, observed throughout the WISE Cryo phase (where all 4 bands were available). Aperture photometry was performed following the same procedure used for the COSINE dataset. The resulting magnitudes were compared with catalog values (\texttt{w1mpro}, \texttt{w1sigmpro} for W1, and \texttt{w2mpro}, \texttt{w2sigmpro} for W2, etc.). As shown in Figure~\ref{Fig08}, our measurements align well with the one-to-one correspondence for magnitudes, confirming the robustness of our approach.

%%%%%%%%%%%%%%%%%%%%%
%%%%%%%%%%%%%%%%%%%%%

We further validated the use of the \texttt{reproject} tool -- modified to accommodate moving targets (Section~\ref{sec:sec23}) -- in reliably deriving S/N from coadded frames. Figure~\ref{Figap2} demonstrates this consistency, showing that the final S/N scales with the square root of the number of stacked images, as expected.
\\

% 3. %%%%%%%%%%%%%%%%%%%%%%%%%%%%%%%%%%%%%%%%%%%%%%%%%%%%%%%%%%%%%%%%%%%%%%%%%%%%%%%%%%%%%%%%%%%%%%%%%%%%%%%%%%%%%
\section{Results and Discussion}
% 3.1 %%%%%%%%%%%%%%%%%%%%%%%%%%%%%%%%%%%%%%%%%%%%%%%%%%%%%%%%%%%%%%%%%%%%%%%%%%%%%%%%%%%%%%%%%%%%%%%%%%%%%%%%%%%%
\subsection{States of the Detected Comets} \label{sec:detected}

Our query of more than a decade of WISE/NEOWISE observations yields the largest uniformly analyzed sample of comets to date. The detected population, introduced in Section~\ref{sec:obsdata}, encompasses a wide range of dynamical classes and activity states. This section presents an overview of their orbital and photometric properties (Sections~\ref{sec:sec311} and \ref{sec:sec312}) and their assessed activity status (Section~\ref{sec:sec313}).

% 3.1.1 %%%%%%%%%%%%%%%%%%%%%%%%%%%%%%%%%%%%%%%%%%%%%%%%%%%%%%%%%%%%%%%%%%%%%%%%%%%%%%%%%%%%%%%%%%%%%%%%%%%%%%%%%%
\subsubsection{Dynamical Classification and Orbital Properties} \label{sec:sec311}

A total of 484 detected comets were assigned to dynamical classes based on their orbital characteristics. While conventional criteria -- orbital periods ($P_{\rm orb}$) near $\sim$20 yrs and the Jovian Tisserand parameter ($T_{\rm J}$)\footnote{A dynamical parameter describing the influence of Jupiter in the restricted three-body problem.} -- form the basis for classification, they proved insufficient for certain borderline cases. We refined the initial grouping provided by Horizons\footnote{Small-Body Database Query: \url{https://ssd.jpl.nasa.gov/tools/sbdb\_query.html}} by introducing supplementary conditions, indicated with asterisks$^\star$ below:
\begin{enumerate}[itemsep=-.01in]
    \item[$\bullet$] Hyperbolic Comets (HCs):
        \subitem{(1)} eccentricity $e$ $>$ 1.
    \item[$\bullet$] Near-Parabolic Comets (NPCs):
        \subitem{(1)$^\star$} $e$ = 1 (hard-coded in Horizons);
        \subitem{(2)$^\star$} $T_{\rm J} \in (2,3)$ and $P_{\rm orb} > 50$ yrs;
        \subitem{(3)} $e < 1$, $T_{\rm J} < 2$, and $P_{\rm orb} > 200$ yrs.
    \item[$\bullet$] Halley-Type Comets (HTCs):
        \subitem{(1)} $T_{\rm J} < 2$ and $P_{\rm orb} \in (20,200)$ yrs.
    \item[$\bullet$] Jupiter-Family Comets (JFCs):
        \subitem{(1)} $T_{\rm J} \in (2,3)$ and $P_{\rm orb} < 50$ yrs;
        \subitem{(2)$^\star$} $T_{\rm J} < 2$ and $P_{\rm orb} < 20$ yrs.
    \item[$\bullet$] Encke-Type Comets (ETCs):
        \subitem{(1)} $T_{\rm J} > 3$ and semi-major axis $a < a_{\rm Jupiter}$.
\end{enumerate}

All classification conditions are applied with logical \texttt{OR}; if a comet meets any one of them, its classification is assigned accordingly. LPCs encompass HCs, NPCs, and HTCs, while SPCs include JFCs and ETCs. Additional criteria introduced under NPCs (2) and JFCs (2) account for comets listed by Horizons as JFCs but exhibiting $e > 0.7$ and $P_{\rm orb} > 50$ yrs, or occasionally exceeding 10$^3$ yrs, traits more consistent with LPCs. To align with the literature and include comets with extended JFC-like orbits, we adopted a $P_{\rm orb}$ threshold of 50~yrs, rather than the classical 20~yrs cutoff.\footnote{Examples include 318P/McNaught-Hartley (20.68 yrs), P/2009 T2 (La Sagra) (20.94 yrs), P/2011 P1 (McNaught) (21.88 yrs), P/2008 Y3 (McNaught) (22.75 yrs), 440P/Kobayashi (25.07 yrs), P/2010 E2 (Jarnac) (25.40 yrs), P/2010 J3 (McMillan) (26.94 yrs), C/2014 W11 (PANSTARRS) (30.67 yrs), C/2008 E1 (Catalina) (34.93 yrs), C/2007 S2 (Lemmon) (44.44 yrs), and C/2021 K1 (ATLAS) (45.49 yrs).} Our classifications agree with Horizons, except for a few comets reclassified from JFCs to NPCs under the revised scheme. As a side note, comet 29P/Schwassmann-Wachmann, a Centaur with a nearly circular orbit around Jupiter and considered transitional between JFCs and trans-Neptunian objects \citep{Kaib2024}, remains classified as a JFC in both our catalog and the Horizons database. Table~\ref{t01} summarizes the adopted dynamical breakdown.

\begin{figure}[htb]
\centering
\includegraphics[width=0.55\textwidth]{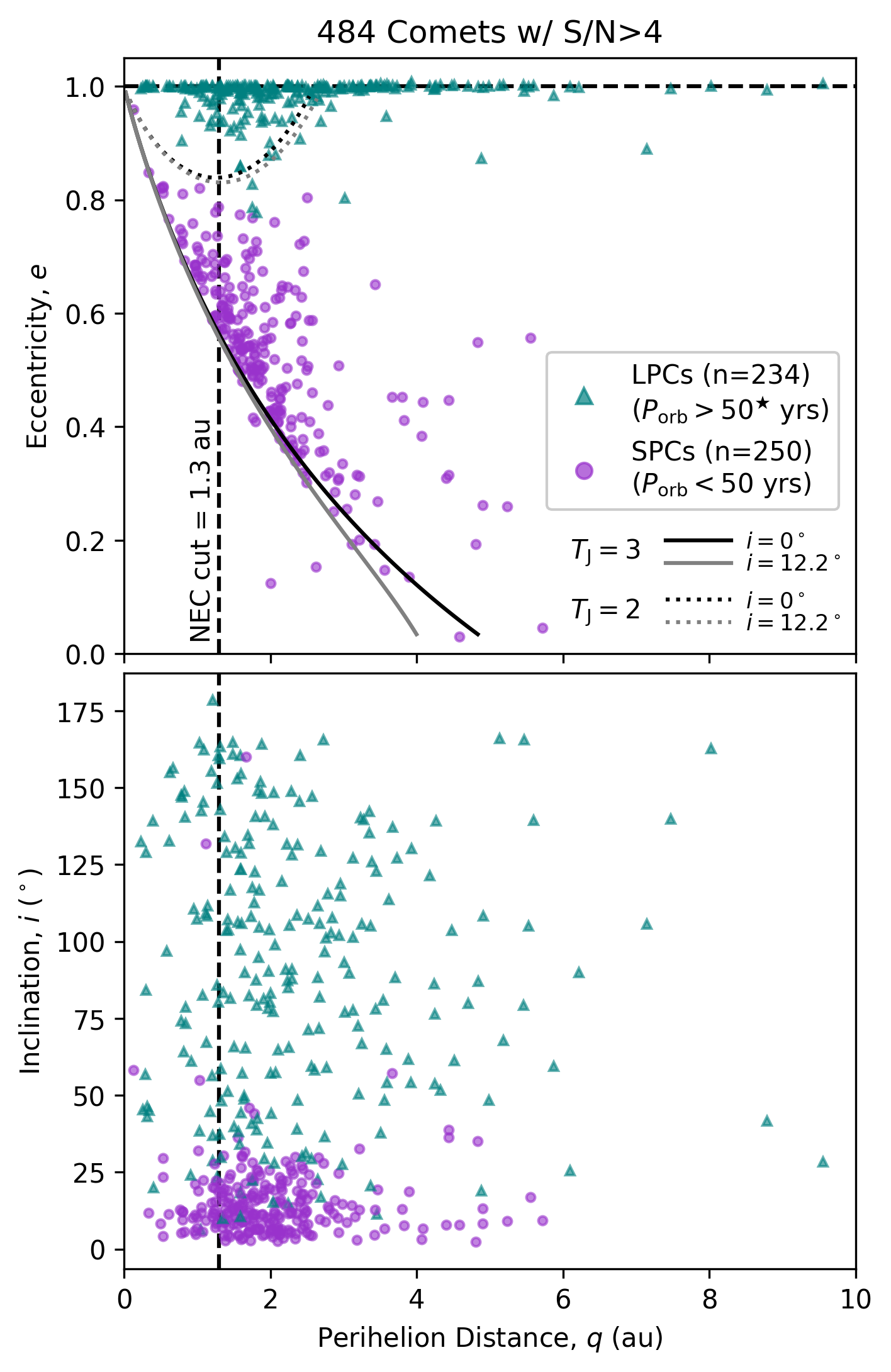}
\caption{Orbital element distribution of 484 detected comets. Solid and dotted black lines indicate $T_{\rm J} = 3$ boundaries; gray lines show $T_{\rm J} = 2$ boundaries for inclinations of 0\arcdeg\ and 12.2\arcdeg\ (median $i$ of SPCs). Vertical dashed lines mark the perihelion threshold ($q = 1.3$ au) separating Near-Earth Comets (NECs) from others. Acronyms for each comet group are defined in the main text.
\label{Fig09}}
\end{figure}

Figure~\ref{Fig09} shows the distribution of orbital elements for the 484 comets. The “50$^\star$ yrs” category under LPCs includes four borderline HTCs with $P_{\rm orb}$ between 20 and 50 yrs: P/2010 JC81 (WISE), C/2010 L5 (WISE), C/2014 W9 (PANSTARRS), and 38P/Stephan-Oterma. As expected, LPCs cluster near $e = 1$ and show an isotropic inclination distribution, consistent with previous studies \citep[e.g.,][]{Levison1996}. In contrast, SPCs are more confined, typically within $2 < T_{\rm J} < 3$, and have lower eccentricities. Their median inclination is 12.3\arcdeg, indicating a strong concentration near the ecliptic in prograde orbits. Two notable SPC outliers are 333P/LINEAR ($i \approx 132.0$\arcdeg) and 389P/Siding Spring ($i \approx 160.1$\arcdeg), both with well-determined orbits (condition code = 0 by Horizons). Despite their retrograde orbits and low $T_{\rm J}$ values (0.418 and $-$0.460, respectively) far outside the typical SPC values, their short periods ($P_{\rm orb} < 15$ yrs; $a < 6$ au) formally place them in the JFC class under our criterion (2).

%%%%%%%%%%%%%%%%%%%%%
%%%%%%%%%%%%%%%%%%%%%
%%%%%%%%%%%%%%%%%%%%%
\begin{deluxetable*}{cc|cc|c|cc}[thb]
\tabletypesize{\footnotesize} %\scriptsize
%\movetableright=-1in
\tablewidth{0pt} 
\tablecaption{Dynamical classification of the 484 detected comets \label{t01}}
\tablehead{
\colhead{} & \colhead{} &
\multicolumn2c{{\bf Long-Period Comets (LPCs)}} & \colhead{} &
\multicolumn2{l}{{\bf Short-Period Comets (SPCs)}} \\ 
\cline{2-7}
& & \# (epochs) & $\supset$NECs & & \# (epochs) & $\supset$NECs \rule[3pt]{0pt}{8pt} %depth & height
}
\startdata
\multirow{3}{*}{\STAB{\rotatebox[origin=c]{90}{Group}}} & HTCs & 26 (42) & 7 (11) & JFCs & 236 (478) & 44 (84) \\
 & NPCs & 140 (244) & N/A & ETCs & 14 (29) & 1 (6) \\
 & HCs & 68 (173) & N/A & & & \\
\hline
 & {\bf Total} & \multicolumn2{c|}{234 (459)} & {\bf Total} & \multicolumn2c{250 (507)} \\
% \hline
% NECs & \multicolumn3{c|}{a} & \multicolumn2c{a} \\
\enddata
\vspace*{-.1in}
\begin{center}
\tablecomments{Near-Earth Comets (NECs) are defined following JPL/Center for Near Earth Object Studies (CNEOS) criteria\footnote{\url{https://cneos.jpl.nasa.gov/about/neo\_groups.html}}: perihelion distance $q < 1.3$ au and orbital period $P_{\rm orb} < 200$ yrs.}
\end{center}
\end{deluxetable*}
%%%%%%%%%%%%%%%%%%%%%

The detected comets span heliocentric distances ($r_{\rm H}$) from 0.996 to 10.804 au (median of 2.548 au). By group, LPCs were observed between 1.081 and 10.804 au (median: 3.203 au), and SPCs between 0.996 and 6.544 au (median: 2.279 au). In terms of true anomaly ($\nu$), LPCs were detected from $-$162.26\arcdeg\ (pre-perihelion) to 165.52\arcdeg\ (post-perihelion), with a median of 0.38\arcdeg; SPCs spanned from $-$177.16\arcdeg\ to 170.30\arcdeg, with a median of 19.43\arcdeg. While LPC detections were nearly symmetric with respect to perihelion ($q$; 228 pre-$q$ vs. 231 post-$q$ epochs), SPCs showed a post-perihelion bias (194 pre-$q$ vs. 313 post-$q$ epochs), consistent with the activity distribution discussed in Section \ref{sec:sec313}.

% 3.1.2 %%%%%%%%%%%%%%%%%%%%%%%%%%%%%%%%%%%%%%%%%%%%%%%%%%%%%%%%%%%%%%%%%%%%%%%%%%%%%%%%%%%%%%%%%%%%%%%%%%%%%%%%%%
\subsubsection{Band Magnitudes and Submissions to MPC} \label{sec:sec312}

Over the course of its mission, WISE/NEOWISE submitted a substantial number of comet observations to the MPC under observatory code \texttt{C51}. Since January 2010, a total of 7,381 frames corresponding to 378 unique comets were reported, comprising 2,861 frames of 170 LPCs and 4,520 frames of 208 SPCs.

To evaluate the brightness distribution of the detected sample and its relationship to MPC submissions, we compared the magnitudes of our comet samples (S/N $>$ 4) against those reported to the MPC. Because coadded images span multiple frames while MPC submissions are based on individual exposures, we matched each submitted detection to the nearest temporal state from our coadded dataset, removing duplicates to enable one-to-one comparisons. Photometric magnitudes were computed using the procedure described in Section~\ref{sec:sec25}. 

\begin{figure}[hbt]
\centering
\includegraphics[width=0.77\textwidth]{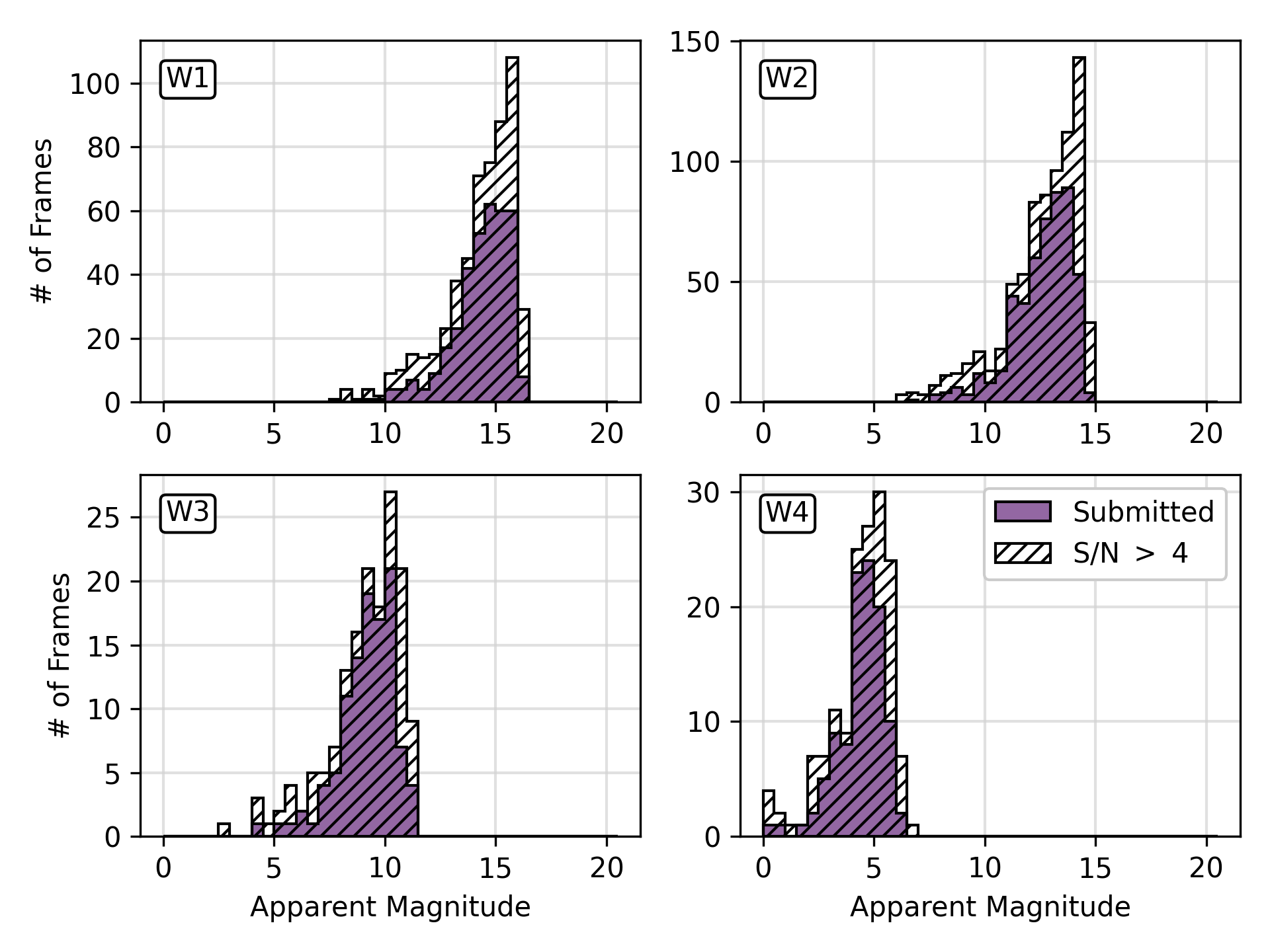}
\caption{Magnitude distributions across the four WISE/NEOWISE bands. Hatched histograms represent fixed-aperture photometry (Section~\ref{sec:sec25}) for detected comets with S/N $>$ 4. Overlaid in purple are magnitudes of comets submitted to the MPC under observatory code \texttt{C51}.
\label{FIG10}} % This ref has to be all CAPS, there is a section header line which references this label, and latex was rendering this as ??. This was because the header line gets rendered as all caps, and this ref was being converted to all caps before finding the associated label.
\end{figure}

As illustrated in Figure~\ref{FIG10}, nearly all MPC-submitted comets fall within our detected sample\footnote{There are 14 comets which were reported to MPC but fell below our S/N threshold of 4. All frames were unable to pass our coaddition selection criteria (Sect. \ref{sec:app0}), except for 162P/Siding Spring, which is visible against a clean background but has an S/N of 3.5.}. The brightness distributions for detected comets in our sample are truncated at approximately 16.3, 14.8, 11.4, and 6.6 mag in the W1, W2, W3, and W4 bands, respectively, due to the imposed S/N $>$ 4 threshold. This results in a non-Gaussian magnitude distribution, with median [25$^{\rm th}$, 75$^{\rm th}$ percentile] values of 14.706 [13.553, 15.496] in W1 (552 frames), 13.006 [11.823, 13.957] in W2 (767 frames), 9.428 [8.401, 10.280] in W3 (155 frames), and 4.764 [3.903, 5.446] in W4 (159 frames). The faint-end tail seen in the MPC-submitted sample is contained in our filtered dataset -- a trend also observed in the corresponding V-band distribution (Fig.~\ref{Figap3}; see Appendix~\ref{sec:app3}).

%%%%%%%%%%%%%%%%%%%%%
%%%%%%%%%%%%%%%%%%%%% 

\begin{figure}[htb]
\centering
\includegraphics[width=1\textwidth]{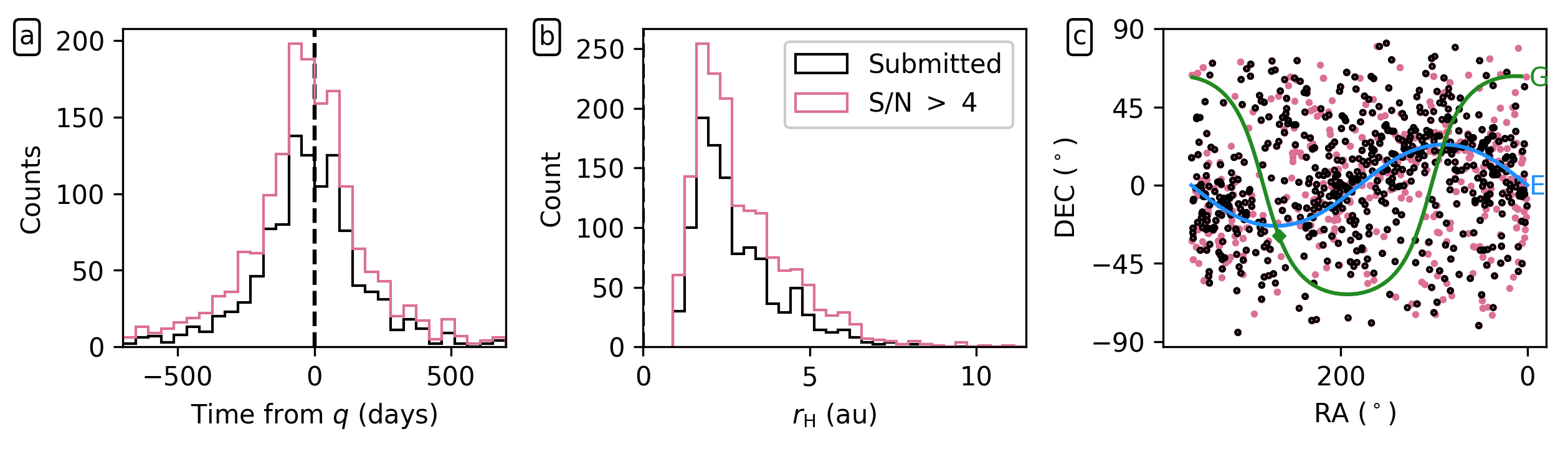}
\caption{Comparison of detected (pink, S/N $>$ 4) and MPC-submitted (black) comet detections as a function of perihelion distance ($q$, Panel a) and heliocentric distance ($r_{\rm H}$, Panel b). Panel c shows their sky distribution in RA and DEC. ``E" and ``G" indicate the Ecliptic and Galactic planes, respectively; the Galactic center is marked by a green diamond at RA = 266.405\arcdeg, DEC = $-$28.936\arcdeg.      
\label{Fig11}}
\end{figure}

Our dataset includes many bright comets (hatched bars in Fig. \ref{FIG10}) that were not reported to the MPC. To assess whether this discrepancy correlates with orbital parameters or observing geometry, we examined the distribution of MPC-submitted comets compared to the full detected sample. Figure \ref{Fig11} summarizes the results. 
Panels a and b show the number of comets as a function of time from perihelion ($q$) and heliocentric distance ($r_{\rm H}$), respectively. Our sample exhibits a consistently higher number of detections across the full $r_{\rm H}$ range, with noticeable differences near perihelion and at small heliocentric distances ($r_{\rm H} \lesssim 2$ au.) Given that comet activity typically peaks near perihelion \citep[e.g.,][]{Womack2021}, this omission likely reflects internal flagging criteria within the WISE/NEOWISE pipeline, which may exclude highly extended, diffuse, or exceptionally high-S/N sources\footnote{WISE Source Catalog Criteria: \url{https://wise2.ipac.caltech.edu/docs/release/allsky/expsup/sec2\_2a.html\#cc\_flags} \\ Moving Object Pipeline Subsystem (WMOPS): \url{https://wise2.ipac.caltech.edu/docs/release/neowise/expsup/sec4\_3.html}}. Panel c presents the spatial distribution of detections in RA-DEC coordinates. MPC-submitted detections are concentrated along the Ecliptic plane and notably sparse near the Galactic plane, likely reflecting the low-inclination orbits of SPCs and increased source confusion in crowded fields. In contrast, our detected sample provides broader coverage across RA–DEC and samples perihelion regions more thoroughly, indicating reduced sensitivity to sky location constraints and offering a more complete view of the comet population.

%%%%%%%%%%%%%%%%%%%%%
%%%%%%%%%%%%%%%%%%%%%

% 3.1.3 %%%%%%%%%%%%%%%%%%%%%%%%%%%%%%%%%%%%%%%%%%%%%%%%%%%%%%%%%%%%%%%%%%%%%%%%%%%%%%%%%%%%%%%%%%%%%%%%%%%%%%%%%%
\subsubsection{Activity} \label{sec:sec313}

The detected comets exhibit diverse activity states (Figs. \ref{Fig05}--\ref{Fig07}), which evolve with their heliocentric distance. While some activation events are triggered by discrete phenomena such as outbursts or impacts (e.g., \citealt{Jewitt2022,Jewitt2025}), cometary activity generally varies smoothly with solar heating. As comets approach the Sun, increased insolation drives the sublimation of volatile ices -- either exposed on the surface or buried beneath a near-nucleus condensation front -- producing extended features such as comae, tails, and trails. These structures evolve on timescales from hours to years and reflect varying gas-to-dust emission ratios (e.g., \citealt{Agarwal2024} and references therein). As such, activity patterns and ejecta morphology offer critical insights into the near-surface composition and thermophysical state of cometary nuclei, otherwise inaccessible through remote sensing. This section presents the overall activity statistics of detected comets, emphasizing trends as a function of true anomaly and dynamical group. 

\begin{figure}[!thb]
\centering
\includegraphics[trim={0.5cm 1.5cm 0.5cm 1.5cm}, clip, width=1\textwidth]{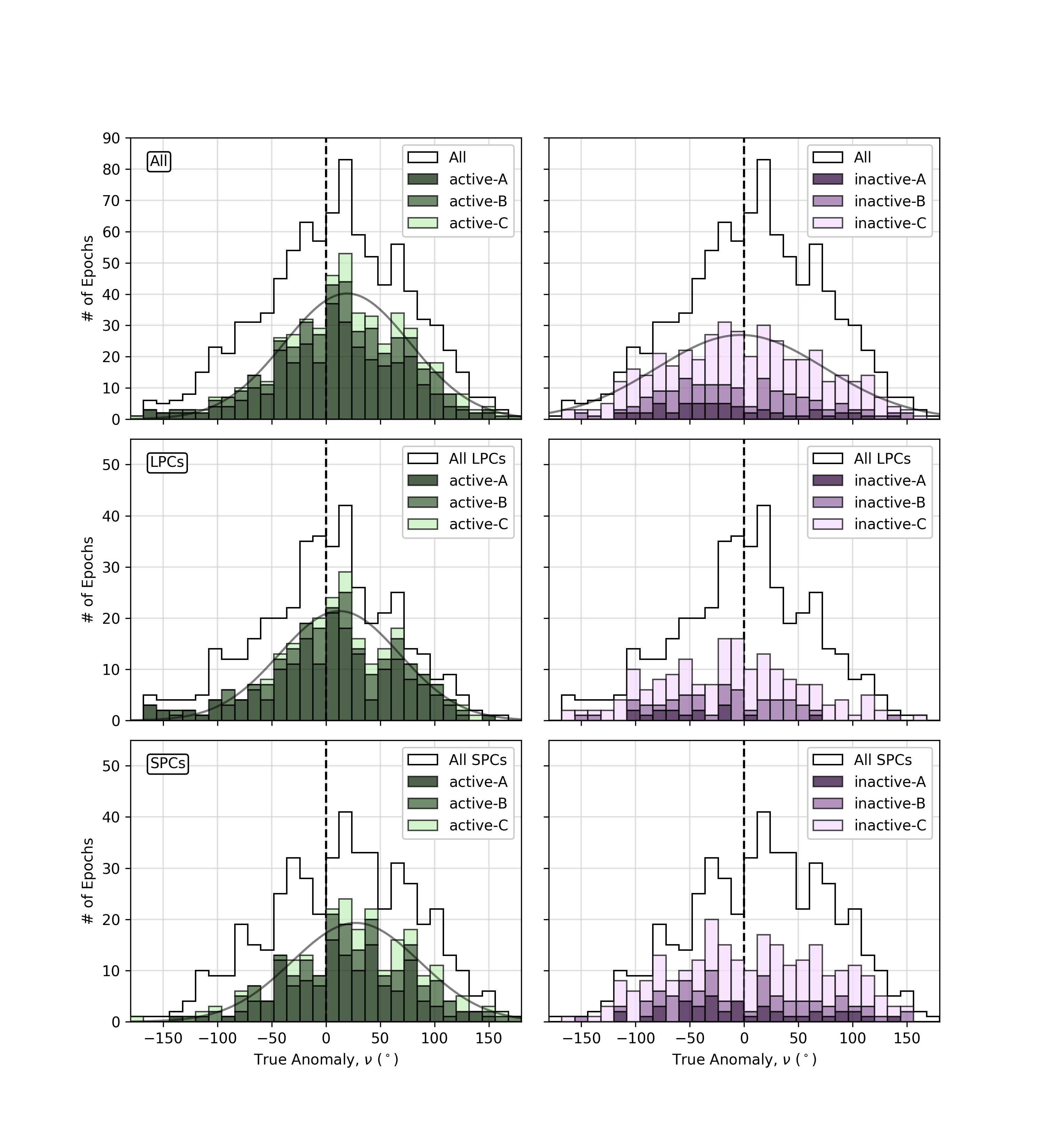} %[trim={left bottom right top},clip]
\caption{Activity distribution across 966 epochs for 484 detected comets (Table \ref{apt01}) as a function of true anomaly ($\nu$). The top, middle, and bottom rows correspond to all (both long- and short-period) comets, LPCs, and SPCs, respectively, with each group divided into active (left) and inactive (right) subsets. Color gradients indicate S/N-based quality grades, with darker shades denoting higher confidence (Grade A = darkest). Gaussian fits are overlaid for approximately symmetric distributions to highlight peak locations. The background pulses seen in the SPCs (bottom row) are artifacts of binning and do not reflect physical features.
\label{Fig12}}
\end{figure}

Figure \ref{Fig12} summarizes the activity status of 484 detected comets across 966 epochs, plotted against true anomaly ($\nu$). This angle provides a consistent reference frame for comparing activation patterns relative to perihelion ($q$) across a broad range of heliocentric distances. Separate distributions are shown for LPCs and SPCs. Activity classifications follow the scheme introduced in Section \ref{sec:sec24}, with lighter shades indicating lower-quality grades (A = darkest, C = lightest). If a comet was detected in multiple bands at a given epoch with differing activation classifications, it was labeled active (``Y") if activity was present in at least one band.

%%%%%%%%%%%%%%%%%%%%%
%%%%%%%%%%%%%%%%%%%%%

% All Active Mean True Anomaly: 16.1  CI(95%) [10.7, 21.4]
% SPC Active Mean True Anomaly: 24.5  CI(95%) [17.0, 32.1]
% LPC Active Mean True Anomaly: 8.1  CI(95%) [1.0, 15.4]
% Active SPC vs Active LPC True Anomaly- KS Test P value: 0.0208
The distributions of all comets (both active and inactive, encompassing LPCs and SPCs), shown in the first row in Fig. \ref{Fig12}, approximately follow Gaussian profiles. Global anisotropy around perihelion ($q$) is more pronounced in active comets, whereas in inactive comets, it is primarily driven by lower-grade N-C and, to some extent, N-B groups and becomes less apparent when considering only N-A comets. We performed Gaussian fitting using \texttt{scipy.optimize.minimize}, estimating peak positions of $\nu$ for each active group, and used the \texttt{bootstrap} module in \texttt{scipy.stats} to compute 95 \% confidence intervals (CI). Results, rounded to the nearest tenth of a degree, are as follows:  
\begin{enumerate}[itemsep=-.01in]
    \item[$\bullet$] Active comets (all types): mean $\nu$ = 16.1\arcdeg, 95 \% CI = [10.7\arcdeg, 21.4\arcdeg].
    \item[$\bullet$] Active LPCs: mean $\nu$ = 8.1\arcdeg, 95 \% CI = [1.0\arcdeg, 15.4\arcdeg].
    \item[$\bullet$] Active SPCs: mean $\nu$ = 24.5\arcdeg, 95 \% CI = [17.0\arcdeg, 32.1\arcdeg].
\end{enumerate}

While both (LPC + SPC) groups show peak activity slightly after $q$, SPCs display a more pronounced post-perihelion skew. This trend, also observed at the individual object level \citep[e.g.,][]{Kelley2008,Epifani2009,Knight2017,Kulyk2018}, likely arises from thermal lag: solar heat absorbed pre-perihelion takes time to propagate inward and sublimate buried ices. This behavior suggests the presence of a refractory surface mantle and is further discussed in Section~\ref{sec:insight}. A Kolmogorov–Smirnov test comparing the $\nu$ distributions of active LPCs and SPCs yields a $p$-value of 0.021, indicating a statistically significant difference.

Inactive comets exhibit a less structured (flatter) $\nu$ distribution, particularly among high-confidence (N-A) cases. While lower-grade inactive comets (N-B and N-C) tend to cluster near $q$, the more uniformly distributed N-A group likely reflects reliable non-detections. The higher apparent inactivity near perihelion in the overall sample may stem from unresolved contamination by circumnuclear signal, primarily driven by low S/N levels (see Sect.~\ref{sec:sec24}). 
\\

% 3.2 Photometry %%%%%%%%%%%%%%%%%%%%%%%%%%%%%%%%%%%%%%%%%%%%%%%%%%%%%%%%%%%%%%%%%%%%%%%%%%%%%%%%%%%%%%%%%%%%%%%%%
\subsection{Photometric Properties} \label{sec:phot}

% A comet’s observed brightness reflects both its intrinsic properties and external influences, such as heliocentric temperature. This section presents the photometric characteristics of the detected comets, including brightness measurements and their statistical distributions.

% 3.2.1 %%%%%%%%%%%%%%%%%%%%%%%%%%%%%%%%%%%%%%%%%%%%%%%%%%%%%%%%%%%%%%%%%%%%%%%%%%%%%%%%%%%%%%%%%%%%%%%%%%%%%%%%%%
\subsubsection{Brightness Distribution} \label{sec:sec321}

Photometry presented in this section was performed using a fixed-size aperture in image space (i.e., not scaled by cometocentric distance), with band-specific radii defined in Section \ref{sec:sec25}. Given the wide range of heliocentric distances ($r_{\rm H}$) and geocentric distances ($\Delta$) sampled over the mission, a fixed aperture ensures uniformity in data processing across all epochs. An analysis of flux variation with aperture size will be presented in a forthcoming paper. 
% We adopted the smallest test aperture -- 2.4 pixels (6.6\arcsec) -- rather than larger options (7 pixels or 12 pixels, equivalently 19.3\arcsec\ or 33.0\arcsec, respectively), as it sufficiently encloses the point-spread function (PSF) for the W1–W3 bands and unbinned W4, while minimizing background contamination and uncertainties from stellar crowding and diffuse noise, especially during the Reactivation phase, when W1 and W2 were more sensitive to stellar background. 
Figure~\ref{FIG10} displays the resulting band magnitude distribution for all detected comets using this aperture size. Histogram statistics, categorized by dynamical class, are summarized in Table~\ref{t02}. 
% As noted in Section~\ref{sec:sec23}, all values are subject to the S/N $>$ 4 threshold, which excludes visually apparent but photometrically unreliable comets. Consequently, the magnitude distributions are truncated rather than Gaussian, with sharp cutoffs corresponding to the detection limit in each band.

% Across all bands, SPCs have narrower magnitude distributions than LPCs, with lower minimum and higher maximum magnitudes. 
Notable trends appear in the W1 and W2 bands, which comprise the bulk of the observations. LPCs are statistically brighter than SPCs, a trend that holds across heliocentric distances (see Section~\ref{sec:sec322}). Among LPCs, HCs are slightly brighter on average than NPCs and HTCs, although all three subgroups share similar brightness ranges. Among SPCs, JFCs are consistently brighter than ETCs. In the W3 and W4 bands, dominated by thermal emission, the statistical differences between groups are less distinct. ETCs appear slightly skewed toward higher brightness compared to JFCs in these bands, but the sample sizes are too small for conclusive interpretation. Given the limited number of detections in W3 and W4, further discussion of brightness trends in these bands is deferred to future work.

%%%%%%%%%%%%%%%%%%%%%
\begin{deluxetable*}{cc|ccc|cc}[thb]
%{c{1cm}|cc{2cm}c{2cm}cc{2cm}c{2cm}}
%>{\RaggedRight}m{6.3cm}}
%{cccc|cc} 
\tabletypesize{\footnotesize} %\scriptsize
%\movetableright=-1in
\tablewidth{0pt} 
\tablecaption{Summary of photometric statistics from 1,633 frames of 484 detected comets in Figure \ref{FIG10}. \label{t02}}
\tablehead{
\colhead{\multirow{2}{*}{Band}} &
\colhead{\multirow{2}{*}{Parameter}} \vline&
\multicolumn3c{LPCs} \vline&
\multicolumn2c{SPCs} \\ 
\cline{3-7}
& & HTCs & NPCs & HCs & JFCs & ETCs
}
\startdata
\multirow{4}{*}{W1} & \# & 23 & 129 & 142 & 245 & 13 \\
 & mean & 14.224 & 14.150 & 13.831 & 14.521 & 15.107 \\
 & median & 14.493 & 14.706 & 14.326 & 14.810 & 15.592 \\
  & 25$^{\rm th}$--75$^{\rm th}$ & [13.502, 15.144] & [13.269, 15.511] & [12.712, 15.361] & [13.931, 15.502] & [14.401, 15.821] \\
\hline
\multirow{4}{*}{W2} & \# & 40 & 185 & 148 & 375 & 19 \\
 & mean & 12.638 & 12.485 & 12.223 & 12.770 & 13.477 \\
 & median & 12.667 & 12.963 & 12.827 & 13.051 & 13.977 \\
  & 25$^{\rm th}$--75$^{\rm th}$ & [11.593, 14.049] & [11.704, 13.959] & [11.123, 13.789] & [12.058, 13.931] & [13.401, 14.161] \\
\hline
\multirow{4}{*}{W3} & \# & 3 & 41 & 11 & 90 & 10 \\
 & mean & 7.240 & 9.237 & 8.663 & 9.168 & 8.879 \\
 & median & 7.758 & 9.791 & 9.581 & 9.453 & 8.725 \\
  & 25$^{\rm th}$--75$^{\rm th}$ & [6.658, 8.081] & [8.522, 10.453] & [8.404, 10.057] & [8.535, 10.279] & [7.870, 9.832] \\
\hline
\multirow{4}{*}{W4} & \# & 3 & 51 & 12 & 85 & 8 \\
 & mean & 3.248 & 4.345 & 3.251 & 4.582 & 3.878 \\
 & median & 3.748 & 4.642 & 3.686 & 4.949 & 3.846 \\
  & 25$^{\rm th}$--75$^{\rm th}$ & [2.755, 3.991] & [3.630, 5.425] & [2.717, 4.788] & [4.220, 5.482] & [2.661, 5.014] \\
\enddata
\vspace*{-.1in}
\end{deluxetable*}
%%%%%%%%%%%%%%%%%%%%%

% 3.2.2 %%%%%%%%%%%%%%%%%%%%%%%%%%%%%%%%%%%%%%%%%%%%%%%%%%%%%%%%%%%%%%%%%%%%%%%%%%%%%%%%%%%%%%%%%%%%%%%%%%%%%%%%%%
\subsubsection{Heliocentric Evolution of Brightness} \label{sec:sec322}

To evaluate how comet brightness as a whole responds to heliocentric temperature variations, we analyzed the distribution of detected magnitudes as a function of heliocentric distance ($r_{\rm H}$), focusing on the W1 and W2 bands. Both bands primarily trace scattered sunlight and provide sufficient statistical coverage for analysis (Table~\ref{t02}). 
In this analysis, we assume that the dominant contributiona to the W1 and W2 band signals arises from sunlight scattered by dust particles, either within the coma/tail or on the nucleus surface. This assumption is supported by the the similarity in the large-scale $r_{\rm H}$-dependent trends between the two bands that will be discussed below. Nonetheless, we acknowledge potential contamination from gas emission features, most notably CO at 4.7~$\mu$m and CO$_2$ at 4.3~$\mu$m in W2 (e.g., \citealt{Reach2009,Reach2013,Bauer2015,Bauer2017}), as well as possible contributions in W1, such as methanol. 
Appendix~\ref{sec:app4-0} provides further discussion on the influence of gas emissions, with Figure~\ref{Filter_trans} showing the central wavelengths of emission lines that may blend with the dust continuum in W1 and W2. A decomposition of gas and dust contributions will be presented in a forthcoming COSINE paper, following the separation of nucleus and extended components. For the present analysis, we therefore combine W1 and W2 data to examine the overall brightness behavior as a function of $r_{\rm H}$.
% While W2 includes a slightly greater contribution from thermal emission than W1, along with possible CO or CO$_2$ gas emission (Fig. \ref{Fig}; e.g., \citealt{Bauer2015,Bauer2017}), preliminary comparisons reveal no substantial differences in their large-scale $r_{\rm H}$ trends. 

We selected 1,354 frames of detected comets -- both active and inactive -- in W1 and W2, and computed their absolute magnitudes, $m(1,1,0)$, standardized to $r_{\rm H} = \Delta = 1$ au and zero phase angle, correcting for effects of varying observing geometry:
\begin{equation}
m(1,1,0) = m_{\rm app} - 5\log_{10}(r_{\rm H}\Delta) - \Phi(\alpha)~.
\label{eq1}
\end{equation}
\noindent Here, $m_{\rm app}$ is the observed apparent magnitude (Section~\ref{sec:sec25}), $r_{\rm H}$ and $\Delta$ are the heliocentric and geocentric distances in au, and $\alpha$ is the phase angle. Constant 5 in the second term corrects for inverse-square flux scaling with distance, isolating intrinsic brightness. The phase function $\Phi(\alpha) = \beta \alpha$ accounts for phase darkening. Due to the lack of published $\beta$ values for cometary dust in W1--W2 wavelengths, we tested a range of $\beta$ from 0 to 0.06 mag/\arcdeg\ (the latter is consistent with extreme values such as for 2P/Encke and 48P/Johnson; \citealt{Fernandez2000,Jewitt2003}). Due to WISE/NEOWISE's low-Earth orbit and a narrow solar elongation of $\sim$$\pm$90\arcdeg\ \citep{Wright2010}, observations sampled the high end of $\alpha$ possible at a given $r_{\rm H}$ (5.05 to 87\arcdeg; Fig. \ref{Figap4}). In particular, $\alpha$ varies rapidly within $\sim$2 au, increasing from $\sim$20\arcdeg\ to $>$80\arcdeg\ as $r_{\rm H}$ decreases, thereby amplifying the importance of applying phase corrections. All non-zero $\beta$ values reduced magnitude scatter in the inner Solar System, with larger $\beta$ values yielding stronger effects. As no qualitative differences emerged across the tested values, we adopted $\beta = 0.035$ mag/\arcdeg\ \citep{Lamy2004}, while acknowledging that this value may not precisely reflect infrared behavior\footnote{The Schleicher-Marcus model \citep{Marcus2007,Schleicher2011} is also widely used for phase correction in optical observations, particularly to account for the non-linear brightening of cometary dust in the back-scattering and forward-scattering regimes. Nevertheless, we chose a simpler linear correction for this work, as most of our data span phase angles $\sim$15--60\arcdeg\ and lie in the infrared spectral range.}.
%Therefore, we adopted a nominal linear phase function, $\Phi(\alpha)$ = $\beta\alpha$, with $\beta$ = 0.035 \arcdeg/mag, commonly used in optical studies \citep{Lamy2004}.

\begin{figure}[htb]
\centering
\includegraphics[width=1\textwidth]{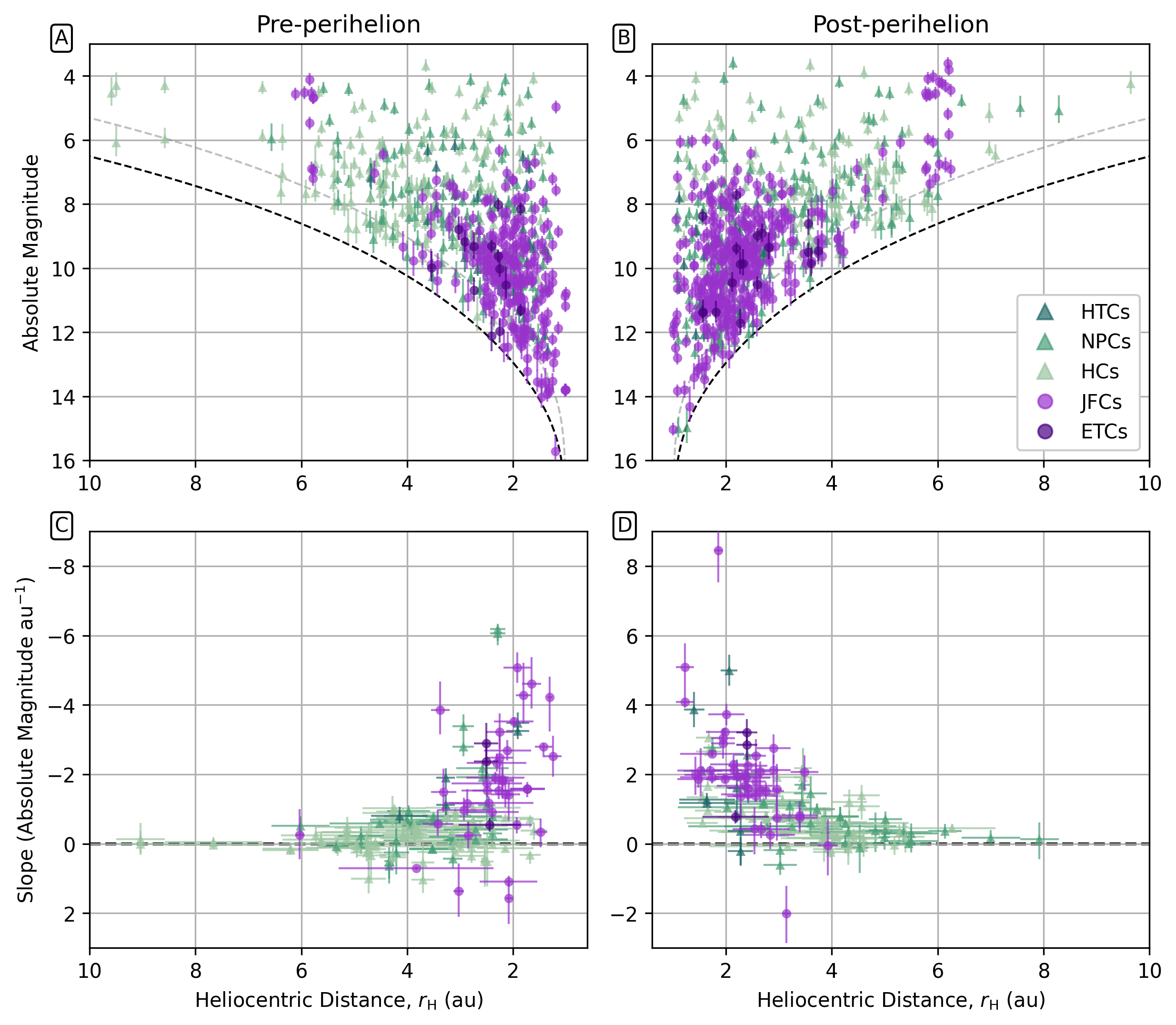}
\caption{Top: Absolute magnitude distribution 1,319 frames of 406 detected comets in the W1 and W2 bands as a function of heliocentric distance ($r_{\rm H}$), shown separately for pre-perihelion (panel a) and post-perihelion (panel b) detections. The dashed curves indicate limiting magnitude trendlines based on the W1-band (16.3 mag, black) and W2-band (14.8 mag, gray) detection thresholds (Fig. \ref{FIG10}), assuming typical WISE/NEOWISE observing geometry ($\sim$90\arcdeg\ solar elongation). Bottom: Linear brightness slopes (mag/au) computed from pairs of consecutive detections in W1 or W2 bands, where two consecutive epochs in the same band occurred either both pre- or both post-perihelion. Panels c and d show slope distributions as a function of $r_{\rm H}$. Horizontal bars indicate the $r_{\rm H}$ separation between the paired detections. Comets are color-coded by dynamical group.
\label{Fig13}}
\end{figure}

The first row of Figure~\ref{Fig13} presents the resulting absolute magnitude distribution (phase-corrected by Eq. \ref{eq1}) as a function of $r_{\rm H}$, separated into pre- (panel a) and post-perihelion (panel b) detections. Excluding 29P/Schwassmann–Wachmann at $r_{\rm H}$ $\sim$6 au, SPCs are primarily detected within 4 au, while LPCs extend to $\sim$10 au. Within 4 au, SPCs occupy the mid-to-faint end of the brightness range, whereas LPCs span a broader and systematically brighter distribution -- up to $\sim$4 magnitudes brighter -- consistent with trends summarized in Table~\ref{t02}. A conspicuous absence of detections in the lower left region of the panels reveals a coverage gap: fainter comets were only observed at smaller $r_{\rm H}$, outlining a brightness envelope similar to that described by \citet{Francis2005}. This pattern likely reflects the WISE/NEOWISE detection threshold, which is well approximated by the overlaid dashed curves. These curves represent the S/N = 4 limits derived from the W1-band (16.3 mag, black) and W2-band (14.8 mag, gray) detection thresholds under typical WISE/NEOWISE observing geometry ($\sim$90\arcdeg\ solar elongation). The broader influence of observing geometry on large-scale photometric trends is discussed further in Appendix \ref{sec:app5}. Post-perihelion detections dominate among SPCs, while LPCs show no strong asymmetry around perihelion. No statistically significant differences were identified among subgroups within either population.

%%%%%%%%%%%%%%%%%%%%%
%%%%%%%%%%%%%%%%%%%%%

We then downselected frames with at least two W1 or W2 detections, requiring that both detections fall either before or after perihelion. Assuming linear brightness evolution between these epochs, we calculated brightness slopes (mag/au), where negative values indicate brightening and positive values indicate fading. The bottom panels of Figure~\ref{Fig13} show these slopes as a function of $r_{\rm H}$, with horizontal bars representing the $r_{\rm H}$ interval used.

Several trends emerge. First, no single slope adequately captures brightness evolution across the full $r_{\rm H}$ range, consistent with prior findings \citep{KrishnaSwamy2010,Holt2024}. Slope variability increases within $\sim$4 au, aligning with the region where water-ice-driven sublimation accelerates \citep{Fulle2020,Ciarniello2022,Kwon2024}. Second, the SPCs' slope distribution is wider than the LPCs. Third, some LPCs (including several HCs) show fading trends between $\sim$5 and 2 au pre-perihelion, consistent with prior reports \citep{Kwon2024,Moreno2025}. Fourth, post-perihelion slopes are steeper on average, suggesting more rapid fading. SPCs, in particular, are underrepresented in the low-slope fading regime ($\lesssim$2 mag/au) near $r_{\rm H} \lesssim 2$ au post-perihelion. Finally, SPCs are virtually absent beyond $\sim$4 au in both pre- and post-perihelion epochs, aside from 29P/Schwassmann–Wachmann near 6 au. Although some SPCs were detected out to $\sim$6.5 au (Fig.~\ref{Fig09}, Table~\ref{apt01}), these were primarily observed during the Cryo phase in W3 and W4 bands, which were excluded from the present slope analysis.
\\

% 3.3 Morphological analysis %%%%%%%%%%%%%%%%%%%%%%%%%%%%%%%%%%%%%%%%%%%%%%%%%%%%%%%%%%%%%%%%%%%%%%%%%%%%%%%%%%%%%
\subsection{Morphological Analysis of Active Comets} \label{sec:sec33}

% This section analyzes the morphology of detected comets, with a particular focus on the orientation of extended features in active cases. The results provide foundational context for the dynamical modeling presented in the subsequent paper of the COSINE series. We also assess how the WISE/NEOWISE observing geometry influenced the sampled comet population and shaped detection patterns across the dataset.

% 3.3.1 %%%%%%%%%%%%%%%%%%%%%%%%%%%%%%%%%%%%%%%%%%%%%%%%%%%%%%%%%%%%%%%%%%%%%%%%%%%%%%%%%%%%%%%%%%%%%%%%%%%%%%%%%%
\subsubsection{Outcomes of Observing Geometry} \label{sec:sec331}

WISE/NEOWISE was designed to observe at a fixed solar elongation of approximately 90\arcdeg\ \citep{Wright2010}. In practice, the scan direction on the trailing side drifted by $\sim$10\arcdeg\ to compensate for orbital precession, while the leading side was held at 90\arcdeg\ due to sun-pointing restrictions\footnote{See Figure 8 in the NEOWISE Explanatory Supplement: \url{https://wise2.ipac.caltech.edu/docs/release/neowise/expsup/sec1\_2.html}}. Before analyzing comet morphology, we evaluated how this restricted viewing geometry -- arising from low-Earth orbit \emph{and} narrow solar elongation -- shaped the sampled comet population. 

We first investigated the positional angles of two key vectors used to characterize extended cometary features: the anti-solar vector ${\bf r}_\odot$ and the negative velocity vector $-{\bf v}$, both of which are commonly employed to infer dust properties and mass-loss histories \citep{Fulle2004,Agarwal2024}. We computed daily variations of these vectors for all 1,335 queryable candidates (Section~\ref{sec:sec21}) across the entire mission span, from the start of the Cryo phase on January 14, 2010, to the end of the Reactivation phase on July 31, 2024. This calculation allowed us to map the regions of the sky WISE/NEOWISE could access, given its observing geometry.

\begin{figure}[htb]
\centering
\includegraphics[width=0.87\textwidth]{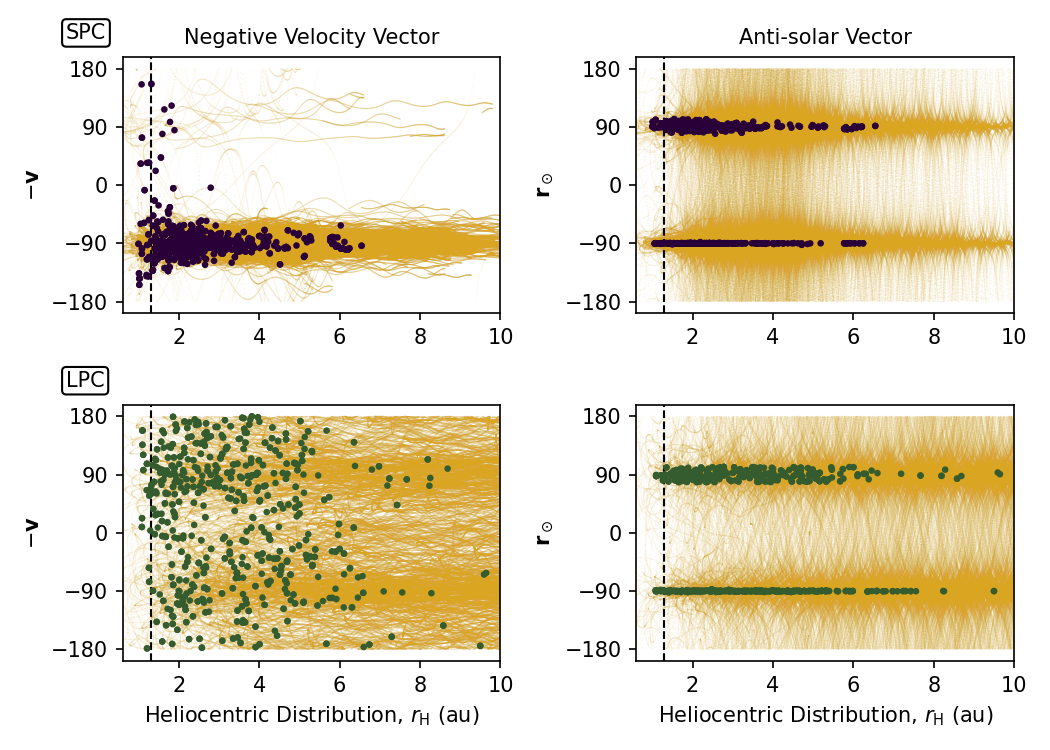}
\caption{Distributions of the anti-solar vector ${\bf r}_\odot$ (right) and negative velocity vector $-{\bf v}$ (left) for 1,335 queryable comets (Sect. \ref{sec:sec21}) as a function of heliocentric distance ($r_{\rm H}$). Daily variations are shown as golden background strands. Colored circles denote the vector orientations and $r_{\rm H}$ values for 966 epochs of 485 detected comets, with SPCs and LPCs color-coded separately. Vertical dashed lines indicate the 1.3 au perihelion threshold for NECs.
\label{Fig14}}
\end{figure}

% mind that the golden rods are not WISE-exclusive. All near/on-ground observatories will follow the same pattern.
Figure \ref{Fig14} displays the results. The golden strands represent the daily position-angle evolution of ${\bf r}_\odot$ and $-{\bf v}$ for each comet, measured counterclockwise from ecliptic north. Superimposed are 966 epochs of detected comets (Table~\ref{apt01}), shown as colored circles and distinguished by dynamical class, illustrating how the observed comets are distributed within the available vector space.

%%%%%%%%%%%%%%%%%%%%%
%%%%%%%%%%%%%%%%%%%%%

The appearance of golden strand structures is not unique to WISE/NEOWISE. Earth-based and near-Earth observations, excluding in-situ missions, naturally exhibit clustering in vector distributions near $\pm$90\arcdeg\ due to geometric projection effects. This clustering arises from the apparent motion of objects as viewed from an Earth-based vantage point.
WISE/NEOWISE’s narrow solar elongation constraint further limited sky accessibility, clearly reflected in the concentration of markers in Figure~\ref{Fig14}. The anti-solar vectors ${\bf r}_\odot$ (right panels) exhibit stronger clustering than the $-{\bf v}$ vectors (left panels), as the former is dictated entirely by observational geometry, while the latter also incorporates the underlying orbital dynamics of the comet population. As a result, the ${\bf r}_\odot$ vectors cluster near $\pm$90\arcdeg, corresponding to trailing and leading scans, respectively, whereas $-{\bf v}$ vectors display a wider spread, modulated by the inclination and directionality of comet orbits. LPCs (HTCs, NPCs, HCs), or so-called nearly isotropic comets \citep{Levison1996}, show a relatively uniform azimuthal distribution of $-{\bf v}$ vectors due to their random orbital orientations, which encompass both prograde and retrograde trajectories. Conversely, SPCs (JFCs, ETCs) are dynamically confined to the ecliptic and generally follow low-inclination, prograde orbits. As a result, their $-{\bf v}$ vectors cluster distinctly near $-90\arcdeg$, consistent with an origin from an ecliptic-aligned source population \citep{Duncan1997,Levison1997}.

These results highlight the need to account for geometric sampling biases in large-scale morphological analyses. Observations from low- or near-Earth vantage points—particularly under narrow solar elongation constraints—produce uneven sky coverage that can obscure or distort the true distribution of extended cometary features. Such limitations may introduce degeneracies in interpreting tail orientations and activity vectors at the \emph{population} level. Consequently, any statistical or global inferences drawn from these datasets must be carefully corrected for these observational biases. Forthcoming surveys with broader elongation coverage, such as the \emph{NEO Surveyor} \citep{Mainzer2023} and ground-based efforts like \emph{LSST} \citep{LSST2021}, are expected to mitigate these observational constraints, enabling more complete and unbiased morphological studies.

% 3.3.2 %%%%%%%%%%%%%%%%%%%%%%%%%%%%%%%%%%%%%%%%%%%%%%%%%%%%%%%%%%%%%%%%%%%%%%%%%%%%%%%%%%%%%%%%%%%%%%%%%%%%%%%%%%
\subsubsection{Relative Angle Distribution of Extended Features} \label{sec:sec332}

Building on the previous section, we investigated the global distribution of the position angles (PAs) of extended features in active comets to identify potential large-scale trends linked to intrinsic dust properties. From the full sample of 484 detected comets, we selected those classified as active with high S/N ($>$10, labeled ``Y-A"), yielding 757 frames for 279 comets across the W1--W4 bands. PAs were measured as described in Section \ref{sec:sec24}. 

To explore potential correlations, we compared each PA with the corresponding directions of the anti-solar vector (${\bf r}_\odot$) and the negative velocity vector ($-{\bf v}$). Alignment toward ${\bf r}_\odot$ may reflect the dominance of small, low-density, and/or carbon-rich grains driven by solar radiation pressure. Conversely, alignment with $-{\bf v}$ is generally associated with larger, denser, or less-volatile particles ejected with higher inertia \citep{Finson1968,Burns1979}. However, no statistically significant correlations from our dataset were found between PAs and either vector, whether considering the full sample or subdividing by dynamical class. This absence of trend likely stems from geometric limitations imposed by the WISE/NEOWISE observing strategy. As illustrated by the circle markers in Figure~\ref{Fig14}, the narrow solar elongation window restricted the available viewing geometry, introducing degeneracies in the PA distribution. Consequently, PA orientations largely reproduce the sampling biases evident in Figure~\ref{Fig14}, rather than reflecting underlying physical differences.

\begin{figure}[tbh]
\centering
\includegraphics[width=0.67\textwidth]{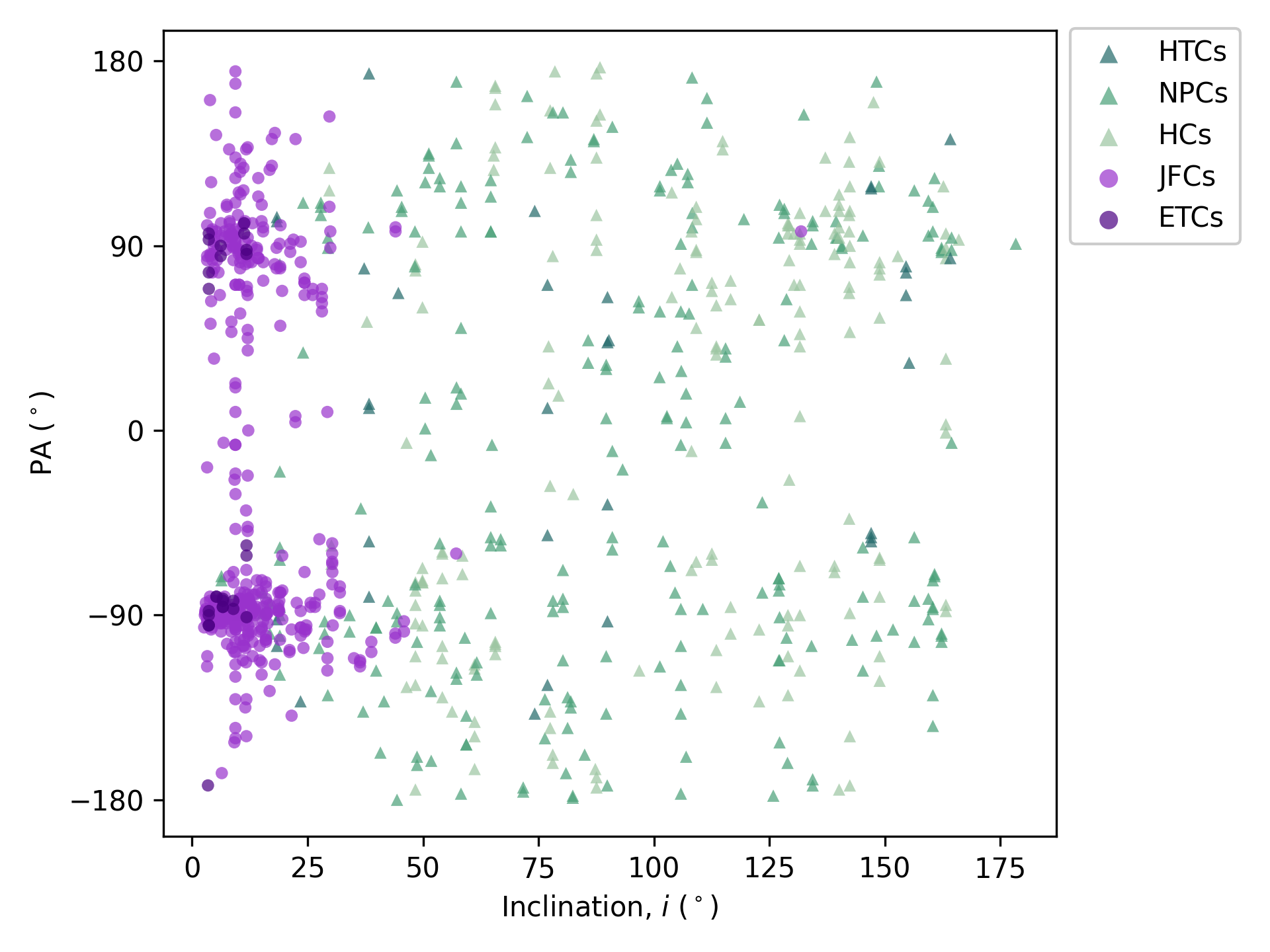}
\caption{Distribution of primary position angles (PAs) for 695 frames of 255 comets classified as “Y-A” (active with S/N $>$ 10) across W1--W4 bands, plotted against orbital inclination ($i$). PA measurements follow the method described in Section~\ref{sec:sec24}. Dynamical group symbols match those in Figure~\ref{Fig14}.
\label{Fig15}}
\end{figure}

Figure \ref{Fig15} reinforces this interpretation by plotting PA against orbital inclination ($i$). Given the distinct $i$ distributions of LPCs and SPCs (Fig. \ref{Fig09}), the color-coded structure in Figure \ref{Fig15} reflects how orbital geometry interacts with observational constraints. For instance, SPCs -- with a median $i$ of 12.2\arcdeg\ in our sample -- show clustering of PAs near $\pm$90\arcdeg, consistent with the restricted viewing geometry illustrated in Figure \ref{Fig14}. This clustering diminishes with increasing $i$ and is largely absent among LPCs, whose broader ${\bf r}_\odot$ and $-{\bf v}$ vector distributions lead to a more dispersed range of observable PAs. Nonetheless, a modest concentration near $\pm$90\arcdeg\ remains, imposed by elongation-driven constraints (Fig. \ref{Fig14}). 

While the \emph{global} distribution of feature alignments appears largely governed by geometric sampling biases and does not reveal significant trends across LPCs, SPCs, or their subgroups, the PA distributions of certain comets are more dispersed compared to their tightly clustered ${\bf r}_\odot$ vectors close to $\pm$90\arcdeg\ distribution (right column of Fig. \ref{Fig14}). For those comets located outside the identified degeneracy zones, alignment angles would provide meaningful insights into the intrinsic properties of the comets. A morphological analysis of these \emph{individual} comets and a discussion of their associated mass-loss histories will be presented in forthcoming COSINE papers.

% Overall, no clear global PA trends were found among LPCs or their subgroups. 
% Taken together, Sections~\ref{sec:sec331} and~\ref{sec:sec332} underscore that \emph{global} analyses of feature orientations must carefully account for the geometry of the observing platform. Nevertheless, for \emph{individual} comets located outside the identified degeneracy zones, or through spatially-resolved analyses of extended structures rather than relying solely on a single dominant PA, alignment measurements can still provide meaningful insight into dust ejection behavior and mass-loss history. These cases will be explored in forthcoming COSINE papers.

%%%%%%%%%%%%%%%%%%%%%
%%%%%%%%%%%%%%%%%%%%%

% 4. %%%%%%%%%%%%%%%%%%%%%%%%%%%%%%%%%%%%%%%%%%%%%%%%%%%%%%%%%%%%%%%%%%%%%%%%%%%%%%%%%%%%%%%%%%%%%%%%%%%%%%%%%%%%%
\section{Initial Insights for the Intrinsic Properties of Comets} \label{sec:insight}

% While comets exhibit broadly similar collective behaviors, notable differences and gradients also emerge both between and within dynamical groups. In this section, we examine the shared photometric characteristics of the dataset and assess how these general trends compare across comet populations. This analysis provides initial insights into the intrinsic physical properties of comets as revealed by WISE/NEOWISE observations.

% % 4.2 %%%%%%%%%%%%%%%%%%%%%%%%%%%%%%%%%%%%%%%%%%%%%%%%%%%%%%%%%%%%%%%%%%%%%%%%%%%%%%%%%%%%%%%%%%%%%%%%%%%%%%%%%%%%
% \subsection{Comparison of Photometric Behaviors Across Dynamical Group of Comets}\label{sec:group_comp}

One of the primary contributions of the COSINE project is to provide population-level insights derived from a large and uniformly analyzed comet sample.
The comets detected by WISE/NEOWISE exhibit broadly uniform observational patterns, such as detection probability, brightness evolution, and activity levels, as a function of geometry, despite substantial individual variation. In general, brightness increases as comets approach the Sun and fades as they recede, with peak brightness typically occurring near perihelion $q$ (Fig. \ref{Fig13}). Similarly, the fraction of active comets rises near $q$, yielding a Gaussian-like distribution in true anomaly centered around $q$ (Fig. \ref{Fig12}). These shared behaviors point to a fundamental, solar-irradiation-driven mechanism governing how comets work across populations.

Despite these overarching similarities, statistical differences between dynamical groups, particularly LPCs and SPCs, are evident. While their brightness distributions overlap, LPCs are systematically brighter than SPCs at comparable heliocentric distances ($r_{\rm H}$), dominating the bright end of the distribution, whereas SPCs populate the faint end (Fig. \ref{Fig13}). 
This finding is consistent with previous studies based on smaller comet samples (\citealt{Bauer2024} and references therein), particularly those reporting systematic differences in optical brightness (e.g., \citealt{Kresak1982,Fernandez1999,Meech2004,Betzler2023,Lacerda2025}) and dust production rates (e.g., \citealt{Solontoi2012,Sarneczky2016}).
The systematic brightness difference likely reflects the diminished activity of SPCs \citep{Meech2004}, attributed to the progressive depletion of near-surface (super-)volatile ices through repeated perihelion passages \citep{Jewitt2002,Prialnik2004}. Such evolutionary effects would also influence dust particle properties, leading comets to have varying sensitivity across bands \citep{Gundlach2020}. Nucleus size may further contribute: \citet{Bauer2017} found a mean LPC-to-JFC size ratio of 1.6 using Cryo and 3-Band WISE/NEOWISE data, consistent with the earlier estimate by \citet{Meech2004}. If LPCs are on average larger, they are more likely detectable at greater distances.

\begin{figure}[!thb]
\centering
\includegraphics[width=0.6\textwidth]{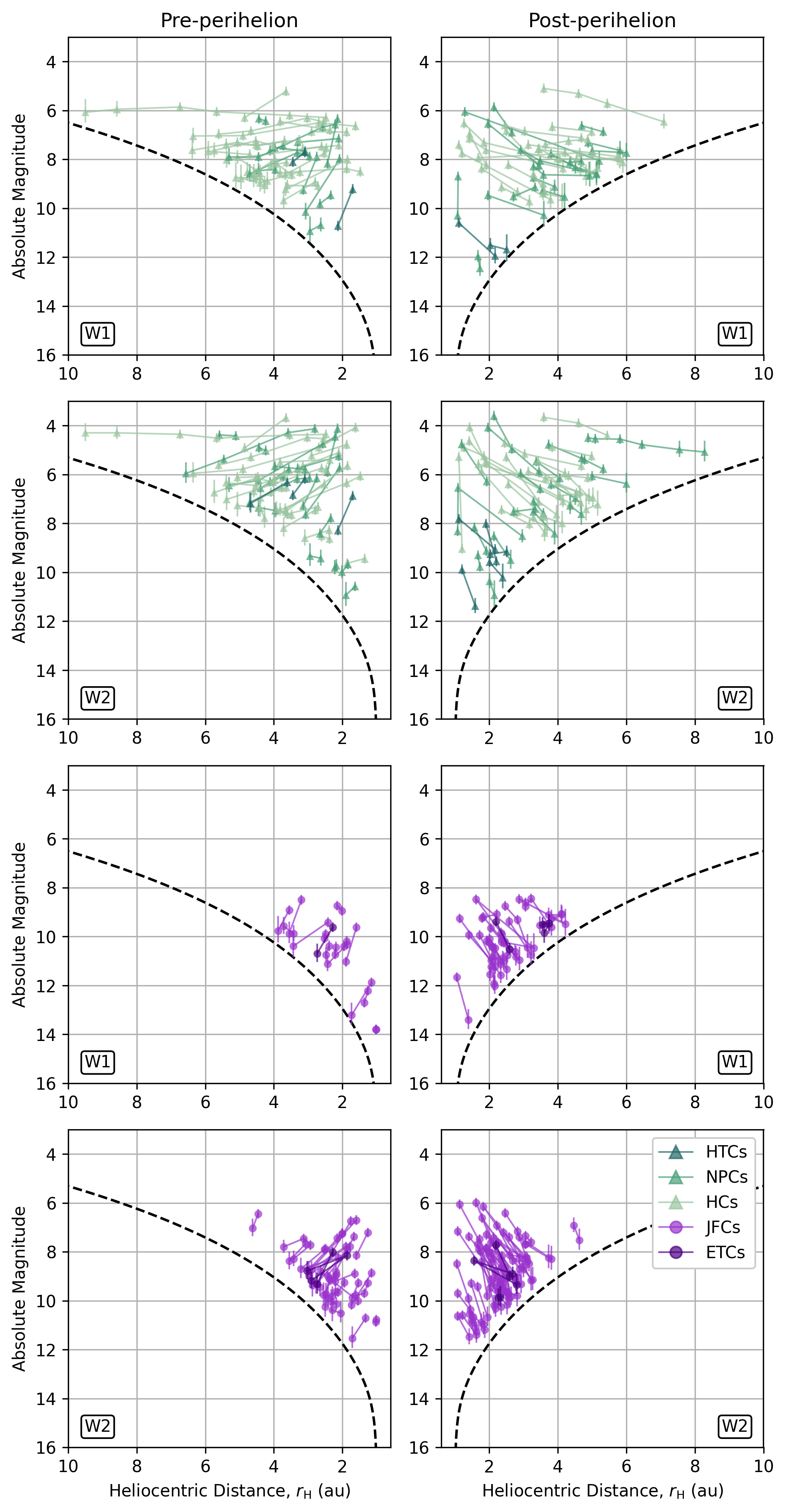}
\caption{Brightness evolution of comets as a function of heliocentric distance, grouped by band and dynamical class. The top two panels and the bottom two panels correspond to LPCs and SPCs, respectively. Only comets observed more than twice were included. Absolute magnitudes (Eq.~\ref{eq1}) are connected by lines. JFC 29P/Schwassmann–Wachmann was excluded due to limited $r_{\rm H}$ coverage. The dashed curves indicate limiting magnitude trendlines based on the W1 and W2 band detection thresholds (16.7 mag and 16.0 mag, respectively; Fig. \ref{FIG10}), assuming typical WISE/NEOWISE observing geometry ($\sim$90\arcdeg\ solar elongation).
\label{Fig16}}
\end{figure}

The above distinctions are more apparent when examining the heliocentric brightness evolution of individual comets, which offer a complementary view to the aggregate trends in the lower panels of Figure~\ref{Fig13}. Figure~\ref{Fig16} presents absolute magnitude evolution (Eq.~\ref{eq1}) for the comets observed at $\ge$two epochs, with consecutive detections occurring either pre- or post-perihelion, grouped by band and dynamical type. W2 magnitudes are typically slightly brighter than W1, but both bands exhibit comparable heliocentric trends. Across all subsets, irrespective of band or dynamical classification, the faint-end of the brightness distribution remains consistently limited by the S/N $= 4$ detection threshold (Sect.~\ref{sec:sec312} and Appendix~\ref{sec:app5}). Figure~\ref{Figap5} shows the dependence of these limiting magnitudes on phase angle. 

%%%
LPCs consistently outshine SPCs. 
For the multi-epoch data, brighter LPCs and those observed at greater $r_{\rm H}$ tend to exhibit more gradual brightening trends, in agreement with earlier studies    \citep{Whipple1978,Meisel1982,Green1995,A'Hearn1995,Lisse2002,Meech2009,Holt2024,Lacerda2025}. 
Within $r_{\rm H} \lesssim 4$ au, where both groups are well sampled, their brightness ranges begin to overlap at the faint end of LPCs and bright end of SPCs, although LPCs overall still show shallower slopes. SPCs brighten and fade more rapidly near perihelion, in contrast to the smoother evolution seen in LPCs, a trend also noted in mid-infrared analyses by \citet{Lisse2002}. A post-$q$ detection bias is apparent as well, with SPCs more frequently detected after $q$ in both bands.

%%%%%%%%%%%%%%%%%%%%%
%%%%%%%%%%%%%%%%%%%%%

Figure \ref{Fig16} also shows that even within a given dynamical group, a single representative photometric behavior is difficult to define. Instead, a continuous gradient spans subgroups. Among LPCs, which cover a wide range in both brightness and $r_{\rm H}$, Halley-Type Comets (HTCs) tend to be fainter and exhibit steeper brightness slopes in both bands -- traits more typical of SPCs. HTCs, likely originating from the Oort Cloud, are dynamically older and have undergone multiple perihelion passages (\citealt{Wang2014} and references therein), while most Hyperbolic Comets (HCs) are thought to be dynamically new and likely on their first passage through the inner solar system \citep{Dybczynski2015}. The prolonged solar exposure of HTCs likely contributes to advanced surface processing, causing them to deviate from their primordial state.
This trend in infrared among LPCs is compatible with the optical analysis of the secular brightness evolution of 272 LPCs \citep{Lacerda2025}.

Taken as a whole, trends in brightness, activity, and photometric slopes suggest a continuum of evolutionary states that closely reflect each comet's dynamical history, particularly the number of perihelion passages since ejection from its source reservoir. From LPCs to SPCs -- and within LPCs, from HCs and NPCs to HTCs -- comets that have spent more time in the inner Solar System tend to be fainter (e.g., \citealt{Epifani2014}), display steeper brightness/fading slopes (\citealt{Lacerda2025} and references therein), and show more pronounced post-perihelion asymmetries. 
The strong pre- and post-perihelion asymmetry in the activity of JFC 67P/Churyumov-Gerasimenko, observed by the Rosetta mission and coordinated ground-based observation campaigns \citep{Guilbert-Lepoutre2014,Keller2015,Snodgrass2016,Opitom2017}, exemplifies the advanced evolutionary state of SPCs seen in this study. 
Covering a complementary range of heliocentric distances from the COSINE sample (mostly within 1 au compared to our coverage greater than 0.99 au), the heliocentric trends in water production rates derived from H Lyman-$\alpha$ emission for 61 comets observed by SOHO/SWAN over multiple decades \citep{Combi2019} highlight comparable evolutionary patterns: comets exhibit systematic variations in water production rates and their heliocentric dependence across dynamical classes, likely modulated by perihelion distance and by extension cumulative exposure to solar insolation, potentially indicative of progressive compositional processing.
% The pronounced pre- and post-perihelion asymmetry in the activity of JFC 67P/Churyumov-Gerasimenko from the Rosetta mission and coordinated ground-based observation campaigns \citep{Guilbert-Lepoutre2014,Keller2015,Snodgrass2016,Opitom2017} and the heliocentric trends in water production rates of the 61 comets observed by SOHO/SWAN over decades are consistent with the large-scale trends in this study \citep{Combi2019}.
This pattern underscores the central role of solar irradiation in shaping cometary evolution. Repeated solar heating depletes near-surface volatiles, pushing the active layer deeper and introducing thermal lag -- a delay between peak insolation and peak activity at a given $r_{\rm H}$ \citep{Epifani2009,Rubin2014,Kulyk2018}. As sublimating gases must traverse increasingly thick, dust-laden layers, mass loss becomes less continuous and more prone to episodic outbursts \citep{Gundlach2020}, consistent with the enhanced post-perihelion activity reported in previous studies (e.g., \citealt{Hughes1990,Ishiguro2016}). Over time, activity-driven fallback and surface migration processes of dust and ice would further enhance surface heterogeneity and amplify seasonal effects (e.g., \citealt{Keller2015,Marschall2020}), potentially explaining the reduced occurrence of super-volatile-driven activity and steeper brightening trends in evolved comets within $\sim$4 au \citep{Fulle2022}.

% 5. %%%%%%%%%%%%%%%%%%%%%%%%%%%%%%%%%%%%%%%%%%%%%%%%%%%%%%%%%%%%%%%%%%%%%%%%%%%%%%%%%%%%%%%%%%%%%%%%%%%%%%%%%%%%%
\section{Summary} \label{sec:summary}

This first paper in the COSINE project (Cometary Object Study Investigating their Nature and Evolution) introduces the project framework, presents new comet observations, and describes the resulting data products. The dataset comprises 1,633 coadded frames spanning 966 epochs for 484 comets with signal-to-noise ratios (S/N) greater than 4, obtained over the full 15-year WISE/NEOWISE mission. This represents the largest uniformly analyzed cometary compilation collected by a single instrument to date. Key findings are summarized below:

\begin{enumerate}
    \item[$\bullet$] From the known population, we selected 1,335 comets with reliable JPL/Horizons orbital solutions that entered within 11.5 au during the WISE/NEOWISE mission. Of these, 484 comets yielded coadded images with S/N $>$ 4, comprising 1.633 frames across 966 epochs used in our analysis.
    
    \item[$\bullet$] The detected sample includes 234 long-period comets (LPCs; HCs, NPCs, HTCs) and 250 short-period comets (SPCs; JFCs, ETCs), observed over heliocentric distances from 0.996 to 10.804 au and true anomalies from $-177.2$\arcdeg\ (pre-perihelion) to $+170.3$\arcdeg\ (post-perihelion). A total of 422 epochs were pre-perihelion and 544 epochs post-perihelion.

    \item[$\bullet$] LPCs are systematically brighter than SPCs in the W1 and W2 bands, which provide the most extensive coverage. No significant brightness differences are observed among subgroups within LPCs or SPCs. Many bright comets near perihelion were not submitted to the MPC, likely due to data quality flagging in WISE/NEOWISE processing.

    \item[$\bullet$] Activity and data quality were assessed using an S/N-based classification: Y-A, Y-B, Y-C (active) and N-A, N-B, N-C (inactive). Activity distributions vs. true anomaly generally follow Gaussian-like profiles centered near perihelion. LPCs show symmetric activity distributions, while SPCs exhibit a post-perihelion bias. Inactive distributions also peak near perihelion, though this is largely driven by N-B and N-C cases; N-A comets show no clear trend against perihelion.

    \item[$\bullet$] Comparisons of absolute magnitudes at fixed $r_{\rm H}$ in W1 and W2 bands highlight systematic differences between LPCs and SPCs. LPCs are detected out to $\sim$10 au, while SPCs (except 29P/Schwassmann–Wachmann) mostly remain within $\sim$4 au. LPCs populate the bright end of the magnitude distribution; SPCs populate the faint end. Among comets detected multiple times, LPCs with higher brightness and at larger $r_{\rm H}$ exhibit shallower slopes. HTCs show steeper slopes and fainter magnitudes, more closely resembling SPCs.

    \item[$\bullet$] Due to WISE/NEOWISE's fixed low-Earth orbit and narrow solar elongation range, observations sample a limited range of anti-solar and negative velocity vector orientations. These geometric constraints, which vary with orbital inclination, introduce strong biases in the distribution of measured feature orientations, underscoring the need to account for observing geometry in global morphological analyses.

    % \item[$\bullet$] The phase curves of inactive comets in W1 and W2 exhibit a consistent brightening trend toward smaller phase angles, reflecting a shared photometric behavior across both bands. This stability holds despite variations in bandpass and a range of nucleus sizes sampled at each phase-angle bin. MCMC modeling of 660 frames from 322 comets with phase angles $\leq$ 45\arcdeg\ yielded best-fit linear slope of $a = 0.187 \pm 0.034$ (W1) and $a = 0.154 \pm 0.034$ (W2). Quadratic fits returned best-fit linear slope of $a = 0.372 \pm 0.103$ and quadratic term of $b = -0.003 \pm 0.002$ (W1) and $a = 0.334 \pm 0.105$, $b = -0.003 \pm 0.002$ (W2).

    \item[$\bullet$] A clear gradient in photometric behaviors across comet populations -- especially between HCs/NPCs, HTCs, and SPCs -- illustrates the influence of solar irradiation on comet evolution. Increased frequency of perihelion passages results in systematically fainter brightness, steeper brightness evolution, and greater post-perihelion asymmetry, consistent with volatile depletion and long-term surface alteration. 
    
\end{enumerate}

This study provides a large statistical characterization of comet populations based on a single-instrument dataset. In future COSINE papers, we will decompose nucleus and coma components and explore the dust ejection history of comets and the physical nature of their nuclei in greater detail.

% Has it really? I'm guessing someone has anlso looking at Catalina and/or PanSTARRS comets. You could say "a large statistical characterization" but try to avoid hyperbole. Also, most of these objects were discovered by a hodgepodge of other optical surveys. For asteroids, this results in a pronounced bias against low albedo objects, which distorts efforts to model the underlying population. What would be the equivalent effects here?

% 8. Acknowledgments %%%%%%%%%%%%%%%%%%%%%%%%%%%%%%%%%%%%%%%%%%%%%%%%%%%%%%%%%%%%%%%%%%%%%%%%%%%%%%%%%%%%%%%%%%%%%
%\doi{10.5281/zenodo.15991}. 
%% Also note that the acknowledgment environment does not support long amounts of text. If you have a lot of people and institutions to acknowledge, do not use this command. Instead, create a new \section{Acknowledgments}.
\begin{acknowledgments}
We thank the anonymous reviewer for the helpful suggestions and warm support for the project.
This publication makes use of data products from the Wide-Field Infrared Survey Explorer, which is a joint project of the University of California, Los Angeles, and the Jet Propulsion Laboratory/California Institute of Technology, funded by the National Aeronautics and Space Administration. This publication also makes use of data products from NEOWISE, which is a project of the Jet Propulsion Laboratory/California Institute of Technology, funded by the Planetary Science Division of the National Aeronautics and Space Administration. This research has made use of the NASA/IPAC Infrared Science Archive, which is funded by the National Aeronautics and Space Administration and operated by the California Institute of Technology. This publication also makes use of software and data products from the NEO Surveyor, which is a joint project of the University of California, Los Angeles and the Jet Propulsion Laboratory/California Institute of Technology, funded by the National Aeronautics and Space Administration.
\end{acknowledgments}


\vspace{5mm}
\facilities{WISE, NEOWISE}


\software{
NumPy \citep{Harris2020}, SciPy \citep{Virtanen2020}, Matplotlib \citep{Hunter2007}, astropy \citep{astropycollab2022}, photutils \citep{Bradley2024}, reproject \citep{Robitaille2020}, Kete \citep{Dahlen2025}
}

% References %%%%%%%%%%%%%%%%%%%%%%%%%%%%%%%%%%%%%%%%%%%%%%%%%%%%%%%%%%%%%%%%%%%%%%%%%%%%%%%%%%%%%%%%%%%%%%%%%%%%%

% APPENDIX %%%%%%%%%%%%%%%%%%%%%%%%%%%%%%%%%%%%%%%%%%%%%%%%%%%%%%%%%%%%%%%%%%%%%%%%%%%%%%%%%%%%%%%%%%%%%%%%%%%%%%%
\appendix

\section{Detailed Procedure for Filtering Ill-Behaved Frames for Frame Selection and Coadding} \label{sec:app0}
\counterwithin{figure}{section}

The selection of frames for coaddition involves a systematic, multi-stage filtering process. For each comet, all acquired frames were grouped by observational epochs to form a candidate pool for generating a single stacked image. Each frame in this pool was then evaluated based on the following criteria:
\begin{enumerate}[itemsep=-.01in]
    \item[$\bullet$] The WISE/NEOWISE frame quality score\footnote{\texttt{qual\_frame}, as defined in the WISE/NEOWISE metadata documentation (\url{https://wise2.ipac.caltech.edu/docs/release/allsky/expsup/sec2\_4g.html\#qual\_frame}), where \texttt{qual\_frame} = 10 indicates the highest quality as graded by the quality assurance process} must be 10.
    \item[$\bullet$] The comet must lie within 11.5 au of the Sun at the time of observation.
    \item[$\bullet$] The frame must not have been taken while the spacecraft was within the South Atlantic Anomaly (SAA), as high radiation levels during these periods often rendered data unusable\footnote{\url{https://wise2.ipac.caltech.edu/docs/release/neowise/expsup/sec2\_1a.html\#saa_sep}}.
    \item[$\bullet$] No sources from the \texttt{ALLWISE} catalog (i.e., the static-source atlas) may lie within 7\arcsec\ of the predicted comet position\footnote{In addition to this automated filter, each individual frame used for coadding was visually inspected to ensure that bright sources, despite being outside the 7\arcsec\ exclusion zone, did not significantly contaminate the comet signal due to their extended influence.}.
    \item[$\bullet$] The image must exhibit an acceptable spacecraft attitude fit position. Specifically, the quadrature sum of the attitude error terms (\verb|ATT_ERRX|, \verb|ATT_ERRY|, and \verb|ATT_ERRZ| in header keywords) must not exceed $8 \times 10^{-5}$; otherwise, the frame is discarded due to measurable PSF smearing. The \verb|ATT_ERR| values are listed in the header of the individual input FITS files.
\end{enumerate}

Frames that passed the initial filtering were further subjected to statistical screening to exclude those compromised by moonlight contamination, cosmic ray hits, high background source density near the Galactic plane or center, or other quality-degrading factors. This statistical analysis proceeded in three stages:
First, the 20$^{\rm th}$ and 80$^{\rm th}$ percentile pixel values were computed for each (non-stacked) individual frame. The 20$^{\rm th}$ percentile, serving as a proxy for the background level, was compared against band-specific thresholds (20/18/75/30 DN for W1--W4, respectively). Frames exceeding these limits were often influenced by proximity to the Moon or regions of high background source density, especially critical in W1 and W2, and thus were excluded from the list of coadding.

Second, the difference between the 80$^{\rm th}$ and 20$^{\rm th}$ percentile values was calculated. Frames with excessive variance (greater than 100/100/4000/2000 DN for W1--W4, respectively) were excluded. This step effectively removed images affected by diffraction spikes, crowded Galactic plane regions, or intense radiation hits.

Third, after these frame-by-frame filters, a population-level (i.e., all images in each pool of coadding) outlier rejection was applied. The median pixel value of each remaining frame was computed. Across all retained frames, the global median and the corresponding Median Absolute Deviation (MAD) were calculated. Any frame whose median pixel value deviated by more than five times the MAD from the overall median was rejected. This removed frames that had passed previous filters, but still have visible artifacts.

For every potential coadd, a diagnostic image was made that displayed all of the input frames, along with the reason for rejection if the frame were to be rejected. These diagnostic images were used to validate that the rejection method was performing as expected. 
% No manual rejection of input images was performed.

Only frames that passed all filtering stages were coadded to produce the final stacked image. The list of contributing frames is recorded in the header of each final coadded FITS file.
\\

\section{Summary of Detected Comets} \label{sec:app1}
\counterwithin{figure}{section}
\counterwithin{table}{section}

\begin{figure}[tbh]
\centering
\includegraphics[width=0.57\textwidth]{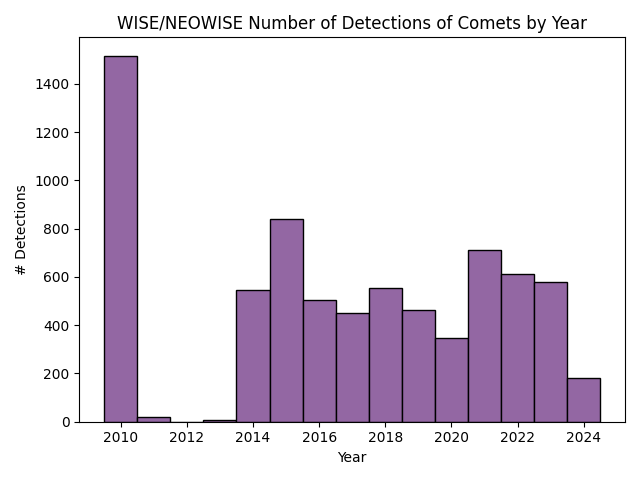}
\caption{Number of detected comets per year. Complete comet profiles are provided in Table~\ref{apt01}.
\label{Figap00}}
\end{figure}

Table \ref{apt01} provides observational details for 1,633 coadded frames corresponding to 484 detected comets across 966 epochs. Each entry includes dynamical classification and activity status at the time of observation. The yearly distribution of detected comets is shown in Figure \ref{Figap00}. Representative false-color images of comets observed during the Cryo phase are displayed in Figure \ref{Figap01}. These images include comets detected with a S/N greater than 4 in at least one band, without enforcing the S/N = 4 threshold across all bands. The W2, W3, and W4 bands are mapped to blue, green, and red, respectively.

%%%%%%%%%%%%%%%%%%%%%
\startlongtable
%\begin{longrotatetable}
\begin{deluxetable*}{c|cc|cccc|cccD}
\tabletypesize{\scriptsize} %\scriptsize
%\movetableright=-1in
\tablewidth{0pt} 
\tablecaption{Journal of observations of 484 all detected comets \label{apt01}}
\tablehead{
\colhead{Comet} \vline& 
\colhead{Subgroup} &
\colhead{NEC} \vline&
\colhead{Phase} &
\colhead{MJD} & 
\colhead{Band} &
\colhead{Activity} \vline& 
\colhead{$r_{\rm H}$ (au)}  &
\colhead{$\Delta$ (au)}  &
\colhead{$\alpha$ ($\arcdeg$)}  &
\multicolumn2c{$\nu$ ($\arcdeg$)}\\ %\multicolumn{2}{c}
Designation & (1) & (2) & (3) & (4) & (5) & (6) & (7) & (8) & (9) & \multicolumn2c{(10)}
}
%\colnumbers
\decimals
\startdata
\multicolumn{12}{c}{{\bf Long-Period comets (LPCs): 234}} \\
\cline{0-11}
12P & HTC & Y & R23 &  60064.09374 & W2 & N-C  &  4.684 &  4.585 &  12.42 & -134.92  \\ 
12P & HTC & Y & R23 &  60171.59677 & W1, W2 & Y-B, Y-A  &  3.613 &  3.380 &  16.22 & -127.23  \\ 
13P & HTC & Y & R24 &  60315.29700 & W1, W2 & N-B, Y-A  &  2.666 &  2.171 &  20.34 & -99.20  \\ 
38P & HTC & N & R18 &  58362.08752 & W1, W2 & Y-A, Y-A  &  1.818 &  1.546 &  33.70 & -43.38  \\ 
38P & HTC & N & R19 &  58574.77516 & W1, W2 & Y-B, Y-A  &  2.335 &  1.784 &  23.57 &  72.04  \\ 
C/2005 L3 & NPC & N & Cryo &  55337.12536 & W3, W4 & Y-B, Y-A  &  8.190 &  8.018 &  7.05 &  68.50  \\ 
C/2006 OF2 & HC & N & Cryo &  55326.60967 & W3, W4 & Y-C, Y-A  &  6.306 &  6.123 &  9.17 &  103.17  \\
\enddata
\tablecomments{Table headers: (1) Subgroup: dynamical classification defined in Section \ref{sec:sec311}. Long-Period Comets (LPCs) contain HCs (Hyperbolic Comets), NPCs (Near-Parabolic Comets), and HTCs (Halley-Type Comets). Short-Period Comets (SPCs) contain JFCs (Jupiter-Family Comets) and ETCs (Encke-Type Comets); (2) NEC: Near-Earth Comets; (3) Phase: WISE/NEOWISE Mission phase (start/end dates in Fig. \ref{Fig01}). RXX denotes Reactivation phase in 20XX (e.g., R19 = Reactivation phase in 2019); (4) MJD: Modified Julian Date of the each midpoint; (5) Band: W1 (centered at 3.4 $\mu$m), W2 (4.6 $\mu$m), W3 (12 $\mu$m), and W4 (22 $\mu$m); (6) Activity: comet activity status. ``Y" (Yes, active) or ``N" (No, inactive) for each band, which are further qualified as Grade A (S/N $\ge$ 20), Grade B (B: 10 $\le$ S/N $<$ 20), and Grade C (C: 4 $\le$ S/N $<$ 10). Detailed criteria are given in Sections \ref{sec:sec24} and \ref{sec:sec313}; (7) $r_{\rm H}$: midpoint heliocentric distance in au used for coadding; (8) $\Delta$: midpoint geocentric distance in au used for coadding; (9) $\alpha$: midpoint phase angle (Sun--comet--observer) in degrees; and (10) $\nu$: midpoint true anomaly in degrees (negative = pre-perihelion, positive = post-perihelion).
}
\end{deluxetable*}
%\end{longrotatetable}
%%%%%%%%%%%%%%%%%%%%%

%%%%%%%%%%%%%%%%%%%%%
%%%%%%%%%%%%%%%%%%%%%

%%%%%%%%%%%%%%%%%%%%%
\begin{figure*}[!b]
\centering
\includegraphics[width=0.90\textwidth]{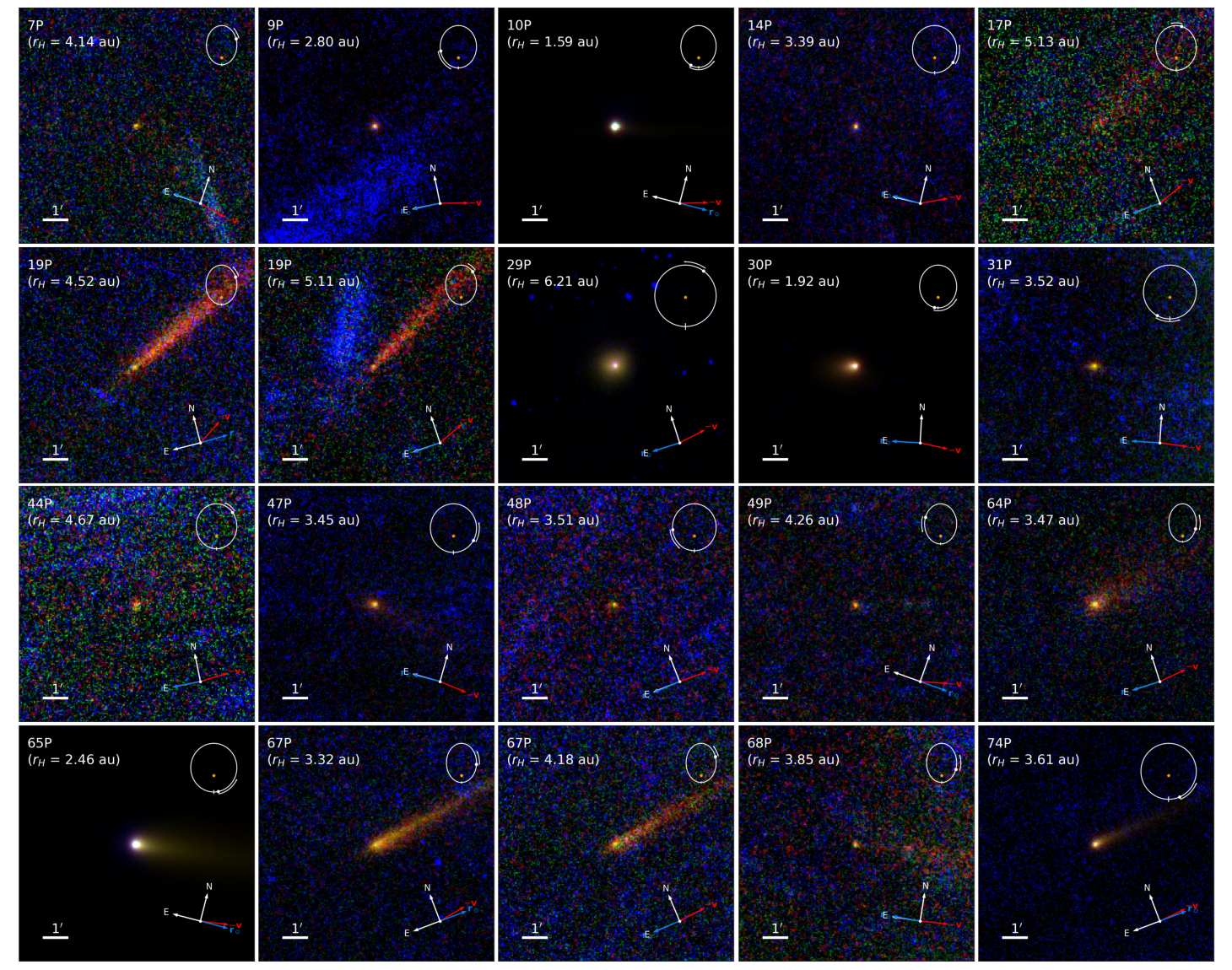}
\caption{False-color composite of 178 comets observed during the Cryo phase. This sample does not represent the complete Cryo-phase dataset. The W2, W3, and W4 bands are mapped to blue, green, and red, respectively, with scaling and normalization consistent with Figure \ref{Fig04}. Each panel includes the comet's heliocentric distance ($r_{\rm H}$ in au) at the time of observation, along with directional vectors indicating Ecliptic North (N), Ecliptic East (E), the negative velocity vector ($-{\bf v}$), and the anti-solar vector (${\bf r}_\odot$). A 1\arcmin\ scale bar is also provided. The object's position in its orbit, given by the true anomaly, is illustrated in the inset diagram at the upper right.
\label{Figap01}}
\end{figure*}
%%%%%%%%%%%%%%%%%%%%%

%\renewcommand{\thefigure}{\arabic{figure} (Cont.)}
\addtocounter{figure}{-1}
%%%%%%%%%%%%%%%%%%%%%
\begin{figure*}[!htb]
\centering
\includegraphics[width=0.90\textwidth]{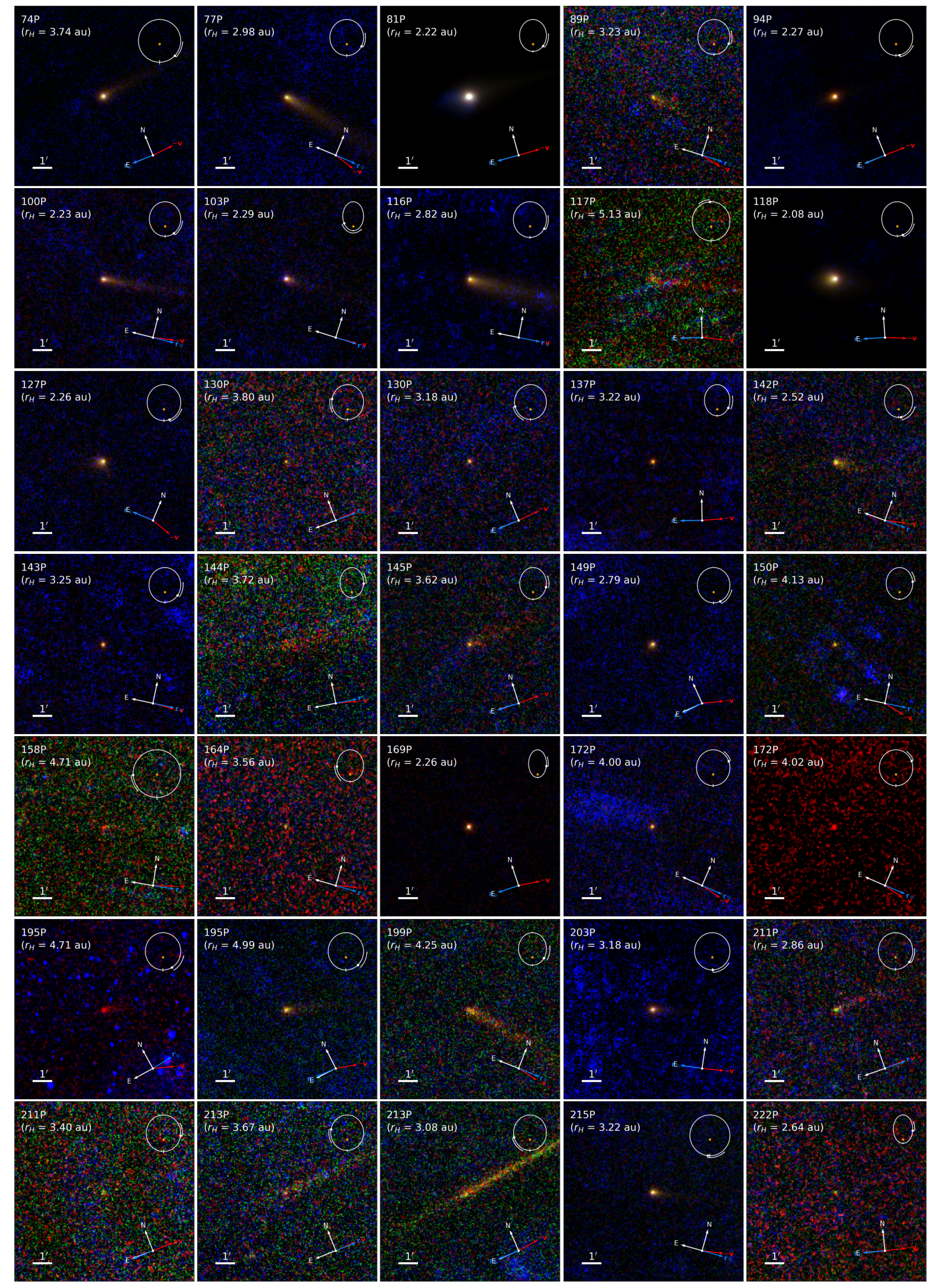}
\caption{Continued.}
\end{figure*}
%%%%%%%%%%%%%%%%%%%%%
\addtocounter{figure}{-1}
%%%%%%%%%%%%%%%%%%%%%
\begin{figure*}[!htb]
\centering
\includegraphics[width=0.90\textwidth]{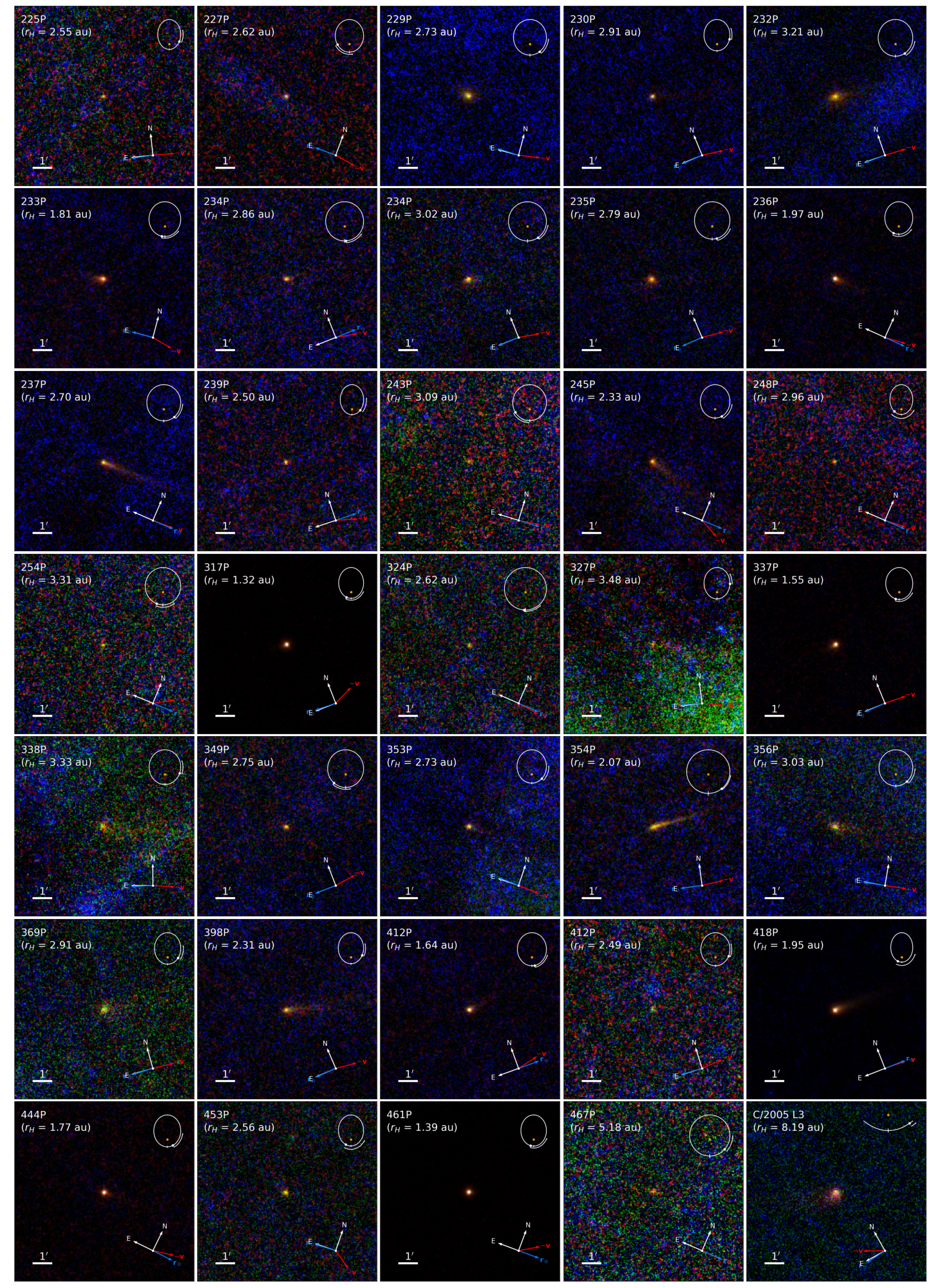}
\caption{Continued.}
\end{figure*}
%%%%%%%%%%%%%%%%%%%%%
\addtocounter{figure}{-1}
%%%%%%%%%%%%%%%%%%%%%
\begin{figure*}[!htb]
\centering
\includegraphics[width=0.90\textwidth]{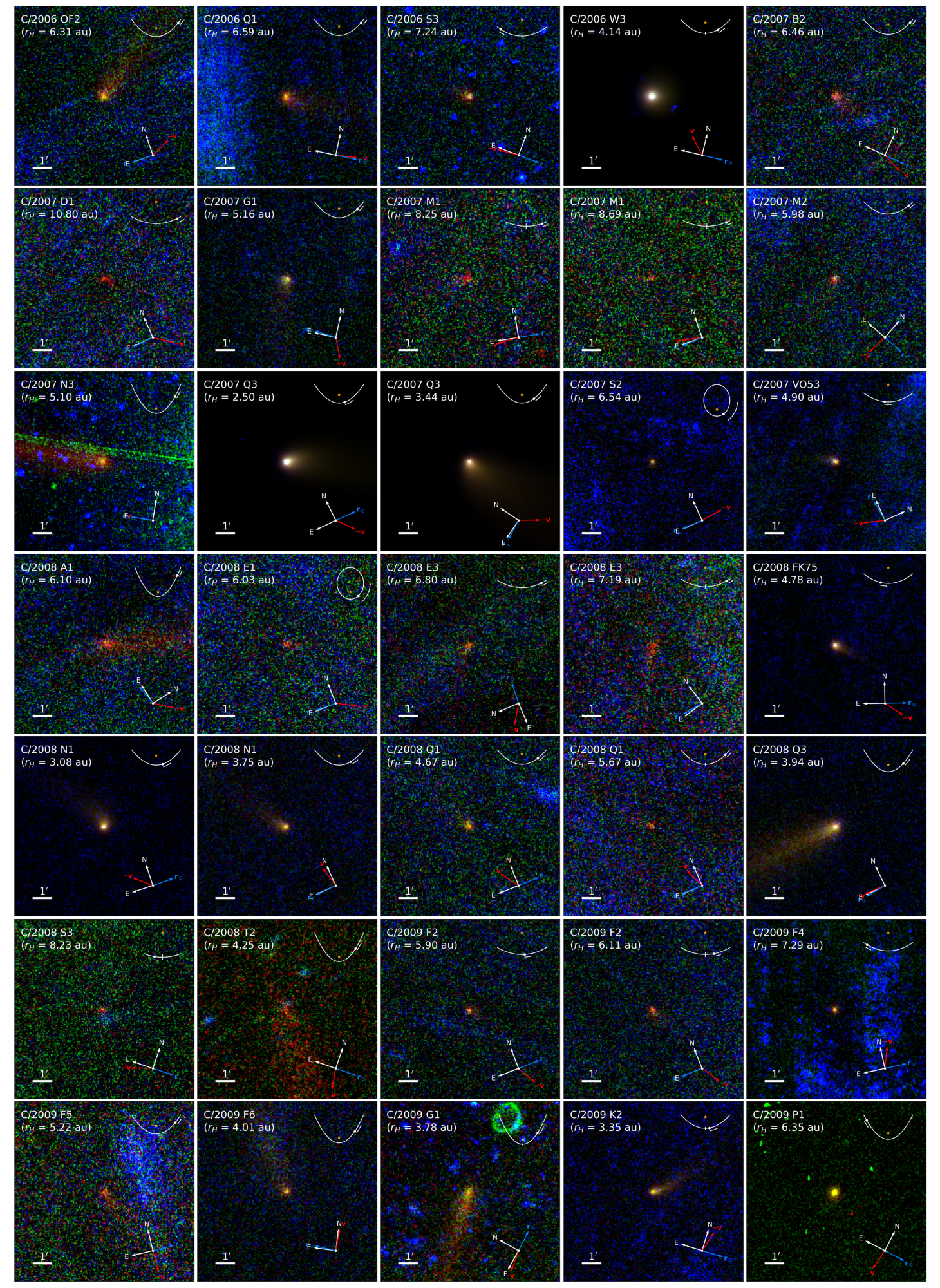}
\caption{Continued.}
\end{figure*}
%%%%%%%%%%%%%%%%%%%%%
\addtocounter{figure}{-1}
%%%%%%%%%%%%%%%%%%%%%
\begin{figure*}[!htb]
\centering
\includegraphics[width=0.90\textwidth]{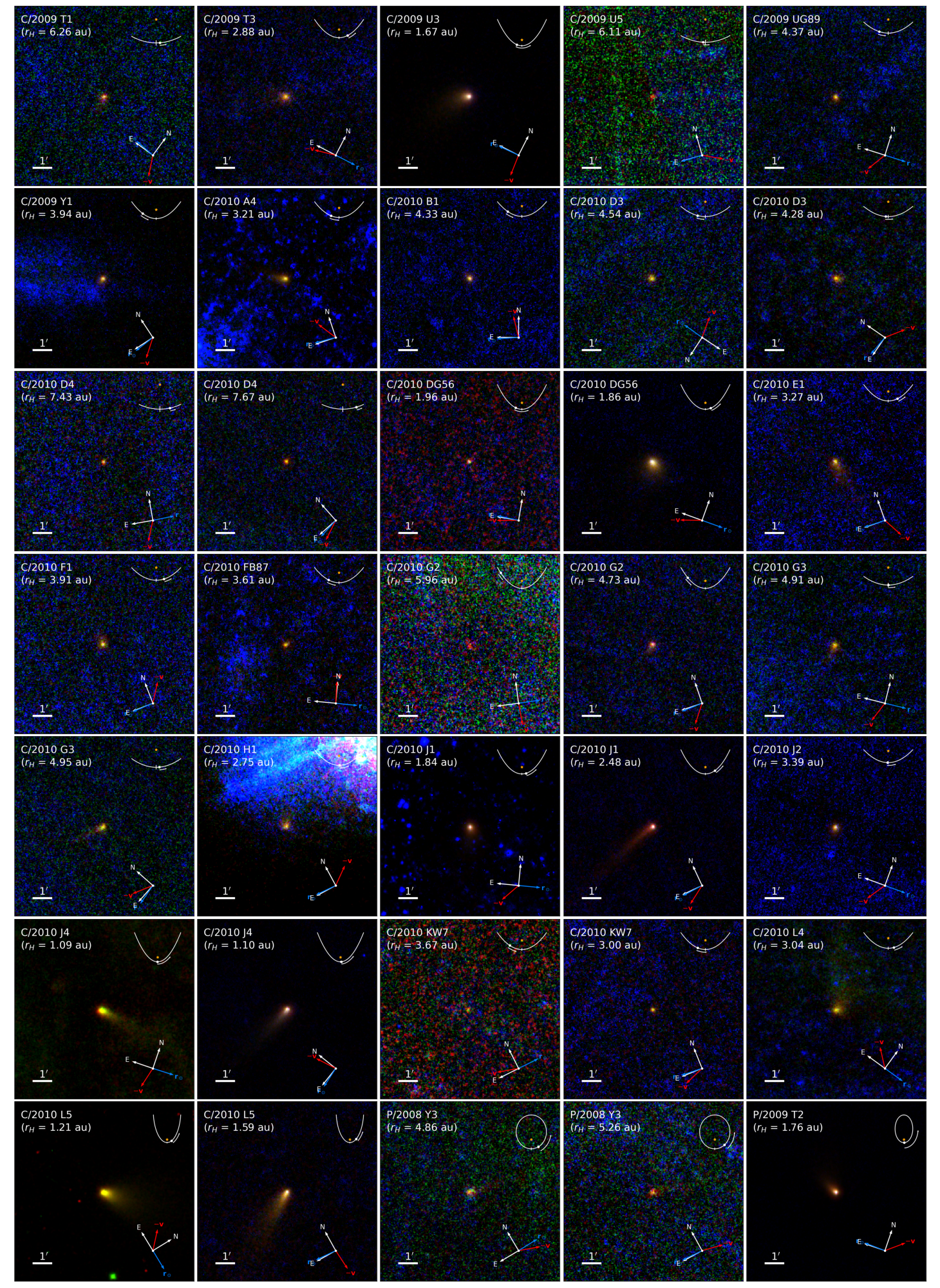}
\caption{Continued.}
\end{figure*}
% %%%%%%%%%%%%%%%%%%%%%
\addtocounter{figure}{-1}
%%%%%%%%%%%%%%%%%%%%%
\begin{figure*}[!htb]
\centering
\includegraphics[width=0.90\textwidth]{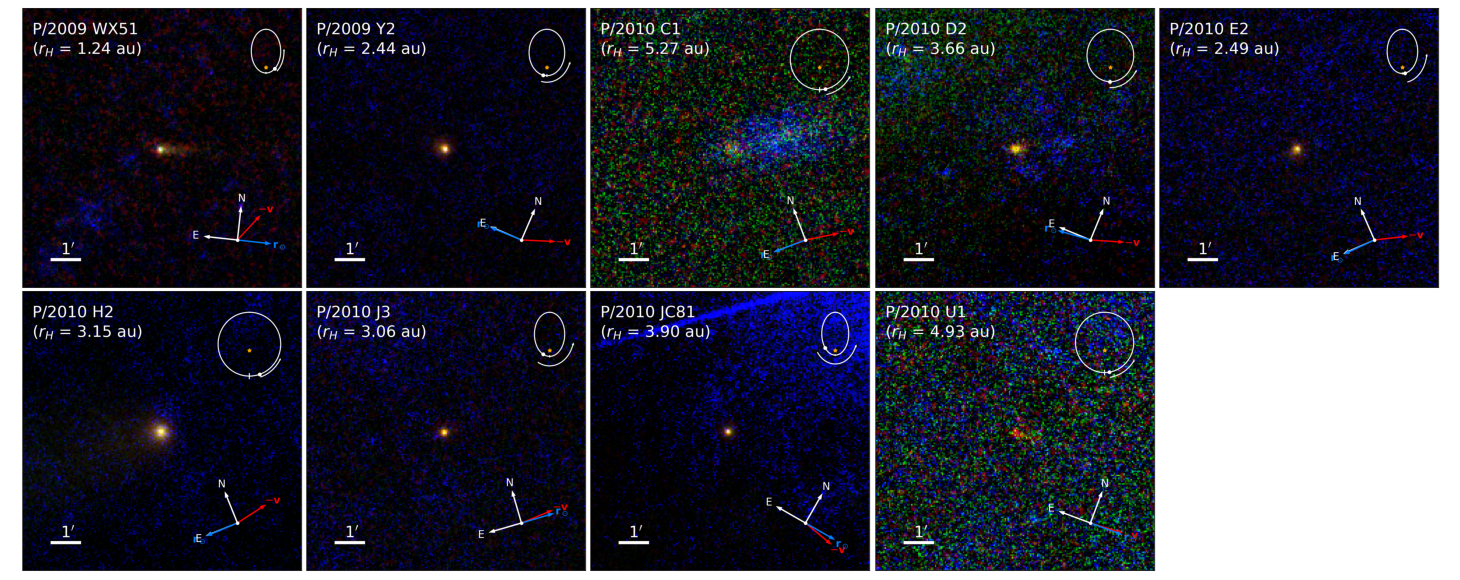}
\caption{Continued.}
\end{figure*}
% %%%%%%%%%%%%%%%%%%%%%
% \renewcommand{\thefigure}{\arabic{figure}}
% % \clearpage

\section{Photometric Procedure for Signal-to-Noise Ratio (S/N) Estimation} \label{sec:app2}
\counterwithin{figure}{section}

The signal-to-noise ratio (S/N) in individual stacked images was estimated through a two-step photometric procedure involving background characterization and source flux measurement. We adopted the standard definition of S/N as outlined by \citet{Howell2012} and implemented in \texttt{astropy}\footnote{\url{https://docs.astropy.org/en/stable/api/astropy.stats.signal\_to\_noise\_oir\_ccd.html}}.

In order to estimate the background, multiple rectangular apertures (patches) were sampled in a circle around the comet's position. Since each patch may have a local increase in background counts due to the extended nature of an active comet, we chose statistical methods that are less sensitive to the outliers caused by the comet's tail/trail. A bootstrap statistical method was used to estimate the background value and its uncertainty. The median of the mean values of selected background patches (Sect. \ref{sec:sec23}) was used to define the background level ($B_{\text{med}}$), while the uncertainty of this background noise ($\sigma_B$) was computed by taking median of the Median Absolute Deviation (MAD), which is a robust statistic of uncertainty that is less sensitive to outliers. The MAD was then multiplied by 1.4826 to provide an estimate of the true standard deviation ($\sigma_{\rm bkg}$).  These parameters served as the foundation for subsequent flux measurements.

The source flux was measured within a circular aperture of radius $r_{\rm ap}$ (see Sects. \ref{sec:sec23}--\ref{sec:sec25}), centered on the comet. Because the aperture includes both source and background flux, the background contribution was first estimated as:
\begin{equation}
    F_{\rm bkg} = B_{\text{med}} \times A_{\rm ap}~,
\label{eq:ap02-1}
\end{equation}
\noindent where $A_{\rm ap}$ is the pixel area of the aperture. The net (background-subtracted) source flux was then:
\begin{equation}
    F_{\rm source} = F_{\rm raw} - F_{\rm bkg}~,
\label{eq:ap02-2}
\end{equation}
\noindent where $F_{\rm raw}$ is the total uncorrected flux measured in the aperture (Sect. \ref{sec:sec25}). 

The associated flux uncertainty, driven by background noise, was estimated as:
\begin{equation}
    \sigma_{\rm F} = \sqrt{F_{\rm raw} +  A_{\rm ap}~ \times (\sigma_{\rm bkg}~^2 + B_{\text{med}})}
\label{eq:ap02-3}
\end{equation}
\noindent The final S/N was computed using:
\begin{equation}
    {\rm S/N} = \frac{F_{\rm source}}{\sigma_{\rm F}}~.
\label{eq:ap02-4}
\end{equation}

This approach ensures robust S/N estimation by systematically incorporating both spatial background variability and instrumental noise. Pixels with \verb|NaN| values were masked prior to analysis. S/N was evaluated along the primary angle (PA), defined as the position angle of the extended feature in active comets, measured from the photocenter (Section~\ref{sec:sec24}).

\begin{figure}[!hbt]
\centering
\includegraphics[width=1\textwidth]{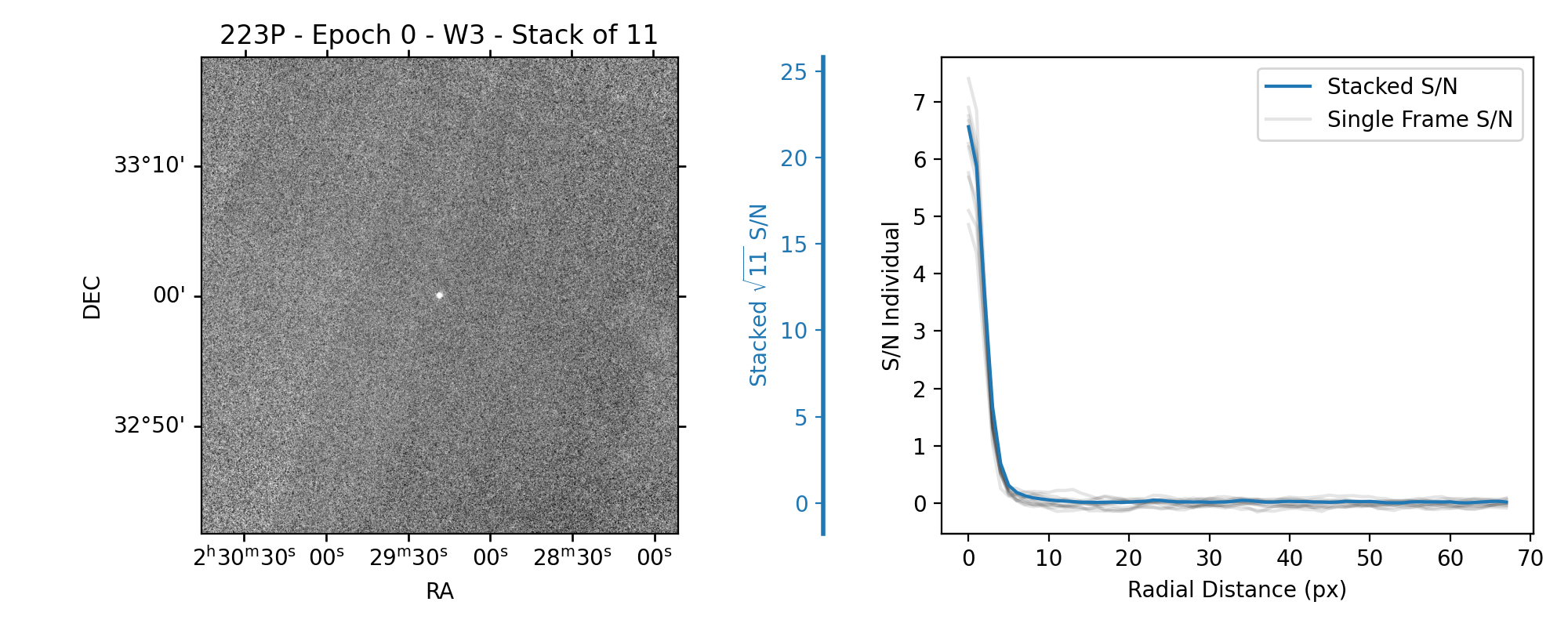}
\caption{Example of comet 223P/Skiff demonstrating the internal consistency of our stacking technique and photometry. Left: Coadded W3-band image from 11 frames taken on MJD 55417.23328 (Cryo phase) at $r_{\rm H}$ = 2.420 au and $\Delta$ = 2.186 au. Right: Radial S/N profiles for the 11 individual frames (black curves) and the final coadded image (blue curve).  Note that the second y-axis (blue) is scaled by exactly $\sqrt{11}$, the number of individual frames used to generate the coadd.
\label{Figap2}}
\end{figure}

Figure \ref{Figap2} illustrates the methodology using comet 223P/Skiff. The left panel shows the coadded W3-band image derived from 11 individual frames taken on MJD 55417.23328 during the Cryo phase ($r_{\rm H} = 2.420$ au, $\Delta = 2.186$ au). The right panel presents radial S/N profiles: individual frames are shown in black, and the final stacked profile is overplotted in blue. The agreement between individual and coadded profiles confirms the internal consistency of our S/N estimation and validates the overall stacking--photometric process. Coadding was performed using a custom implementation of \texttt{reproject}, optimized for tracking moving targets (Sections~\ref{sec:sec23} and \ref{sec:sec25}).

%%%%%%%%%%%%%%%%%%%%%
%%%%%%%%%%%%%%%%%%%%%

\section{V-band magnitude Distribution of Comets in Figure ~\ref{FIG10}} \label{sec:app3}
\counterwithin{figure}{section}

This section presents the V-band magnitude distribution for the comets displayed in Figure \ref{FIG10}, providing a reference for astronomers planning follow-up observations in the visible regime for targets initially identified in infrared surveys. 

For all detected comets in Figure \ref{FIG10}, we retrieved the \texttt{$M$1} and \texttt{$K$1} photometric parameters from JPL/Horizons. The total V-band magnitude at the time of observation was then computed using the standard relation:
\begin{equation}
\begin{matrix}
    M_{\rm V} = M{\rm 1} + K{\rm 1} \times \log_{10}(r_{\rm H}) + 5 \times \log_{\rm 10}(\Delta) + \beta\alpha~,
\end{matrix}
\label{eq:eqapp3}
\end{equation}
\noindent where $r_{\rm H}$ is the heliocentric distance in au, $\Delta$ is the observer--comet distance in au, $\alpha$ is the phase angle in degrees, and $\beta$ is the phase correction coefficient. A value of $\beta$ = 0.035 mag/deg was adopted, consistent with the assumption used in Section~\ref{sec:sec322}. The calculated magnitudes were verified to be consistent with those provided directly by Horizons.

\begin{figure}[!hbt]
\centering
\includegraphics[width=0.6\textwidth]{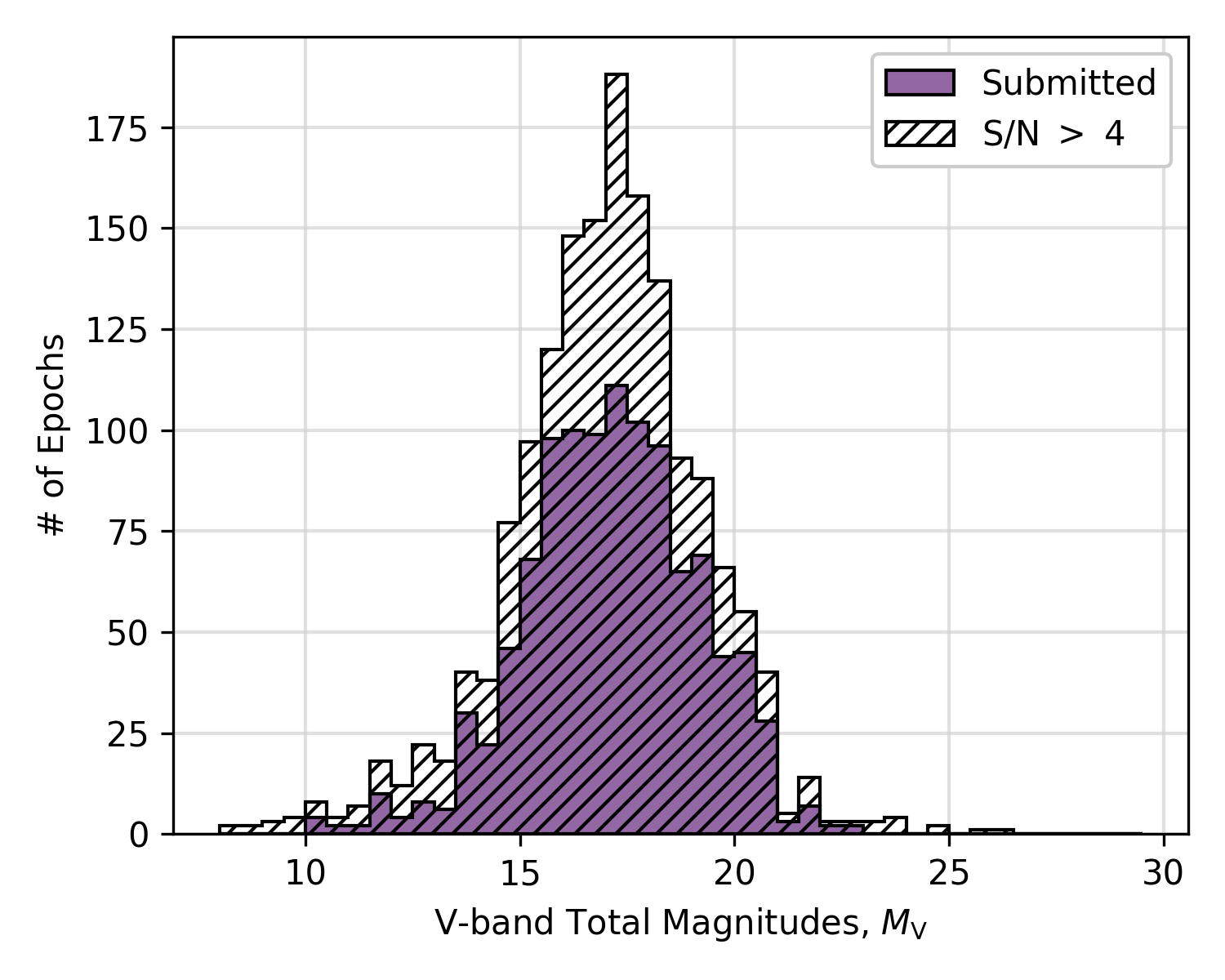}
\caption{Number of epochs as a function of estimated V-band magnitude for the comets shown in Figure \ref{FIG10}, using the same color scheme.
\label{Figap3}}
\end{figure}

Figure~\ref{Figap3} illustrates the resulting distribution of V-band magnitudes for the comets in Figure~\ref{FIG10}, plotted by the number of observing epochs. The same color scheme is used for consistency.

%%%%%%%%%%%%%%%%%%%%%
%%%%%%%%%%%%%%%%%%%%%

\section{Potential Gas Emission Features Within the W1 and W2 Bandpasses} \label{sec:app4-0}
\counterwithin{figure}{section}

In this analysis, we have assumed that the W1 and W2 signals are primarily dominated by sunlight scattered by dust particles, either in the coma/tail, or on the nucleus surface (Sect.~\ref{sec:sec322}). However, we recognize the potential contribution of gas emission features within these bandpasses. Such contributions may reach up to several tens of percent of the total flux in some cases (e.g., \citealt{Ootsubo2012}).

\begin{figure}[!hbt]
\centering
\includegraphics[width=0.6\textwidth]{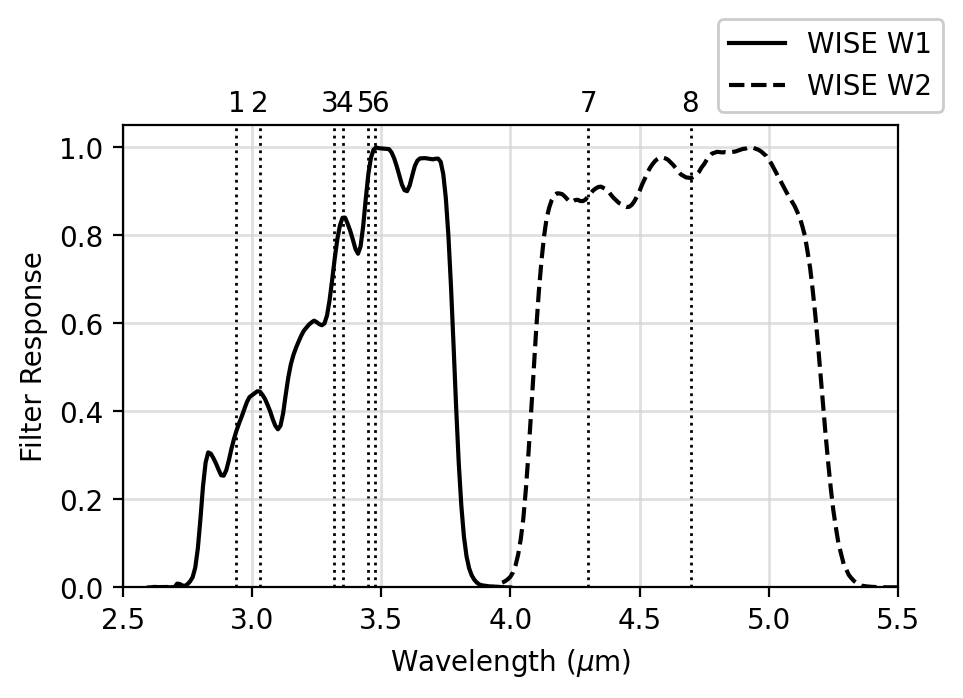}
\caption{Filter transmission curves for the W1 and W2 bands from the WISE Explanatory Supplement, overlaid with the mean central wavelengths of prominent cometary gas emission features \citep{DelloRusso2011,Reach2013,Paganini2015}: 1 - H$_2$O at $\sim$2.94 $\mu$m; 2 - HCN at $\sim$3.03 $\mu$m; 3 - CH$_4$ at $\sim$3.32 $\mu$m; 4 - C$_2$H$_6$ $\nu$7 at $\sim$3.35 $\mu$m; 5 - C$_2$H$_6$ $\nu$5 at $\sim$3.45 $\mu$m); 6 - CH$_3$OH at $\sim$3.48 $\mu$m; 7 - CO$_2$ at $\sim$4.3 $\mu$m; 8 - CO at $\sim$4.7 $\mu$m.}
\label{Filter_trans}
\end{figure}

Figure~\ref{Filter_trans} overlays the W1 and W2 filter response curves with the mean central wavelengths of major gas emission bands frequently observed in active comets (e.g., \citealt{Mumma2011,Ootsubo2012}). The mean wavelengths, compiled from \citet{DelloRusso2011}, \citet{Reach2013}, and \citet{Paganini2015}, include H$_2$O ($\sim$2.94 $\mu$m), HCN ($\sim$3.03 $\mu$m), CH$_4$ ($\sim$3.32 $\mu$m), C$_2$H$_6$ ($\nu$7 band at $\sim$3.35 $\mu$m and $\nu$5 band at $\sim$3.45 $\mu$m), CH$_3$OH ($\sim$3.48 $\mu$m), CO$_2$ ($\sim$4.3 $\mu$m), and CO ($\sim$4.7 $\mu$m). Filter response profiles for the WISE bands are available in the WISE Explanatory Supplement\footnote{\url{https://wise2.ipac.caltech.edu/docs/release/prelim/expsup/sec4\_3g.html\#WISEZMA}}.

%%%%%%%%%%%%%%%%%%%%%
%%%%%%%%%%%%%%%%%%%%%

These molecular bands can affect non-negligibly to the broadband fluxes measured in W1 and W2, with their influence depending on heliocentric distance ($r_{\rm H}$). In general, CO is a known driver of cometary activity at $r_{\rm H} \gtrsim 6$~au, CO$_2$ dominates around 4~au, and H$_2$O sublimation governs most activity within 4~au \citep{Womack2017,Fulle2022,Fraser2024}. Consequently, the gas-to-dust ratio in these bands is expected to vary with the observing circumstances, and without spectrally resolved data, disentangling gas emission from dust-scattered light remains challenging, limiting precise interpretation of W1 and W2 fluxes and their ratios.

For instance, \citet{Reach2013} analyzed 23 comets using Spitzer observations and found elevated 4.5-$\mu$m/3.6-$\mu$m flux ratios in some comets with spherical morphologies, consistent with significant CO+CO$_2$ emission. They categorized comets as CO$_2$-rich or -poor based on this ratio. Similarly, \citet{Ootsubo2012} performed 2.5--5 $\mu$m spectroscopy of 18 comets with AKARI, and, along with \citet{Reach2013}, investigated the CO$_2$/H$_2$O production rate as a function of $r_{\rm H}$. (In Spitzer data, similar to WISE/NEOWISE, CO$_2$ and CO bands fall into a wide-band filter and thus assumed CO$_2$ is dominant to CO). While no strong trends were found within 2.5~au, the CO$_2$/H$_2$O ratio increases at larger distances due to suppressed H$_2$O activity.
%The median value of CO$_2$/H$_2$O ratio observed was 17 \% \citep{Ootsubo2011}. 
\citet{Bauer2015} also reported favorable CO+CO$_2$ observations in LPCs at larger r$_{\rm H}$, suggesting that LPCs may retain more CO, whereas CO$_2$ abundances appear similar between LPCs and SPCs.

These examples underscore the need for caution in interpreting broadband fluxes, as multiple volatile and dust components may contribute to the observed signals in W1 and W2.

\section{Phase Angle Variation During the WISE/NEOWISE Mission} \label{sec:app4}
\counterwithin{figure}{section}

\begin{figure}[!hbt]
\centering
\includegraphics[width=0.5\textwidth]{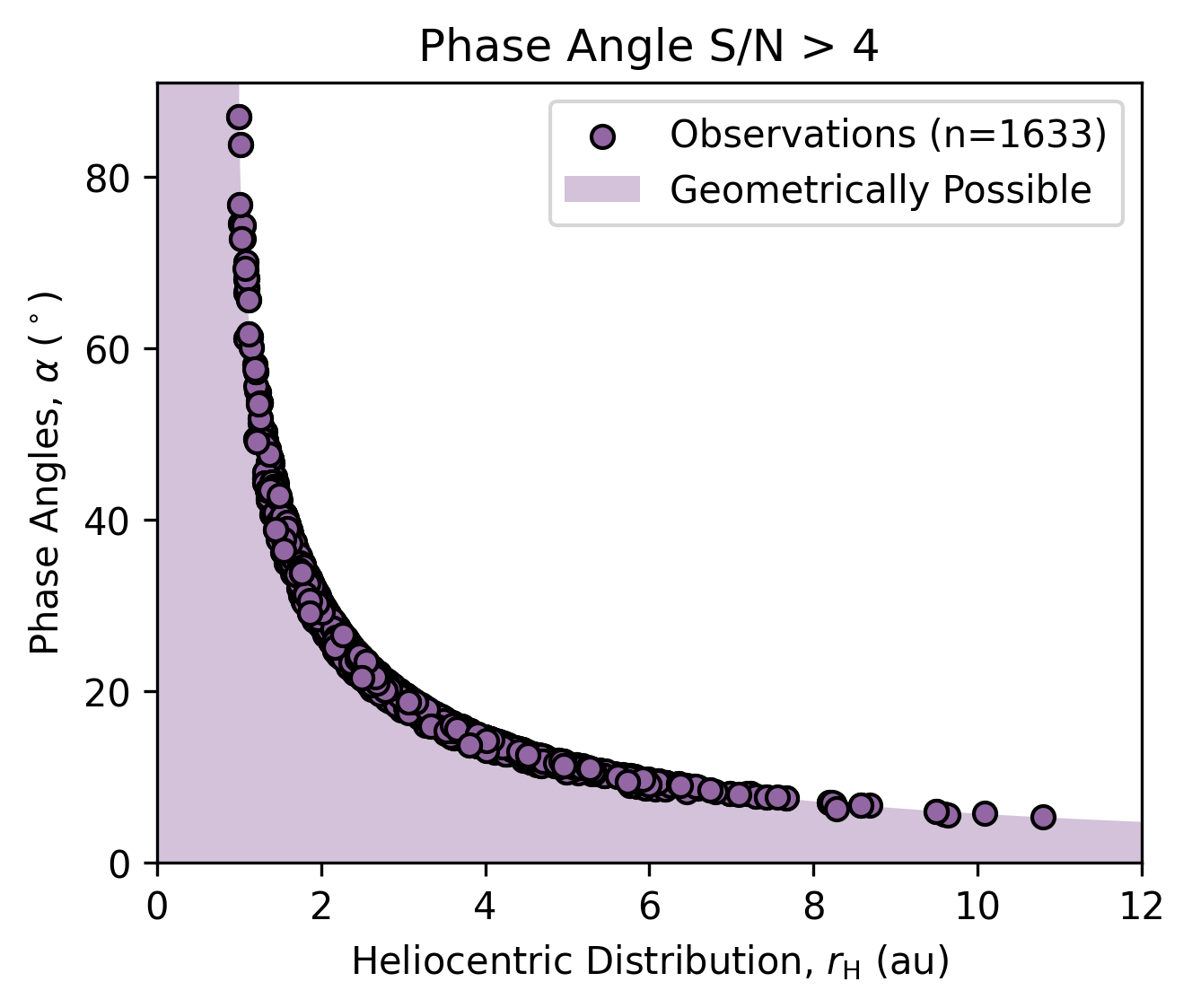}
\caption{Phase-angle distribution of 1,633 comet detection frames (circle symbols) from WISE/NEOWISE. The purple-shaded region indicates the geometrically allowed range of phase angles at a given heliocentric distance ($r_{\rm H}$), assuming the comet is located at that $r_{\rm H}$. Owing to WISE/NEOWISE's near-Earth orbit and its fixed $\sim$90\arcdeg\ solar elongation constraint, the mission preferentially sampled the upper end of the available phase-angle range.
\label{Figap4}}
\end{figure}

Figure \ref{Figap4} presents the distribution of phase angles for all 1,633 comet detection frames at their respective observation times, as listed in Table \ref{apt01}. The plot illustrates how the observing geometry of the WISE/NEOWISE mission influenced the phase-angle sampling of cometary observations.

%%%%%%%%%%%%%%%%%%%%%
%%%%%%%%%%%%%%%%%%%%%

\section{Impact of WISE/NEOWISE Detectability on Large-Scale Photometric Trends} \label{sec:app5}
\counterwithin{figure}{section}

The brightness distribution of detected comets as a function of heliocentric distance ($r_{\rm H}$), shown in Figure \ref{Fig13}, displays a distinct curvature at the faint end, even after geometric correction (dashed curves). This curvature suggests the presence of an intrinsic detection threshold imposed by the WISE/NEOWISE instrumentation, consistent with the behavior of infrared observatories described in \citet{Epifani2009}. The S/N = 4 detection limits in Figures \ref{Fig13} and \ref{Fig16} were derived using our measured band magnitude limits (16.3, 14.8, 11.4, and 6.6 mag for W1, W2, W3, and W4, respectively; Fig. \ref{FIG10}). These values were combined with the observing geometry of the spacecraft -- assumed to be located 1 au from the Sun with a fixed solar elongation of $\sim$90\arcdeg. Using simple trigonometric calculations, we computed the expected range of absolute magnitudes as a function of $r_{\rm H}$ for each band. The resulting detectability thresholds well encompass the faint-end distribution of the observed comets.

To further investigate how this detection limit influences large-scale photometric trends, we analyzed the phase function: the variation of a comet's reduced magnitude, ($m(1,1,\alpha)$), corrected for $r_{\rm H}$ and $\Delta$, with respect to the phase angle $\alpha$. This function is fundamental for characterizing the surface scattering properties of cometary nuclei. We modeled a global phase function using comets classified as inactive (activity label ``N") in all four bands. For each, we computed:
\begin{equation}
    m(1,1,\alpha) = m_{\rm app} - 5\log_{10}(r_{\rm H}\Delta)~,
\label{eq2}
\end{equation}
\noindent with variables defined as in Equation \ref{eq1}. Restricting the sample to inactive comets ensures that $m(1,1,\alpha)$ reflects only phase-angle-dependent scattering effects, excluding contributions from coma or other transient activity.

\begin{figure}[tbh]
\centering
\includegraphics[width=0.77\textwidth]{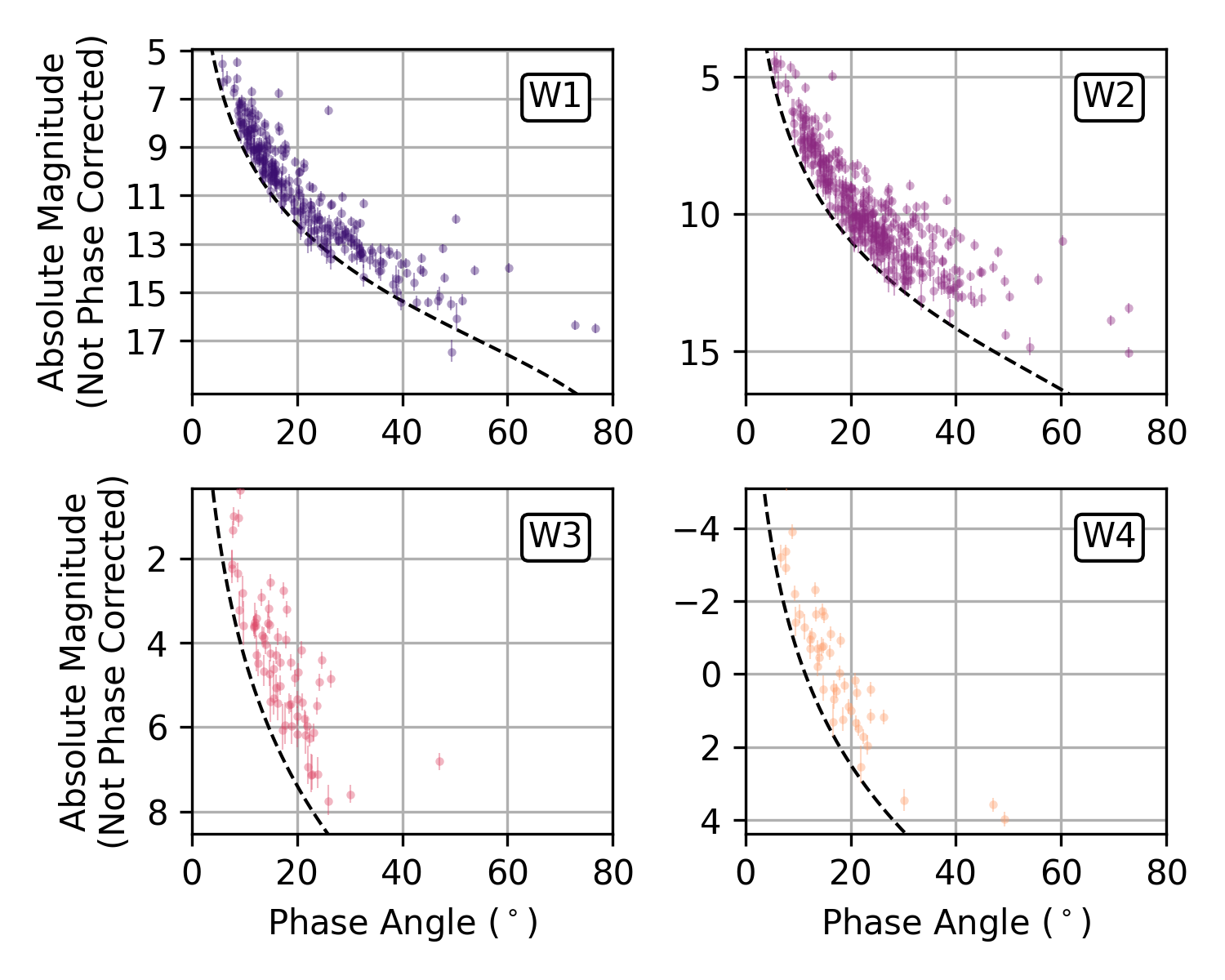}
\caption{Distributions of inactive comets in all 4 bands as a function of phase angle. The data points represent $m(1,1,\alpha)$ values derived per frame, and S/N = 4 trendlines are overlaid for each band.
\label{Figap5}}
\end{figure}

Figure \ref{Figap5} presents the resulting distributions in both bands. A statistically significant brightening trend is observed with decreasing $\alpha$. Assuming comparable albedos across the population, the observed 2--4 mag spread at a given $\alpha$ can be attributed primarily to variations in nucleus size, with brighter objects corresponding to larger nuclei. No discernible differences were found among the inactivity grades (N-A, N-B, N-C), and we therefore combined all of the inactive frames per band for this analysis. The overall morphology of the magnitude-phase angle distribution is clearly bounded at the faint end by an upward truncation. The dashed curves here represent the same detectability limits as those shown in Figures \ref{Fig13} and \ref{Fig16}, but plotted against phase angle and \emph{without} applying a phase correction. The predicted detection threshold closely matches the observed cutoff in faint-end magnitudes.

%%%%%%%%%%%%%%%%%%%%%
%%%%%%%%%%%%%%%%%%%%%

% These results emphasize the importance of accounting for instrumental sensitivity and geometric constraints when interpreting population-level photometric trends. Although this limitation affects global inferences, it can be mitigated through analysis of individual comets with repeated observations spanning a range of phase angles. Such object-specific phase curve studies will be conducted in future installments of the COSINE project.

\end{document}